\documentclass[journal,twocolumn,romanappendices]{IEEEtran}

\usepackage[T1]{fontenc}
\usepackage{url}
\usepackage{ifthen}
\usepackage{cite}
\usepackage[cmex10]{amsmath} 
\usepackage{longtable}
\usepackage{makecell}
\usepackage{algorithm,algorithmicx}
\usepackage[noend]{algpseudocode}
\usepackage{color}
\usepackage[normalem]{ulem}
                             
\usepackage{amssymb,amsthm}
\usepackage{mathtools,bbm}  
\usepackage{xifthen,xparse}
\usepackage{hyperref}  
\usepackage{float}
\usepackage{hhline}
\usepackage{caption}
\usepackage{subcaption}
\usepackage{rotating}

\usepackage{stmaryrd}

\usepackage{booktabs,tabularx}
\newcolumntype{C}{>{\centering\arraybackslash}X}

\usepackage[table]{xcolor}

\definecolor{dkgreen}{rgb}{0,0.6,0}
\definecolor{gray}{rgb}{0.5,0.5,0.5}
\definecolor{mauve}{rgb}{0.58,0,0.82}

\usepackage{pgfplots}
  \pgfplotsset{compat=newest}
  \usetikzlibrary{plotmarks}
  \usetikzlibrary{arrows.meta}
  \usepgfplotslibrary{patchplots}
  \usepackage{grffile}
  \usepackage{amsmath}

\usepackage{tikz,pgfplots,tkz-graph,color}
\pgfplotsset{compat=newest}
\pgfplotsset{plot coordinates/math parser=false}
\usetikzlibrary{decorations.markings,calc,positioning,matrix}
\usepackage{quantikz}

\usepackage{textcomp}

\allowdisplaybreaks

\newtheorem{theorem}{Theorem}
\newtheorem{corollary}[theorem]{Corollary}
\newtheorem{remark}[theorem]{Remark}
\newtheorem{lemma}[theorem]{Lemma}
\newtheorem{conjecture}[theorem]{Conjecture}
\newtheorem{example}{Example}

\newenvironment{example*}
  {\addtocounter{example}{-1}\example}
  {\endexample}

\newif\ifnotes
\notestrue

\newcommand{\llbr}{[\![}
\newcommand{\rrbr}{]\!]}

\begin{document}

\title{Tailoring Fault-Tolerance to\\ Quantum Algorithms}

 \author{%
   \IEEEauthorblockN{Zhuangzhuang Chen and Narayanan Rengaswamy}%
   \thanks{Accepted to IEEE Journal on Selected Areas in Information Theory, vol. 6, pp. 311-324, 2025 (DOI: 10.1109/JSAIT.2025.3602446).}
   \thanks{Z. Chen and N. Rengaswamy are with the
           Department of Electrical and Computer Engineering,
           University of Arizona, Tucson, Arizona 85721, USA.
           Email: \{ zhuangzhuangchen , narayananr \}@arizona.edu}%
    \thanks{Part of this work was presented at the 2024 IEEE International Conference on Quantum Computing and Engineering~\cite{Chen-qce24}.}
  }

{\maketitle}

\begin{abstract}
The standard approach to universal fault-tolerant quantum computing is to develop a general purpose quantum error correction mechanism that can implement a universal set of logical gates fault-tolerantly.
Given such a scheme, any quantum algorithm can be realized fault-tolerantly by composing the relevant logical gates from this set.
However, we know that quantum computers provide a significant quantum advantage only for specific quantum algorithms.
Hence, a universal quantum computer can likely gain from compiling such specific algorithms using tailored quantum error correction schemes.
In this work, we take the first steps towards such algorithm-tailored quantum fault-tolerance.
We consider Trotter circuits in quantum simulation, which is an important application of quantum computing.
We develop a solve-and-stitch algorithm to systematically synthesize physical realizations of Clifford Trotter circuits on the well-known $\llbr n,n-2,2 \rrbr$ error-detecting code family.
Our analysis shows that this family implements Trotter circuits with essentially optimal depth under reasonable assumptions, thereby serving as an illuminating example of tailored quantum error correction.
We achieve fault-tolerance for these circuits using flag gadgets, which add minimal overhead.
Importantly, the solve-and-stitch algorithm has the potential to scale beyond this specific example, as illustrated by a generalization to the four-qubit logical Clifford Trotter circuit on the $\llbr 20,4,2 \rrbr$ hypergraph product code, thereby providing a principled approach to tailored fault-tolerance in quantum computing.
\end{abstract}


\section{Introduction}
\label{sec:intro}


\IEEEPARstart{Q}{uantum} error correction (QEC) is fundamental to the realization of scalable fault-tolerant quantum computers.
In recent years, QEC has moved from theory to practice where there have been several demonstrations of small error corrected systems~\cite{Postler-arxiv23,Bluvstein-nature24,DaSilva-arxiv24,Ali-arxiv24}.
The next frontier is the development of \emph{scalable} QEC schemes that enable significant quantum advantages for \emph{specific} problem domains, when compared to the best classical supercomputers.
The common approach is to pursue universal fault-tolerant quantum computing where a general-purpose QEC scheme is shown to fault-tolerantly realize a universal set of logical operations on the encoded information~\cite{Cohen-sciadv22,Wang-arxiv23,Zhu-arxiv23}.
Given such a scheme, one can execute any quantum algorithm on the logical qubits by \emph{composing} elements of this fault-tolerant universal set.
In parallel, quantum algorithms continue to be developed for various problems.
However, it is widely expected that significant quantum advantage will be achieved only for some targeted problems and applications~\cite{Montanaro-npjqi16,AuYeung-arxiv23,Dalzell-arxiv23}.
Hence, even in a universal fault-tolerant quantum computer, there are likely gains to be achieved by compiling such specific algorithms using \emph{tailored} QEC mechanisms.
More specifically, rather than composing \emph{individual} gates from the universal set, it could be much more efficient to directly design fault-tolerant realizations of the \emph{logical circuit as a whole}.
Such exciting opportunities form the primary motivation for this work.
To the best of our knowledge, this work constitutes the first such exploration.

\begin{figure}[t]
\centering
\includegraphics[scale=1,keepaspectratio]{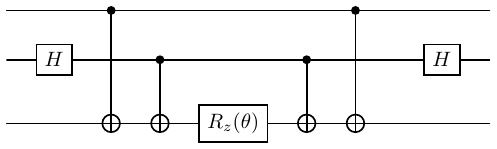}

\caption{Quantum Simulation Kernel (QSK) with 3 qubits.}
\label{fig:qsk_3qubit}

\end{figure}

At the outset, it is unclear how one can pose the goal of tailoring QEC schemes to logical algorithms as a systematic mathematical problem.
When can we say that a QEC scheme is ``well-matched'' to execute an algorithm?
In this paper, we work with the objective of achieving a depth-optimal fault-tolerant realization of a given logical circuit.
Since quantum simulation is a key motivation and application of quantum computers, we pursue the goal of executing \emph{Trotter circuits} fault-tolerantly with optimal depth~\cite{Whitfield-molphy11,Daley-nature22}.
A Trotter circuit is also referred to as the \emph{quantum simulation kernel (QSK)}~\cite{Li-asplos22}, so we will use these terminologies interchangeably.
An example QSK circuit for $3$ qubits is shown in Fig.~\ref{fig:qsk_3qubit}, where the $Z$-rotation angle $\theta$ depends on the specifics of the simulation algorithm.
For this work, we set $\theta = \frac{\pi}{2}$ throughout, i.e., consider \emph{Clifford}~\cite{Dehaene-pra03} QSK (C-QSK) circuits, to take the first steps towards the above goal. 
We develop a principled approach of deriving the physical realization of the logical C-QSK circuit on the well-known error-detecting family of $\llbr n,n-2,2 \rrbr$ codes~\cite{Gottesman-phd97,Chao-npjqi18}.
The methods have the potential to generalize beyond this family, as we illuminate through a hypergraph product code, and also be amenable to optimization with respect to compilation constraints.

In prior work, we developed the \emph{Logical Clifford Synthesis (LCS)} algorithm~\cite{Rengaswamy-tqe20}%
\footnote{Implementations: \url{https://github.com/nrenga/symplectic-arxiv18a} (MATLAB), \url{https://github.com/AparnaGupta0301/Logical-Clifford-Synthesis} (Python)}
that systematically constructs all physical Clifford circuits that realize a given logical Clifford circuit on any specified \emph{stabilizer code}~\cite{Gottesman-phd97,Calderbank-it98}.
The input logical Clifford circuit and stabilizer code are unrestricted, so the algorithm is very general.
The LCS algorithm is computationally efficient because it leverages the $2n \times 2n$ \emph{binary symplectic matrix} representation of Clifford circuits~\cite{Dehaene-pra03,Aaronson-pra04}, rather than their $2^n \times 2^n$ unitary matrix representation, and has complexity polynomial in $n$.
Hence, this algorithm can be directly applied to obtain physical realizations of logical Clifford QSK circuits on the error-detecting code family.
However, there are two key disadvantages of the LCS algorithm:
\begin{enumerate}
    \item For an $\llbr n,k,d \rrbr$ stabilizer code, the number of symplectic matrix solutions that realize a given logical Clifford circuit is $2^{r(r+1)/2}$, where $r = n-k$, ignoring stabilizer degrees of freedom.
    Hence, despite the computationally efficient nature of the algorithm, the search space grows super-exponentially in the dimension of the stabilizer group.
    For codes with constant rate~\cite{Tillich-it13,Panteleev-stoc22,Leverrier-focs22}, i.e., $k = \Theta(n)$, the solution space is super-exponential in $n$.
    Currently, there is no known method to systematically obtain only ``good'' solutions from this space, although recent work has tailored solutions to the hardware~\cite{Kuehnke-arxiv25}.

    \item The physical circuits output by the algorithm are neither necessarily fault-tolerant nor optimized for specific objectives such as depth or number of two-qubit gates.
    Guaranteeing fault-tolerance for a generic code while retaining the unitary nature of the solutions is hard.
    The linear algebraic approach of the algorithm does not enable one to track circuit properties, such as depth, through the steps of the algorithm.
\end{enumerate}

Due to these drawbacks, we consider a \emph{bottom-up} approach to synthesizing physical realizations of logical C-QSK circuits on the error-detecting code family.
The key idea is to characterize the logical circuit via input-output Pauli tracking constraints (similar to the LCS algorithm), solve for separate small circuits to satisfy each of these constraints (the ``solve'' step), and then suitably merge these circuits to simultaneously satisfy all constraints (the ``stitch'' step), thereby realizing the logical circuit.
We refer to this as the \emph{solve-and-stitch} method.
For this code family, we achieve fault-tolerance by adding flag gadgets~\cite{Chao-npjqi18} to the two-qubit gates in the solutions and proving that any single fault in the circuit remains detectable at the end.
Most excitingly, by leveraging some properties of the code family, we prove that the depth of these fault-tolerant physical circuits only grows \emph{linearly in the number of logical qubits $k$}.
Since the logical QSK circuit itself has depth $2k+1$ (from $2(k-1)$ CNOTs, the $Z$-rotation, and two layers of single-qubit gates), the depth of these physical circuits is optimal.
While $k=n-2$ for this family, which means that the depth is effectively linear in $n$, the proof involves calculations primarily in terms of the variable $k$.
Therefore, the proof strategy has the potential to extend beyond this specific code family.

Overall, our results suggest that the error-detecting code family is indeed well-matched to efficiently realize logical C-QSK circuits fault-tolerantly.
Obviously, the code doesn't allow one to correct errors, so we will consider extending the solve-and-stitch approach to more non-trivial code families in future work.
In such cases, decoders can also be tailored to these circuits since the error propagation has more structure.

The remainder of the paper is organized as follows.
Section~\ref{sec:prelim} provides necessary background to understand the details of this work.
Section~\ref{sec:transversal_qsk} explores transversal implementations of the logical C-QSK circuits and shows that this appears to be impossible using stabilizer codes.
Section~\ref{sec:qsk_distance_2_family} develops the solve-and-stitch method to physically realize logical C-QSK circuits on the error-detecting code family.
Section~\ref{sec:depth_analysis} provides the depth analysis for the circuits produced in Section~\ref{sec:qsk_distance_2_family}.
Section~\ref{sec:discussion} provides several examples where the solve-and-stitch construction remains valid beyond the 
$\llbr n,n-2,2 \rrbr$ code family. In particular, we highlight variations involving odd numbers of logical qubits, isolated logical gates, and a representative case on a hypergraph product code.
Section~\ref{sec:ft_qsk_flags} designs flagged versions of the circuits from Section~\ref{sec:qsk_distance_2_family} and establishes their fault-tolerance.
Finally, Section~\ref{sec:conclusion} concludes the work and highlights opportunities for future work related to our results.
Some additional details related to the results and further insights developed during this work are provided in the appendices.

\section{Preliminaries}
\label{sec:prelim}

In Appendix~\ref{sec:prelim_contd} we review standard definitions related to the Pauli group, stabilizer codes, and the $\llbr n,n-2,2 \rrbr$ code family.

\subsection{Quantum Simulation Kernel (QSK)}
\label{sec:qsk}

In quantum (Hamiltonian) simulation, a common strategy is to expand the Hamiltonian in the orthonormal basis of Hermitian Pauli operators $E_j$ to write $\mathbb{H} = \sum_j \alpha_j E_j$, where $\alpha_j \in \mathbb{R}$~\cite{Whitfield-molphy11}.
The goal in simulation is to start with an initial state $\ket{\phi}$ and compute the state at a future time instant $t$ as $\ket{\phi(t)} = \exp(-\imath \mathbb{H} t) \ket{\phi}$ (at least when $\mathbb{H}$ is time-independent), where $\imath \coloneqq \sqrt{-1}$ and $t \in \mathbb{R}$.
Since $\{ E_j \}$ may not mutually commute, one can leverage the \emph{Trotter-Suzuki formula}~\cite{Hatano-qaoom05} to approximate $\exp(-\imath \mathbb{H} t)$ by repeated applications of $\prod_j \exp(-\imath \alpha_j E_j t/T)$ for $T$ (an integer) times.
The circuit to implement such an exponentiated Pauli operator $\exp(\imath \theta E)$ is called a \emph{Trotter} circuit or a \emph{quantum simulation kernel (QSK)} circuit. 
An example for $3$ qubits is shown in Fig.~\ref{fig:qsk_3qubit}, which corresponds to the case $E_j = Z_1 X_2 Z_3$. 
The structure generalizes for any $k$-qubit Pauli: start with single-qubit gates on those qubits that have a non-$Z$ component in $E_j$ (Hadamard $H$ for $X$ or $H_y \coloneqq \frac{1}{\sqrt{2}} \begin{bsmallmatrix}
1 & -\imath \\ \imath & -1 \end{bsmallmatrix}$ for $Y$), then perform CNOTs from the first $(k-1)$ qubits (control) to the $k$-th qubit (target), then perform a $Z$-rotation $R_z(\theta)$ of the $k$-th qubit, and finally apply the gates before the rotation in the reverse order.
The circuit depth is $(2k + 1)$, i.e., linear in $k$.

We only consider Clifford QSK (C-QSK) circuits, so we set $\theta = \frac{\pi}{2}$ (i.e., $R_z(\theta) = P \coloneqq \sqrt{Z}$).
In this case the circuit is completely characterized by how it maps Pauli operators at its input to Pauli operators at its output using standard Pauli conjugation relations for Clifford gates.
For the $k=3$ circuit in Fig.~\ref{fig:qsk_3qubit}, one can check the following input-output mappings:
\begin{IEEEeqnarray}{rClCrCl}
X_1 & \mapsto & Y_1 X_2 Z_3 & \quad , \quad & Z_1 & \mapsto & Z_1 \ , \nonumber \\
X_2 & \mapsto & X_2 & \quad , \quad & Z_2 & \mapsto & - Z_1 Y_2 Z_3 \ , \nonumber \\
X_3 & \mapsto & Z_1 X_2 Y_3 & \quad , \quad & Z_3 & \mapsto & Z_3 \ .
\end{IEEEeqnarray}
Hence, for any $\llbr n,k,d \rrbr$ stabilizer code, the physical realization of the logical C-QSK circuit must satisfy the above relations for its $k$ \emph{logical} Pauli operators.
Besides, the physical circuit must also preserve the stabilizer group $\mathcal{S}$ under conjugation.

\subsection{Model for Circuit Depth}
\label{sec:depth}

Throughout this work, we define \textit{circuit depth} as the number of sequential layers of gates, where each layer may contain multiple gates acting on disjoint qubit subsets. In particular, each two-qubit gate contributes one unit of depth, and layers of single-qubit gates are either grouped into one depth unit if they act in parallel, or counted as separate units depending on gate structure. Our depth model is independent of hardware connectivity or scheduling constraints and thus reflects the logical circuit structure rather than physical layout. The primary goal is to analyze asymptotic scaling and structural optimizations in fault-tolerant circuit synthesis. Hardware-aware optimization, parallel gate scheduling, and mapping to physical architectures are beyond the scope of this work, but these may further improve our solutions.

\section{Fault-Tolerant QSK via Transversality}
\label{sec:transversal_qsk}

A natural fault-tolerant implementation of logical circuits is via a \emph{transversal} physical operation, i.e., one that splits into a Kronecker product of single-qubit gates.
Any fault in such a physical circuit does not propagate into other qubits.
While the Eastin-Knill theorem~\cite{Eastin-prl09} prevents an error-detecting stabilizer code from implementing a \emph{universal} set of gates fault-tolerantly, there might exist a stabilizer code that realizes the specific logical C-QSK circuit transversally.
In this section, we pursue this possibility and consider transversal $H$ or $P$ gates for implementing the logical C-QSK circuit.

\begin{theorem}
\label{thm:transversal_H}
There exists no stabilizer code where transversal Hadamard realizes the logical C-QSK circuit.
\begin{IEEEproof}
\normalfont
See Appendix~\ref{sec:transversal_H_proof} for the proof.
\end{IEEEproof}
\end{theorem}

\begin{theorem}
\label{thm:transversal_P}
There exists no stabilizer code where transversal Phase realizes the logical C-QSK circuit.
\begin{IEEEproof}
\normalfont
See Appendix~\ref{sec:transversal_P_proof} for the proof.
\end{IEEEproof}
\end{theorem}

In our proofs, we have refrained from specifying any particular code, leading us to the compelling conclusion that finding a stabilizer code capable of realizing the logical C-QSK circuits via transversal combinations of Hadamard and Phase gates \emph{appears} inherently impossible. 
Note that this encompasses all transversal Clifford circuits since Hadamard and Phase gates generate the single-qubit Clifford group.
Remarkably, based on the properties of C-QSK circuits, this conclusion can likely be extended to C-QSK circuits of arbitrary size and combinations of $H$ and $H_y$ gates. 
Our analysis suggests that leveraging stabilizer degrees of freedom is unlikely to alter this conclusion.
Of course, all this remains to be proven rigorously, hence we leave it as a conjecture.

\begin{conjecture}
There exists no stabilizer code where transversal Clifford gates realize the logical C-QSK circuit for any non-trivial exponentiated Pauli operator.    
\end{conjecture}

\section{Realizing C-QSK on $\llbr n,n-2,2 \rrbr$ Code Family}
\label{sec:qsk_distance_2_family}

The discussion in Sec.~\ref{sec:transversal_qsk} implies that transversal implementations of logical C-QSK circuits seem inherently impossible.
Hence, we must incorporate two-qubit gates in the construction of physical circuits. 
This, however, raises a crucial question: 

\emph{Given specific input-output Pauli mapping rules dictated by the logical C-QSK circuit on a stabilizer code, what can we infer about the structure of the physical circuit satisfying these Pauli constraints, even without considering fault-tolerance?}

Our goal in this section is to address this question by developing a principled approach to circuit synthesis that allows us to track structural information during the synthesis.
Naturally, such an approach provides a feasible solution that upper bounds depth, number of two-qubit gates etc.

The \emph{Logical Clifford Synthesis (LCS)} algorithm~\cite{Rengaswamy-tqe20} systematically synthesizes physical (Clifford) circuits satisfying Pauli constraints imposed by the logical (Clifford) circuit and code structure. 
This includes the constraints to ensure that the stabilizer group is preserved under conjugation by the physical Clifford circuit.
The algorithm formulates the Pauli constraints as linear equations on a target binary symplectic matrix representing the desired physical Clifford circuit.
Then it systematically solves for all feasible symplectic solutions by using symplectic transvections.
Finally, it decomposes each solution into elementary Clifford gates.
Despite its value, the LCS algorithm falls short of elucidating circuit properties and structure \emph{during the construction of solutions}.
This is because it directly finds the symplectic representation~\cite{Dehaene-pra03} of the circuit, which hides the circuit structure until it is decomposed into elementary gates.
Therefore, one must determine all solutions before identifying the most desirable circuit.
Unfortunately, since the number of solutions is $2^{r(r+1)/2}$, where $r = n-k$, the computational efficiency decreases super-exponentially with the dimension of the stabilizer group. 

We circumvent these issues of the LCS algorithm by developing a \emph{bottom-up} approach to synthesize physical circuits.
The key idea is to ``solve'' for a small circuit satisfying one Pauli constraint by identifying a \emph{root qubit}.
Such a qubit is involved in both the input and output Pauli operator, but the Pauli acting on that qubit changes from input to output.
Then these small circuits for different constraints are ``stitched'' together appropriately to satisfy all constraints simultaneously.
By employing this approach for logical C-QSK circuits on $\llbr n, n-2, 2 \rrbr$ codes, we can track circuit properties, such as depth, \emph{during the construction of the circuit}.
Notably, this method yields results comparable to the optimal circuits generated by LCS for this code family, \emph{but without enumerating all solutions}. 
The following section illustrates such realization of logical C-QSK circuits under different scenarios.

\vspace*{-10pt}

\subsection{Circuit Synthesis via Solve-and-Stitch Approach}
\label{sec:qsk_distance2_codes}

\subsubsection{Even number of Hadamards for $\llbr 6,4,2 \rrbr$ code}
\label{sec:qsk_642_evenH}

We introduced and described the procedure in detail in~\cite{Chen-qce24}.
For completeness, we include it in Appendix~\ref{sec:Example of [[6,4,2]] even}.
In what follows, we will focus on the challenging cases of odd number of Hadamards and the presence of $H_y$ gates, since these were not explained in~\cite{Chen-qce24}.

\subsubsection{Even number of Hadamards for $\llbr 8,6,2 \rrbr$ code}
\label{sec:qsk_862_evenH} 

See Appendix~\ref{sec:Example of [[8,6,2]] even} for details.

\subsubsection{Odd number of Hadamards for $\llbr 6,4,2 \rrbr$ code}
\label{sec:qsk_642_oddH}

\begin{figure}[h]
\centering

\includegraphics[scale=0.98,keepaspectratio]{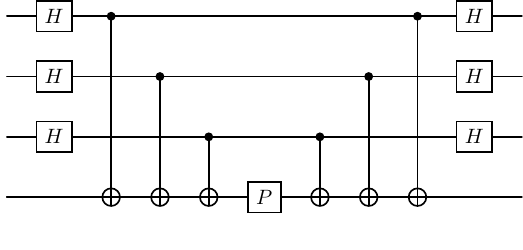}

\vspace*{-10pt}
\caption{\label{fig:qsk_4qubit_oddH}C-QSK Circuit with odd number of Hadamard gates.}
\end{figure}

In the case of logical C-QSK circuits with an odd number of Hadamard gates, we encounter some variations in the established pattern, necessitating the consideration of stabilizer effects during the physical circuit construction. 
While there are distinct differences, the fundamental idea of root qubits and the solve-and-stitch mechanism continues to prove valuable. 
In this example, we present the implementation of a $4$-qubit C-QSK circuit with $H$ applied to the first three qubits, as shown in Fig.~\ref{fig:qsk_4qubit_oddH}.

The mapping of each Pauli constraint is listed below:
\begin{IEEEeqnarray}{rClCrCl}
\label{eq:logical_qsk_4qubit_oddH_constraints}
X_1 X_2 & \mapsto & X_1 X_2 & \quad , \quad & Z_2 Z_6 & \mapsto & - X_1 Y_2 X_3 X_4 Z_5 \ , \nonumber \\
X_1 X_3 & \mapsto & X_1 X_3 & \quad , \quad & Z_3 Z_6 & \mapsto & - X_1 X_2 Y_3 X_4 Z_5 \ , \nonumber \\
X_1 X_4 & \mapsto & X_1 X_4 & \quad , \quad & Z_4 Z_6 & \mapsto & - X_1 X_2 X_3 Y_4 Z_5 \ , \nonumber \\
X_1 X_5 & \mapsto & X_2 X_3 X_4 Y_5 Z_6 & \quad , \quad & Z_5 Z_6 & \mapsto & Z_5 Z_6 \ .
\end{IEEEeqnarray}

\begin{figure}[t]
\centering

\includegraphics[scale=1,keepaspectratio]{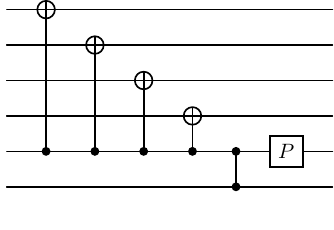}

\vspace*{-10pt}
\caption{\label{fig:XIIIX_IXXXYZ}$X_{1}{X_{5}}\mapsto X_{2}X_{3}X_{4}Y_{5}Z_{6}$}

\vspace*{-10pt}
\end{figure}

In examining the logical-$X$ Pauli constraints, we observe that only the mapping of $\overline{X_{4}}$ is non-trivial, necessitating the construction of only a single rooted circuit for the logical-$X$ component. 
For this particular Pauli constraint, we designate the fifth qubit as the root. 
Unlike the case of even number of Hadamards, observe that the output of this mapping results in the \emph{disappearance} of $X_{1}$. 
To accommodate this, we first introduce $\text{CNOT}_{5 \rightarrow 1}$ into the rooted circuit, as CNOT transforms $XX$ to $XI$. 
For the remaining qubits, we once again apply CNOT or CZ gates based on the input-output Pauli relations.
As always, we relocate the Phase gate to the end. 
Importantly, the mappings of $\overline{X_{1}}$, $\overline{X_{2}}$, and $\overline{X_{3}}$ in the logical-$X$ component are trivial and continue to be satisfied in the constructed rooted circuit, which is shown in Fig.~\ref{fig:XIIIX_IXXXYZ}.

\begin{figure}[t]
\centering

\vspace*{5pt}

\begin{subfigure}[t]{0.35\textwidth}
\centering

\includegraphics[scale=0.9,keepaspectratio]{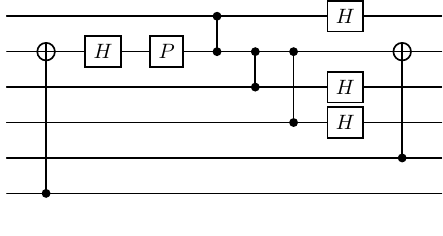}

\vspace*{-12pt}
\caption{\label{fig:IZIIIZ_XYXXZ}$Z_{2}{Z_{6}}\mapsto X_{1}Y_{2}X_{3}X_{4}Z_{5}$ (ignoring sign)}
\end{subfigure}

\vspace*{5pt}

\begin{subfigure}[t]{0.35\textwidth}
\centering

\includegraphics[scale=0.9,keepaspectratio]{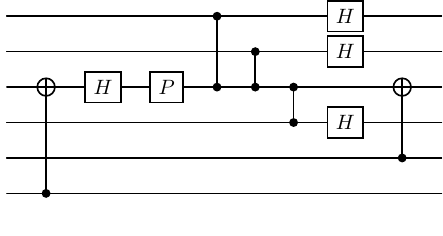}

\vspace*{-12pt}
\caption{\label{fig:IIZIIZ_XXYXZ}$Z_{3}{Z_{6}}\mapsto X_{1}X_{2}Y_{3}X_{4}Z_{5}$ (ignoring sign)}
\end{subfigure}

\vspace*{5pt}

\begin{subfigure}[t]{0.35\textwidth}
\centering

\includegraphics[scale=0.9,keepaspectratio]{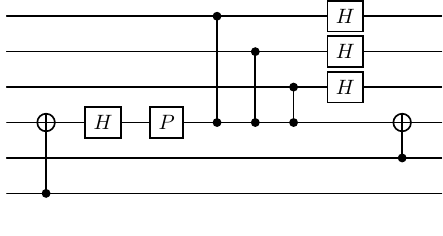}

\vspace*{-12pt}
\caption{\label{fig:IIIZIZ_XXXYZ}$Z_{4}{Z_{6}}\mapsto X_{1}X_{2}X_{3}Y_{4}Z_{5}$ (ignoring sign)}
\end{subfigure}

\caption{\label{fig:local_circuit_Z_642_oddH} Rooted circuits for logical-$Z$ for odd number of Hadamards in C-QSK circuit on $\llbr 6,4,2 \rrbr$ code.}

\vspace*{-12pt}

\end{figure}

In the logical-$Z$ components, the mappings of $\overline{Z_{1}}$, $\overline{Z_{2}}$, and $\overline{Z_{3}}$ are non-trivial. 
Notably, in each of these mappings, the gate $X_{1}$ \emph{emerges} while the gate $Z_{6}$ \emph{disappears} in the output configuration. 
To address this, first we designate the second, third, and fourth qubits as the root qubits, respectively. 
The disappearance of $Z_{6}$ at the outputs introduces additional processing in the rooted circuits. 
Therefore, in each rooted circuit shown in Fig.~\ref{fig:local_circuit_Z_642_oddH} we introduce CNOT gates at the beginning, specifically $\text{CNOT}_{6\rightarrow2}$, $\text{CNOT}_{6\rightarrow3}$, and $\text{CNOT}_{6\rightarrow4}$, respectively. 
The designated root qubit serves as the target qubit for these additional CNOT gates because CNOTs transform $ZZ$ into $IZ$. 
Subsequently, these additional CNOT gates are followed by the $H$-$P$-CZ-$H$ structure involving appropriate qubits. 
Observe that some CZ gates involve the first qubit due to the emergence of $X_{1}$ in the output, a feature not observed in scenarios with an even number of Hadamard gates in the logical C-QSK circuit. 
To complete each rooted circuit, an additional CNOT gate is added at the end to produce $Z_5$, which is already present in the logical-$X$ rooted circuit.

\begin{figure}[h]
\centering

\includegraphics[scale=0.77,keepaspectratio]{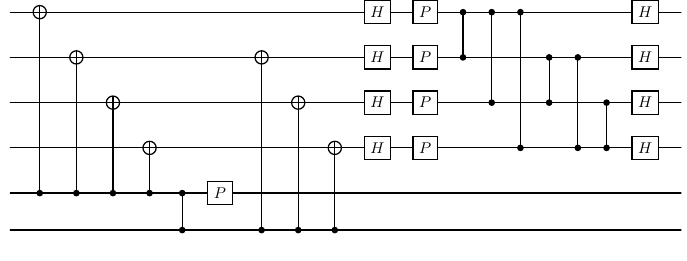}


\caption{\label{fig:solve-stitch_642_qsk_oddH_invalid_stabilizers} Physical realization of Fig.~\ref{fig:qsk_4qubit_oddH} on the $\llbr 6,4,2 \rrbr$ code produced by the solve-and-stitch approach using root qubits. But this does not preserve the two stabilizer generators.}

\vspace*{-5pt}

\end{figure}

The construction of the complete physical circuit in Fig.~\ref{fig:solve-stitch_642_qsk_oddH_invalid_stabilizers} closely mirrors the scenario with an even number of Hadamard gates on the logical circuit. 
The process involves stitching together individual rooted circuits, validating Pauli relations, and avoiding duplicated gates. 
Note that an additional Hadamard and Phase are introduced on the first qubit to retain $X_1 X_2, X_1 X_3$, and $X_1 X_4$ unchanged.
This also forms a symmetric $H$-$P$-CZ-$H$ special structure on the first four qubits. 
However, while the logical constraints on the complete circuit are satisfied, the two stabilizers fail to be preserved.

To rectify this, additional gates $P_{6}$ and $\text{CNOT}_{6\rightarrow 1}$ are incorporated into the circuit, which play a crucial role in preserving stabilizers without impacting the logical constraints on the full circuit.
Together with $H_1$ and $P_1$, these constitute non-trivial modifications not present in the rooted circuits.

\begin{figure}[h]
\centering

\includegraphics[scale=0.72,keepaspectratio]{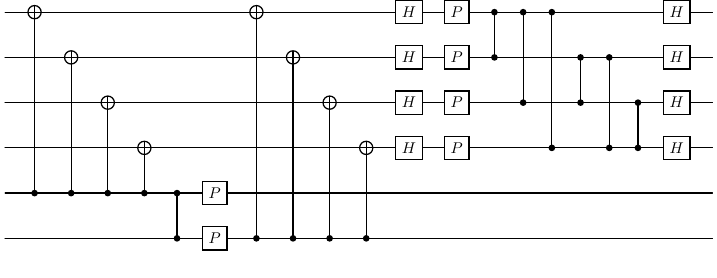}


\caption{\label{fig:solve-stitch_642_qsk_oddH} Physical realization of Fig.~\ref{fig:qsk_4qubit_oddH} on the $\llbr 6,4,2 \rrbr$ code produced by the solve-and-stitch approach using root qubits. Signs for $\overline{Z_i}$ can be fixed with Pauli gates at the end.}

\vspace*{-10pt}

\end{figure}

\subsubsection{Mixed Hadamards and $H_y$ gates on logical circuits}
\label{sec:qsk_Hy}
We have already established the physical implementation of the Quantum Simulation Kernel (QSK) logical circuit for different parities of Hadamard gates positioned at the beginning and end of the circuit on the $\llbr n,n-2,2 \rrbr$ code. This solve-and-stitch approach, however, is not limited to circuits with only Hadamard gates; it can also be extended to QSK circuits that incorporate a combination of Hadamard and $H_y$ gates. The procedure begins by substituting all $H_y$ gates with Hadamard gates in the QSK circuit and then applying Algorithm~\ref{alg:solve-and-stitch} to generate the corresponding physical circuit. Following this, specific modifications are made to the generated circuit to accommodate the varying parities of the Hadamard and $H_y$ gates in the original QSK logical circuit. This ensures that all Pauli constraints are satisfied simultaneously, even when $H_y$ gates are present. 
We illustrate this process using the simplest case—when both the number of Hadamard and $H_y$ gates is even—as a representative example below. 


\begin{figure}
  \centering

\includegraphics[scale=0.92,keepaspectratio]{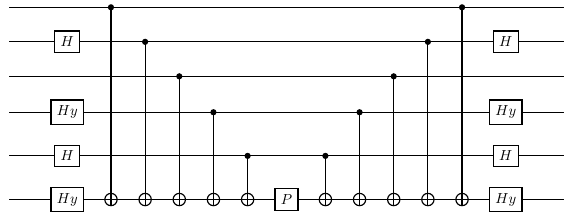}




\caption{\label{fig:even_H_even_Hy_logical} C-QSK circuit with even number of Hadamard gates and even number of $H_y$ gates.}

\end{figure}

The C-QSK circuit in Fig.~\ref{fig:even_H_even_Hy_logical} dictates the following input-output mappings of logical Pauli operators:
\begin{IEEEeqnarray}{rClCrCl}
\label{eq:even_H_even_Hy_logical}
\overline{X_1} & \mapsto & \overline{Y_1} \, \overline{X_2} \, \overline{Z_3} \, \overline{Y_4} \, \overline{X_5} \, \overline{Y_6} & \quad , \quad & \overline{Z_1} & \mapsto & \overline{Z_1} \ , \nonumber \\
\overline{X_2} & \mapsto & \overline{X_2} & \quad , \quad & \overline{Z_2} & \mapsto & - \overline{Z_1} \, \overline{Y_2} \, \overline{Z_3} \, \overline{Y_4} \, \overline{X_5} \, \overline{Y_6} \ , \nonumber \\
\overline{X_3} & \mapsto & \overline{Z_1} \, \overline{X_2} \, \overline{Y_3} \, \overline{Y_4} \, \overline{X_5} \, \overline{Y_6} & \quad , \quad & \overline{Z_3} & \mapsto & \overline{Z_3} \ , \nonumber \\
\overline{X_4} & \mapsto & \overline{Z_1} \, \overline{X_2} \, \overline{Z_3} \, \overline{Z_4} \, \overline{X_5} \, \overline{Y_6} & \quad , \quad & \overline{Z_4} & \mapsto & - \overline{Z_1} \, \overline{X_2} \, \overline{Z_3} \, \overline{X_4} \, \overline{X_5} \, \overline{Y_6} \ , \nonumber \\ 
\overline{X_5} & \mapsto & \overline{X_5} & \quad , \quad & \overline{Z_5} & \mapsto & - \overline{Z_1} \, \overline{X_2} \, \overline{Z_3} \, \overline{Y_4} \, \overline{Y_5} \, \overline{Y_6} \ , \nonumber \\
\overline{X_6} & \mapsto & -\overline{Z_1} \, \overline{X_2} \, \overline{Z_3} \, \overline{Y_4} \, \overline{X_5} \, \overline{Z_6} & \quad , \quad & \overline{Z_6} & \mapsto & - \overline{Z_1} \, \overline{X_2} \, \overline{Z_3} \, \overline{Y_4} \, \overline{X_5} \, \overline{X_6} \ . \nonumber 
\end{IEEEeqnarray}

\begin{figure}
\centering

\includegraphics[scale=1,keepaspectratio]{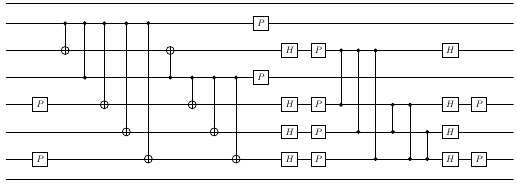}
\caption{\label{fig:even_H_even_Hy}Physical implementation of logical C-QSK circuit in Fig.~\ref{fig:even_H_even_Hy_logical}.}

\end{figure}

Substituting for the logical operators of the $\llbr 8,6,2 \rrbr$ code, we obtain the following mappings of physical Pauli operators: 
{\small
\begin{IEEEeqnarray}{rClCrCl}
\label{eq:even_H_even_Hy}
X_1 X_2 & \mapsto & X_1 Y_2 X_3 Z_4 Y_5 X_6 Y_7 & \quad , \quad & Z_2 Z_8 & \mapsto & Z_2 Z_8 \ , \nonumber \\
X_1 X_3 & \mapsto & X_1 X_3 & \quad , \quad & Z_3 Z_8 & \mapsto & Z_2 Y_3 Z_4 Y_5 X_6 Y_7 Z_8 \ , \nonumber \\
X_1 X_4 & \mapsto & X_1 Z_2 X_3 Y_4 Y_5 X_6 Y_7 & \quad , \quad & Z_4 Z_8 & \mapsto & Z_4 Z_8 \ , \nonumber \\
X_1 X_5 & \mapsto & X_1 Z_2 X_3 Z_4 Z_5 X_6 Y_7 & , & Z_5 Z_8 & \mapsto & Z_2 X_3 Z_4 X_5 X_6 Y_7 Z_8 \ , \nonumber \\
X_1 X_6 & \mapsto & X_1 X_6 & \quad , \quad & Z_6 Z_8 & \mapsto & Z_2 X_3 Z_4 Y_5 Y_6 Y_7 Z_8 \ , \nonumber \\
X_1 X_7 & \mapsto & X_1 Z_2 X_3 Z_4 Y_5 X_6 Z_7 & \quad , \quad & Z_7 Z_8 & \mapsto & Z_2 X_3 Z_4 Y_5 X_6 X_7 Z_8 \ . \nonumber
\end{IEEEeqnarray}
}
To obtain the physical circuit that satisfies all these Pauli constraints, replace the $H_y$ gates with Hadamard gates in the logical circuit in Fig.~\ref{fig:even_H_even_Hy_logical}, run Algorithm~\ref{alg:solve-and-stitch} to get the corresponding physical circuit, then add $P_5$ and $P_7$ at the beginning and at the end of this physical circuit. This new physical circuit is a valid solution and is shown in Fig.~\ref{fig:even_H_even_Hy}.

The remaining three parity combinations for $H$ and $H_y$ gates (even-odd, odd-even, odd-odd) are treated analogously using Algorithm~\ref{alg:solve-and-stitch-Hy}, and their corresponding physical circuits are provided in Appendix~\ref{sec:Hy examples}. Although our depth analysis (in the next section) does not consider $H_y$ gates for simplicity, this method highlights the versatility and robustness of the solve-and-stitch approach in constructing fault-tolerant quantum circuits under varying gate configurations.

\subsection{Solve-and-Stitch Approach on all $\llbr n,n-2,2 \rrbr$ Codes}
\label{sec:general n,n-2,2 code}

By delving into the constructions of physical circuits utilizing the root qubit idea in Sec.~\ref{sec:qsk_distance2_codes}, we discerned both the similarities and differences in the patterns of these physical circuits. 
Our analysis indicates that these patterns are generalizable to all members of the $\llbr n,n-2,2 \rrbr$ family. 
We argue that the characteristics of these patterns stem from the inherent properties of the codes and the C-QSK circuits themselves. 
Hence, the solve-and-stitch algorithm constitutes a tailored design to realize C-QSK circuits on these codes.

In this section, we begin by establishing the logical Pauli mappings for C-QSK circuits with an arbitrary number of qubits. 
Note that a C-QSK circuit with odd number of qubits can be embedded into one with even number of qubits by introducing a qubit at the top which starts and ends in the $\ket{0}$ state.
Hence, we will restrict ourselves to C-QSK circuits with $n-2$ qubits where $n$ is even, as is necessary for the code family.
Our results below will indicate the minimal overhead incurred by this assumption.
Given the logical Pauli operators of the $\llbr n,n-2,2 \rrbr$ codes, we translate the logical Pauli mappings to physical Pauli constraints.
Then we show that the stitching procedure always preserves all prior Pauli constraints.
Finally, we provide the general algorithm to realize the logical C-QSK circuit on the $\llbr n,n-2,2 \rrbr$ code family.

Define the set $[k] \coloneqq \{1,2,3,...,k\}$. 
Let $i,j \in [k]$ index (logical) qubits of the C-QSK circuit.
Let $I_h \subseteq [k]$ be the index set of qubits where Hadamard gates appear in the C-QSK circuit. For Trotter circuits with $H_y$ gates, let $I_{hy}$ represent the index set of qubits where $H_y$ gates appear in the QSK circuit. We also define $I_e$ as the index set of qubits where neither Hadamard nor $H_y$ gates are applied, i.e., $I_e=[k]\backslash(I_h \cup I_{hy})$. It is worth mentioning that $I_h$ and $I_{hy}$ do not overlap, i.e., $I_h \cap I_{hy} = \emptyset$. Unless otherwise specified, the following lemmas, theorems, corollaries, remarks pertain to the C-QSK circuit without $H_y$ gates.

\begin{lemma}
\label{lem:qsk_logical_Pauli_mappings}
The logical Pauli mappings of the C-QSK circuit on $k$ qubits can be expressed as follows:
\begin{enumerate}

\item If $i \notin I_{h}$, then
\begin{align}
\overline{X_{i}} & \mapsto \overline{Y_{i}} \left( \prod_{j\in I_{h}} \overline{X_{j}} \right)\left( \prod_{j \in [k] \setminus (I_{h} \cup \{ i \})} \overline{Z_{j}} \right), \\
\overline{Z_{i}} & \mapsto \overline{Z_{i}}.
\end{align}

\item If $i \in I_{h}$, then
\begin{align}
\overline{X_{i}} & \mapsto \overline{X_{i}}, \\    
\overline{Z_{i}} &\mapsto - \overline{Y_{i}} \left( \prod_{j \in I_{h} \setminus \{ i \}}\overline{X_{j}} \right) \left(\prod_{j \in [k] \setminus I_{h}} \overline{Z_{j}} \right).    
\end{align}

\end{enumerate}
\begin{IEEEproof}
\normalfont
See Appendix~\ref{Proof of qsk_logical_Pauli_mapping} for the proof.
\end{IEEEproof}
\end{lemma}

Let us now substitute for the logical Pauli operators in the above result.
For the $\llbr n,n-2,2 \rrbr$ codes, logical Pauli operators are defined as $\overline{X_{i}} = X_{1} X_{i+1}, \overline{Z_{i}} = Z_{i+1} Z_{n}$, and 
\begin{align}
\overline{Y_{i}} &= \imath \overline{X_{i}} \, \overline{Z_{i}} \\
  &= \imath X_{1} X_{i+1} Z_{i+1} Z_{n} \\
  &= X_{1} \imath (X_{i+1} Z_{i+1}) Z_{n} \\
  &= X_{1} Y_{i+1} Z_{n}. 
\end{align}
Given these physical operators, the physical Pauli mappings for logical C-QSK circuits on these codes are provided in Corollary~\ref{cor:qsk_physical_Pauli_mappings_evenH} and Corollary~\ref{cor:qsk_physical_Pauli_mappings_oddH}, for the cases of even and odd number of Hadamards, respectively. 
The primary distinction lies in the parity of the number of elements in $I_{h}$, which governs the number of terms in the products in Lemma~\ref{lem:qsk_logical_Pauli_mappings}. 

\begin{corollary}    
\label{cor:qsk_physical_Pauli_mappings_evenH}
Any realization of the logical C-QSK circuit with even number of Hadamards on the $\llbr n,n-2,2 \rrbr$ codes must satisfy the following mappings of physical Pauli operators:
\begin{enumerate}

\item If $i \notin I_{h}$, then
\begin{align}
\label{eq:qsk_physical_Pauli_mappings_evenH_1X}
& X_{1}X_{i+1} \nonumber \\
& \mapsto X_{1} Y_{i+1} \left( \prod_{j \in I_{h}} X_{j+1} \right) \left( \prod_{j \in [n-2] \setminus (I_{h} \cup \{ i \})} Z_{j+1} \right), \\   
\label{eq:qsk_physical_Pauli_mappings_evenH_1Z}
& Z_{i+1} Z_{n} \mapsto Z_{i+1} Z_{n}. 
\end{align}

\item If $i \in I_{h}$, then
\begin{align}
\label{eq:qsk_physical_Pauli_mappings_evenH_2X}
& X_{1}X_{i+1} \mapsto X_{1} X_{i+1}, \\   
\label{eq:qsk_physical_Pauli_mappings_evenH_2Z}
& Z_{i+1} Z_{n} \nonumber \\
& \mapsto - Y_{i+1} \left( \prod_{j \in I_{h} \setminus \{ i \}} X_{j+1} \right) \left( \prod_{j \in [n-2] \setminus I_{h}} Z_{j+1} \right) Z_{n}.
\end{align}

\end{enumerate}

\end{corollary}

\begin{corollary}
\label{cor:qsk_physical_Pauli_mappings_oddH}
Any realization of the logical C-QSK circuit with odd number of Hadamards on the $\llbr n,n-2,2 \rrbr$ codes must satisfy the following mappings of physical Pauli operators:  
\begin{enumerate}

\item If $i \notin I_{h}$, then
\begin{align}
\label{eq:qsk_physical_Pauli_mappings_oddH_1X}
& X_{1}X_{i+1} \nonumber \\
& \mapsto Y_{i+1} \left( \prod_{j \in I_{h}} X_{j+1} \right) \left( \prod_{j \in [n-2] \setminus (I_{h} \cup \{ i \})} Z_{j+1} \right) Z_{n}, \\   
\label{eq:qsk_physical_Pauli_mappings_oddH_1Z}
& Z_{i+1} Z_{n} \mapsto Z_{i+1} Z_{n}. 
\end{align}

\item If $i \in I_{h}$, then
\begin{align}
\label{eq:qsk_physical_Pauli_mappings_oddH_2X}
& X_{1}X_{i+1} \mapsto X_{1} X_{i+1}, \\   
\label{eq:qsk_physical_Pauli_mappings_oddH_2Z}
& Z_{i+1} Z_{n} \nonumber \\
& \mapsto - X_{1} Y_{i+1} \left( \prod_{j \in I_{h} \setminus \{ i \}} X_{j+1} \right) \left( \prod_{j \in [n-2] \setminus I_{h}} Z_{j+1} \right).
\end{align}

\end{enumerate}

\end{corollary}

From the above expressions, it is clear that for each mapping corresponding to $\overline{X_i}$ or $\overline{Z_i}$, qubit $i+1$ is always the root.

Armed with these general expressions for desired Pauli mappings, we can investigate the solve-and-stitch algorithm carefully.
The ``solve'' phase of the algorithm solves for small rooted circuits, one for each mapping, by identifying root qubits.
It is important to observe that each rooted circuit has controlled gates only with the root as the control (resp. target) for logical-$X$ (resp. logical-$Z$) mappings.
These are $\text{CNOT}_{i+1 \rightarrow j+1}$ and $\text{CZ}_{i+1, j+1}$.
Let us focus on the logical-$X$ mappings first and validate the stitching procedure.

\begin{lemma}
\label{lem:logical_X_stitching}
The rooted circuits for logical-$X$ constraints in Lemma~\ref{lem:qsk_logical_Pauli_mappings} can be stitched by concatenating them and dropping duplicate CZ gates.
The stitched circuit satisfies all the logical-$X$ constraints simultaneously.
\begin{IEEEproof}
\normalfont
See Appendix~\ref{Proof of logical_X_stitching} for the proof.
\end{IEEEproof}
\end{lemma}

Next, consider the stitching of logical-$Z$ rooted circuits.

\begin{lemma}
\label{lem:logical_Z_stitching}
The rooted circuits for logical-$Z$ constraints in Lemma~\ref{lem:qsk_logical_Pauli_mappings} can be stitched by concatenating them and dropping duplicate CZ gates.
The stitched circuit satisfies all the logical-$Z$ constraints simultaneously.
\begin{IEEEproof}
\normalfont
See Appendix~\ref{Proof of logical_Z_stitching} for the proof.
\end{IEEEproof}
\end{lemma}

\begin{remark}
\label{rem:logical_Z_stitching_CNOTs_first}
The $\text{CNOT}_{j+1 \rightarrow i+1}$ gates to produce $Z_{j+1}$ for $j \notin I_h$ can as well be moved to the beginning of the rooted circuit for each logical-$Z$ constraint.
In this case, the $Z_{i+1}$ first produces $Z_{i+1} Z_{j+1}$ before transforming itself to $Y_{i+1}$, which does not affect the arguments made above for stitching.
Furthermore, these are identical to the CNOTs used for stitching logical-$X$ rooted circuits in Lemma~\ref{lem:logical_X_stitching}.
\end{remark}

Now, we are ready to prove the main result of this section: the construction of the complete physical circuit for logical C-QSK on $\llbr n,n-2,2 \rrbr$ codes by stitching the logical-$X$ and logical-$Z$ stitched circuits, as shown in Fig.~\ref{fig:qsk_distance2_physical_full}.


\begin{figure*}
\centering
\scalebox{0.92}{
\begin{tikzpicture}

\node[draw,rectangle,minimum height=3cm, minimum width=0.2cm,align=center] (CZij) at (0.3,0) {$\text{CZ}_{i+1,j+1}$\\ gates\\ $i \notin I_h$\\ $j \notin I_h$};

\node[draw,rectangle,minimum height=3cm, minimum width=0.2cm,align=center,fill=gray!40] (CZin) at (2,0) {$\text{CZ}_{i+1,n}$\\ gates\\ $i \notin I_h$};

\node[draw,rectangle,minimum height=3cm, minimum width=0.2cm,align=center] (Pi) at (3.9,0) {\colorbox{gray!40}{$P_n$ or $P_n^{\dagger}$},\\ $P_{i+1}$\\ gates\\ $i \notin I_h$};

\node[draw,rectangle,minimum height=3cm, minimum width=0.2cm,align=center] (CNOTij) at (6.1,0) {$\text{CX}_{i+1 \rightarrow j+1}$\\ gates\\ $i \notin I_h$\\ $j \in I_h$};
    
\node[draw,rectangle,minimum height=3cm, minimum width=0.2cm,align=center,fill=gray!40] (CNOTi1) at (8.1,0) {\colorbox{gray!40}{$\text{CX}_{n \rightarrow 1}$},\\ $\text{CX}_{i+1 \rightarrow 1}$\\ gates\\ $i \notin I_h$};
    
\node[draw,rectangle,minimum height=3cm, minimum width=0.2cm,align=center,fill=gray!40] (CNOTni) at (9.9,0) {$\text{CX}_{n \rightarrow i+1}$\\ gates\\ $i \in I_h$};
    
\node[draw,rectangle,minimum height=3cm, minimum width=0.2cm,align=center] (Pi) at (11.85,0) {\colorbox{gray!40}{$H_{1}$-$P_{1}$},\\ $H_{i+1}$-$P_{i+1}$\\ gates\\ $i \in I_h$};

\node[draw,rectangle,minimum height=3cm, minimum width=0.2cm,align=center] (CZij2) at (13.83,0) {\colorbox{gray!40}{$\text{CZ}_{i+1,1}$},\\ $\text{CZ}_{i+1,j+1}$\\ gates\\ $i,j \in I_h$};

\node[draw,rectangle,minimum height=3cm, minimum width=0.2cm,align=center] (Pi) at (15.83,0) {\colorbox{gray!40}{$H_1$}, $H_{i+1}$\\ gates\\ $i \in I_h$};

\end{tikzpicture}
}

\caption{\label{fig:qsk_distance2_physical_full} The sequence of gates in the full physical circuit realizing the logical C-QSK (i.e., Clifford Trotter) circuit on the $\llbr n,n-2,2 \rrbr$ codes. The set $I_h \subseteq [n-2]$ consists of the indices of logical qubits on which the logical circuit applies a Hadamard. The white blocks and non-highlighted gates are common to the cases of even and odd number of Hadamards, i.e., $|I_h|$, but the gray blocks and the highlighted gates are needed only for the case of odd number of Hadamards. From Corollary~\ref{cor:qsk_depth_distance2_logical_id_H}, there are additional optimizations possible to reduce the gate count and circuit depth, which are described in Algorithm~\ref{alg:solve-and-stitch}.}
\end{figure*}

\begin{theorem}
\label{thm:logical_X_Z_stitching}
The logical C-QSK (i.e., Clifford Trotter) circuit can be realized on the $\llbr n,n-2,2 \rrbr$ codes by concatenating the logical-$X$ and logical-$Z$ stitched circuits from Lemmas~\ref{lem:logical_X_stitching} and~\ref{lem:logical_Z_stitching}, then dropping duplicate CNOT gates.
If the number of Hadamards in the logical circuit is odd, then CNOTs and Phase gates can be added to preserve the stabilizer group.
\begin{IEEEproof}
\normalfont
See Appendix~\ref{Proof of logical_X_Z_stitching} for the proof.
\end{IEEEproof}
\end{theorem}

\begin{remark}
\label{rem:logical Hy}
For a logical C-QSK circuit on the $\llbr n,n-2,2 \rrbr$ code that includes a combination of Hadamard and $H_y$ gates, the physical construction closely resembles that of a Hadamard-only C-QSK circuit, with modifications depending on the parities of the Hadamard and $H_y$ gates in the logical circuit; refer to Algorithm~\ref{alg:solve-and-stitch-Hy} for the complete procedure.
\end{remark}

\begin{algorithm}
\begin{algorithmic}[1]

\caption{\label{alg:solve-and-stitch} Solve-and-Stitch algorithm to realize logical C-QSK (i.e., Clifford Trotter) circuits on $\llbr n,n-2,2 \rrbr$ codes}

\State \textbf{Input:} Logical C-QSK circuit on an even number, $k$, of qubits implementing an exponentiated Pauli operator $\exp\left( -\imath\frac{\pi}{4} E_1 E_2 \cdots E_k \right)$, where $E_i \in \{ X,Z \}$

\State \textbf{Output:} Physical circuit on $n=k+2$ qubits efficiently realizing the logical C-QSK circuit on the $\llbr n,n-2,2 \rrbr$ code, as shown in Fig.~\ref{fig:qsk_distance2_physical_full} (optimizations from Cor.~\ref{cor:qsk_depth_distance2_logical_id_H})

\State \textbf{Initialization:} $n \coloneqq k+2,\ I_h \coloneqq \{ i \in [n-2] \colon E_i = X \}$
\State \textbf{Initialization:} Set $\texttt{logical\_Hadamard\_all} = 0$

\If{$|I_h| > (n-2)/2$}
\State Set $\texttt{logical\_Hadamard\_all} = 1$
\State $E_i \gets H E_i H$; \ $I_h \gets I_h^c = [n-2] \setminus I_h$
\State Apply $H_t$ for $t \in [n]$, SWAP qubits $1$ and $n$ 

\Comment{This implements the logical Hadamard on all $k = n-2$ logical qubits, which will be repeated at the end of the circuit to retain the desired functionality}
\EndIf

\Comment{The sequence of gates below follows Fig.~\ref{fig:qsk_distance2_physical_full}}

\State Apply $\text{CZ}_{i+1,j+1}$ gates for all pairs $i,j \notin I_h$ 

\State Apply $\text{CZ}_{s,t}, P_s$ gates for all pairs $s,t \in [n]$ and cancel the CZ gates from above to reduce CZ count \\ \Comment{These CZ and $P$ gates together implement the logical identity on the code, so they do not affect functionality} \\ \Comment{Besides these cancellations, the complement in Line 7 greatly reduces the number of CZ gates in Line 25 too}

\If{$|I_h|$ is odd}
\State Apply $\text{CZ}_{i+1,n}$ gates for $i \notin I_h$
\EndIf

\State Apply $P_{i+1}$ for each $i \notin I_h$

\If{$|I_h|$ is odd}
\State Apply $P_n$ \Comment{Sometimes $P_n^{\dagger}$ may be appropriate}
\EndIf

\State Apply $\text{CNOT}_{i+1 \rightarrow j+1}$ for each $i \notin I_h, j \in I_h$

\If{$|I_h|$ is odd}
\State Apply $\text{CNOT}_{i+1 \rightarrow 1}$ gates for $i \notin I_h$
\State Apply $\text{CNOT}_{n \rightarrow 1}$
\State Apply $\text{CNOT}_{n \rightarrow i+1}$ gates for $i \in I_h$
\EndIf

\algstore{myalg}

\end{algorithmic}

\end{algorithm}

\begin{algorithm}
\begin{algorithmic}[1]

\algrestore{myalg}

\State Apply $H_{i+1}$ followed by $P_{i+1}$ for each $i \in I_h$

\If{$|I_h|$ is odd}
\State Apply $H_1$ followed by $P_1$
\EndIf

\State Apply $\text{CZ}_{i+1,j+1}$ gates for all pairs $i,j \in I_h$

\If{$|I_h|$ is odd}
\State Apply $\text{CZ}_{i+1,1}$ for each $i \in I_h$
\EndIf

\State Apply $H_{i+1}$ for each $i \in I_h$

\If{$|I_h|$ is odd}
\State Apply $H_1$
\EndIf

\Comment{This completes the sequence of gates from Fig.~\ref{fig:qsk_distance2_physical_full}}

\State Check for all the mappings in Corollary~\ref{cor:qsk_physical_Pauli_mappings_evenH} ($|I_h|$ even) or Corollary~\ref{cor:qsk_physical_Pauli_mappings_oddH} ($|I_h|$ odd)

\If{some signs in the mappings are not satisfied}
\State Apply an appropriate Pauli operator to fix all the signs

\Comment{Such an operator can be found efficiently using the binary representation of Pauli operators, e.g., see~\cite{Rengaswamy-tqe20}}
\EndIf

\If{$\texttt{logical\_Hadamard\_all} = 1$}
\State Apply $H_t$ for $t \in [n]$, SWAP qubits $1$ and $n$ \\ \Comment{This complements the action of Line 8}
\EndIf

\State \textbf{Return:} Physical circuit constructed by the procedure

\end{algorithmic}
\end{algorithm}

\begin{algorithm}
\begin{algorithmic}[1]

\State \textbf{Input:} Logical C-QSK circuit on an even number, $k$, of qubits implementing an exponentiated Pauli operator $\exp\left( -\imath\frac{\pi}{4} E_1 E_2 \cdots E_k \right)$, where $E_i \in \{ X,Y,Z \}$

\State \textbf{Output:} Physical circuit on $n=k+2$ qubits efficiently realizing the logical C-QSK circuit on the $\llbr n,n-2,2 \rrbr$ code

\State \textbf{Initialization:} $n \coloneqq k+2,\ I_h \coloneqq \{ i \in [n-2] \colon E_i = X \}, \ I_{hy} \coloneqq \{ i \in [n-2] \colon E_i = Y \}, \ I_e \coloneqq \{ i \in [n-2] \colon E_i = Z \}$

\State Replace all $H_y$ gates into Hadamard gates, run Algorithm ~\ref{alg:solve-and-stitch}, generate physical circuit.

\If{$|I_h|$ is even, $|I_{hy}|$ is even}
\State Apply $\text{P}_{I_{hy}+1}$ at the beginning and the end of the physical circuit generated in Line 4.
\EndIf

\If{$|I_h|$ is odd, $|I_{hy}|$ is odd}
\State Apply $\text{P}_{I_{hy}+1}$ at the beginning and the end of the physical circuit generated in Line 4, then apply $\text{CZ}_{n,I_e+1}$, $\text{CNOT}_{n\rightarrow{I_h+1}}$, $\text{CNOT}_{n\rightarrow{I_{hy}+1}}$, $\text{CZ}_{n,I_{hy}+1}$ at the end.
\EndIf

\If{$|I_h|$ is odd, $|I_{hy}|$ is even}
\State Apply $\text{CZ}_{I_{hy}+1,I_e+1}$ at the beginning of the physical circuit generated in Line 4, then apply additional $\text{P}_{I_{hy}+1}$ at the position of original Phase gates in the physical circuit from Line 4, and duplicate these Phase gates at the end of the circuit as well.
\EndIf

\If{$|I_h|$ is even, $|I_{hy}|$ is odd}
\State Delete all $\text{CZ}$ and $\text{P}$ in logical-$X$ component of the physical circuit generated in Line 4, then delete all gates involving in the last qubit, then apply $\text{CZ}_{I_{hy}+1,I_e+1}$ at the beginning, then apply additional $\text{P}_{I_{hy}+1}$ just before logical-$Z$ component of the circuit, and duplicate these Phase gates at the end of the circuit as well.
\EndIf

\State Check for all the Pauli constraints

\If{some signs in the mappings are not satisfied}
\State Apply an appropriate Pauli operator to fix all the signs

\Comment{Such an operator can be found efficiently using the binary representation of Pauli operators, e.g., see~\cite{Rengaswamy-tqe20}}
\EndIf

\State \textbf{Return:} Physical circuit constructed by the procedure

\caption{\label{alg:solve-and-stitch-Hy} Solve-and-Stitch algorithm to realize logical C-QSK (i.e., Clifford Trotter) circuits with mixed Hadamard and $H_y$ gates on $\llbr n,n-2,2 \rrbr$ codes}
\end{algorithmic}
\end{algorithm}

\section{Depth Analysis and Circuit Optimization}
\label{sec:depth_analysis}

In our endeavor of constructing physical circuits for realizing logical C-QSK circuits on the $\llbr n,n-2,2 \rrbr$ code family, we have successfully demonstrated the effectiveness of the idea of root qubits. 
However, a pertinent question arises: can we optimize the circuit to reduce the number of two-qubit gates and, consequently, minimize the circuit depth? 
In this section we show that this is indeed possible.
First, we prove that the circuit depth grows \emph{quadratically} with $k$, independent of the number of Hadamard gates in the logical circuit.
To reduce the depth, we introduce a promising optimization strategy: the inclusion of ``complementary'' CZ gates, which together with $P$ gates on all $n$ qubits only realizes the logical identity on the code, thereby retaining the functionality of the physical circuits. 
This approach ensures that the circuit depth grows only \emph{linearly} with $k$, which is optimal since the logical circuit itself has linear depth (see Section~\ref{sec:qsk}). 

Let $h \coloneqq |I_h|$ denote the number of Hadamard gates in the logical C-QSK circuit implementing $\exp\left( -\imath\frac{\pi}{4} E_1 E_2 \cdots E_k \right)$, where $E_i \in \{ X,Z \}$, i.e., $h = |\{ i \in [n-2] \colon E_i = X \}|$.
We will use the sequence of gates in Fig.~\ref{fig:qsk_distance2_physical_full} for all arguments.
The validity of this sequence to implement the logical C-QSK circuit has been established in Theorem~\ref{thm:logical_X_Z_stitching}.
For all depth calculations, we assume that each two-qubit gate increases the depth by $1$, even though this may not be the case for some hardware that can perform gates in a parallel fashion.
A layer of identical single-qubit gates on multiple qubits is also considered depth-$1$.
Hence, our expressions can safely be considered as upper bounds to the true depth in practice.

\begin{theorem}
\label{thm:qsk_depth_distance2}
Consider the physical circuit constructed by the solve-and-stitch algorithm in Theorem~\ref{thm:logical_X_Z_stitching} to realize a logical C-QSK circuit on the $\llbr n,n-2,2 \rrbr$ codes. 
The depth of this circuit is upper bounded by $\frac{k(k-1)}{2} + 5$ when $h$ is even and $\frac{(k+2)(k+1)}{2} + 5$ when $h$ is odd, where $k = n-2$.
\begin{IEEEproof}
\normalfont
See Appendix~\ref{Proof of qsk_depth_distance2} for the proof.
\end{IEEEproof}
\end{theorem}

\begin{figure}
\centering

\includegraphics[scale=1,keepaspectratio]{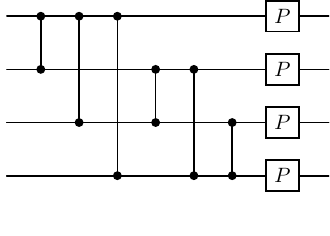}

\caption{\label{fig:logical_identity} Logical identity gadget for the $\llbr 4,2,2 \rrbr$ code.}
\end{figure}

We have demonstrated that the depth of physical circuits grows quadratically with the dimension of the $\llbr n,n-2,2 \rrbr$ codes. 
To mitigate this, we propose the incorporation of ``complementary'' CZ gates at the beginning of the physical circuit.
When combined with $P$ gates on all $n$ qubits, these CZ gates only implement the logical identity on the code, thereby not affecting the logical functionality of the physical circuit.
Figure~\ref{fig:logical_identity} illustrates an example of this logical identity circuit for the $\llbr 4,2,2 \rrbr$ code, which can be extended to any member of the $\llbr n,n-2,2 \rrbr$ code family. 
Indeed, this CZ-$P$ gadget induces the mappings $\overline{Z_{i}} = Z_{i+1} Z_{n} \mapsto Z_{i+1} Z_{n}$ and $\overline{X_{i}} = X_{1} X_{i+1} \mapsto X_{1} X_{i+1}$, which are trivial, indicating that all logical Pauli operators remain unchanged. 
It is easily checked that the stabilizers are also preserved.

Therefore, by implementing the logical identity at the beginning of the physical circuit in Fig.~\ref{fig:qsk_distance2_physical_full}, we can cancel the $\text{CZ}_{i+1,n}$ gates for all $i \notin I_h$. 

Importantly, the depth increase scales as $\mathcal{O}(kh)$, where $k = n - 2$ is the number of logical qubits and $h$ is the number of Hadamard gates in the logical Trotter circuit. In many physically motivated Hamiltonians, such as Ising or Heisenberg or Fermi-Hubbard models, the number of Hadamard gates is small compared to $k$ and can be treated as a fixed constant. Consequently, the physical depth scales essentially linearly with $k$. Since the logical Clifford Trotter circuit itself has depth that grows linearly with $k$, we expect any reasonable fault-tolerant physical implementation to exhibit the same scaling trend. Thus, achieving a physical circuit depth of $\mathcal{O}(k)$—even with added gadgets for fault tolerance—can be considered optimal in scaling within this context.
The following theorem provides expressions for the depth of the physical circuit incorporating these cancellations, asserting that the circuit's depth grows only linearly with $k$, albeit quadratically with $h$.






\begin{theorem}
\label{thm:qsk_depth_distance2_logical_id}
Consider the physical circuit constructed by the solve-and-stitch algorithm in Theorem~\ref{thm:logical_X_Z_stitching} to realize a logical C-QSK circuit on the $\llbr n,n-2,2 \rrbr$ codes. 
After integrating the logical identity into this circuit, the depth is at most
\begin{align}
\begin{cases}
(2+2h)k-h^2 - h + 6 & \text{for\ even}\ h, \\
(2+2h)k - h^2 + h + 7 & \text{for\ odd}\ h.
\end{cases}
\end{align}
\begin{IEEEproof}
\normalfont
See Appendix~\ref{Proof of qsk_depth_distance2_logical_id} for the proof.
\end{IEEEproof}
\end{theorem}

To illustrate the gains from the optimized circuits, in Fig.~\ref{fig:solve-stitch_depth_scaling} we plot the relationship between $k$ and the depth when $h=2$. 
The depth of the optimized physical circuit is $6k-1$ in this illustration. 
Observe that for lower values of $k$, the original circuit from Fig.~\ref{fig:qsk_distance2_physical_full} has a smaller depth. 
However, when $k \geq 14$, the optimized circuit from Theorem~\ref{thm:qsk_depth_distance2_logical_id} proves advantageous. 
It is also natural to consider the setting of fixed $k$ and increasing $h$.
Here, we can leverage the structure of the problem, specifically the exponentiated Pauli operator $\exp\left( -\imath\frac{\pi}{4} E_1 E_2 \cdots E_k \right)$, where $E_i \in \{ X,Z \}$.
We realize that the operator can be sandwiched within transversal Hadamard gates on the $k$ qubits to swap $X$s with $Z$s.
Such a logical operation can be implemented on the code by performing physical transversal Hadamard gates and swapping qubits $1$ and $n$.
From Section~\ref{sec:distance_2_family} it is clear that this satisfies the necessary mappings on the logical Pauli operators and preserves the stabilizers.
The operation itself adds depth $3$, split into $2$ at the beginning and $1$ at the end of the sandwich, since Fig.~\ref{fig:qsk_distance2_physical_full} already contains a layer of Hadamards at the end.
Intuitively, when $h > k/2$, we can use this strategy to fall back to the case of $h < k/2$ and hence achieve a lower depth circuit. 
The following corollary establishes this rigorously.

\begin{corollary}
\label{cor:qsk_depth_distance2_logical_id_H}
When $h > k/2$, the logical C-QSK circuit can be realized on the $\llbr n,n-2,2 \rrbr$ codes with depth at most 
\begin{align}
\begin{cases}
k(k+1) - h^2 + h + 9 & \text{for\ even}\ h, \\
k(k+3) - h^2 - h + 10 & \text{for\ odd}\ h.
\end{cases}
\end{align}
These are respectively smaller than the depths in Theorem~\ref{thm:qsk_depth_distance2_logical_id}.
\begin{IEEEproof}
\normalfont
See Appendix~\ref{Proof of qsk_depth_distance2_logical_id_H} for the proof.
\end{IEEEproof}
\end{corollary}

Algorithm~\ref{alg:solve-and-stitch} carefully describes the complete procedure to synthesize Hadamard-only logical C-QSK circuits on the $\llbr n,n-2,2 \rrbr$ codes, while Algorithm~\ref{alg:solve-and-stitch-Hy} describes the procedure to synthesize logical C-QSK circuits with mixed Hadamard and $H_y$ gates on the $\llbr n,n-2,2 \rrbr$ codes based on Algorithm~\ref{alg:solve-and-stitch}. We emphasize that Algorithm~\ref{alg:solve-and-stitch-Hy} was not already introduced in~\cite{Chen-qce24}.

\begin{figure}[t]
\centering
\vspace*{-10pt}

\hspace*{-15pt}
\scalebox{0.65}{


\begin{tikzpicture}

\begin{axis}[%
width=4.521in,
height=3.566in,
at={(0.758in,0.481in)},
scale only axis,
xmin=0,
xmax=30,
xlabel style={font=\color{white!15!black}},
xlabel={Number of logical qubits, $k$},
ymin=0,
ymax=450,
ylabel style={font=\color{white!15!black}},
ylabel={Depth of complete physical circuit},
xmajorgrids, ymajorgrids,
axis background/.style={fill=white},
title style={font=\bfseries},
legend style={at={(0.025,0.85)}, anchor=south west, legend cell align=left, align=left, draw=white!15!black}
]
\addplot [color=black, mark size=1.8pt, mark=square, mark options={solid, black}]
  table[row sep=crcr]{%
2	5\\
4	10\\
6	19\\
8	32\\
10	49\\
12	70\\
14	95\\
16	124\\
18	157\\
20	194\\
22	235\\
24	280\\
26	329\\
28	382\\
30	439\\
};
\addlegendentry{Circuit from Theorem~12 (shown in Fig.~6)}

\addplot [color=red, mark size=2.5pt, mark=asterisk, mark options={solid, red}]
  table[row sep=crcr]{%
2	11\\
4	23\\
6	35\\
8	47\\
10	59\\
12	71\\
14	83\\
16	95\\
18	107\\
20	119\\
22	131\\
24	143\\
26	155\\
28	167\\
30	179\\
};
\addlegendentry{Optimized circuit from Theorem~13}

\end{axis}

\begin{axis}[%
width=5.833in,
height=4.375in,
at={(0in,0in)},
scale only axis,
xmin=0,
xmax=1,
ymin=0,
ymax=1,
axis line style={draw=none},
ticks=none,
axis x line*=bottom,
axis y line*=left
]
\end{axis}
\end{tikzpicture}%
}

\vspace*{-8pt}
\caption{Depth of original circuit from Theorem~\ref{thm:qsk_depth_distance2} and the optimized circuit from Theorem~\ref{thm:qsk_depth_distance2_logical_id}, where $h=2$ is fixed.}
\label{fig:solve-stitch_depth_scaling}

\vspace*{-10pt}
\end{figure}


\begin{remark}
Throughout our analysis, we have only considered logical C-QSK circuit on even number of qubits, since $k = n-2$ is even for the code family.
To construct the logical circuit for odd number of qubits, one can transform this to the even setting as follows. 
Add a qubit at the top of the logical circuit, initialize it to $\ket{0}$, and perform CNOTs to the last qubit as always.
The modification clearly doesn't alter the functionality, but the resulting circuit has even number of qubits as necessary for our constructions.
The depth overhead from this procedure is minimal in that it does not alter the linear scaling with $k$.
\end{remark}

\section{Discussion on Solve and Stitch}
\label{sec:discussion}

We have systematically explored the construction of physical circuits for logical C-QSK circuits on $\llbr n,n-2,2 \rrbr$ codes. 
However, it is essential to note that in certain cases, while the root qubit idea and the construction of rooted circuits still apply, some principles may vary. 
We contend that the root qubit idea could potentially lead to a further reduction in circuit depth compared to algorithms such as LCS, warranting further exploration and consideration.
In this section, we briefly illustrate this under a few different settings.

\subsection{Logical C-QSK Circuit on Odd Number of Qubits}
\label{sec:logical_qsk_odd}

\begin{figure}
\centering

\includegraphics[scale=0.87,keepaspectratio]{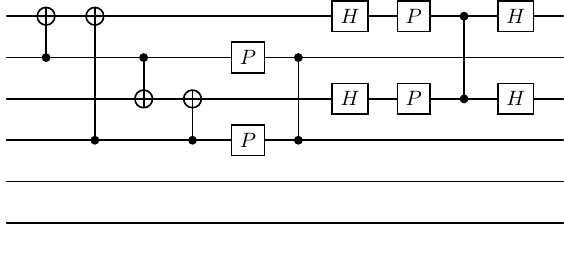}

\vspace*{-15pt}
\caption{\label{fig:solve-stitch_642_qsk_3qubit_no-swap} Physical realization of Fig.~\ref{fig:qsk_3qubit} on the $\llbr 6,4,2 \rrbr$ code produced by the solve-and-stitch approach using root qubits. }

\vspace*{-12pt}
\end{figure}

Let us consider the $3$-qubit logical C-QSK circuit in Fig.~\ref{fig:qsk_3qubit}, which we will realize on the $\llbr 6,4,2 \rrbr$ code. 
In this context, we treat the fourth logical qubit as idle. 
The Pauli mapping rules in this scenario are listed below:
\begin{IEEEeqnarray}{rClCrCl}
\label{eq:logical_qsk_3qubit_constraints_physical}
X_1 X_2 & \mapsto & Y_{2} X_{3} Z_{4} & \quad , \quad & Z_2 Z_6 & \mapsto & Z_2 Z_6 \ , \nonumber \\
X_1 X_3 & \mapsto & X_1 X_3 & \quad , \quad & Z_3 Z_6 & \mapsto & - X_{1} Z_{2} Y_{3} Z_{4} Z_{6} \ , \nonumber \\
X_1 X_4 & \mapsto & Z_{2} X_{3} Y_{4} & \quad , \quad & Z_4 Z_6 & \mapsto & Z_4 Z_6 \ , \nonumber \\
X_1 X_5 & \mapsto & X_1 X_5 & \quad , \quad & Z_5 Z_6 & \mapsto & Z_5 Z_6 \ .
\end{IEEEeqnarray}
For the construction of logical-$X$ rooted circuits, the methodology is similar: selecting root qubits and avoiding the use of the same gates more than once when building the physical circuit. 
In contrast, for logical-$Z$, there is only one non-trivial rooted circuit, which reveals only half of the $H$-$P$-CZ-$H$ structure. 
However, as we construct the physical circuit, a complete and whole structure emerges, preserving the stabilizers (Fig.~\ref{fig:solve-stitch_642_qsk_3qubit_no-swap}). 
This illustrates that even if certain patterns in the physical circuit don't explicitly appear in the rooted circuits, we can still construct the physical circuit based on previous knowledge about the circuit and the code.












\subsection{Solve-and-Stitch for Single Logical Hadamard}
\label{sec:solve-and-stitch_logical_H}

\begin{figure}[t]
    \centering
    \vspace*{-10pt}
    \includegraphics[scale=1,keepaspectratio]{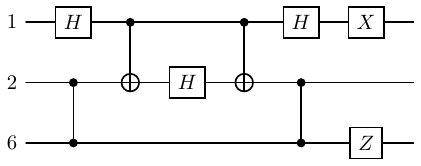}
    \caption{Physical circuit proposed in~\cite{Chao-npjqi18} for realizing $\overline{H_{1}}$.}
    \label{fig:Chao_642_singleH}
\end{figure}

\begin{figure}[t]
    \centering
    \includegraphics[scale=0.9,keepaspectratio]{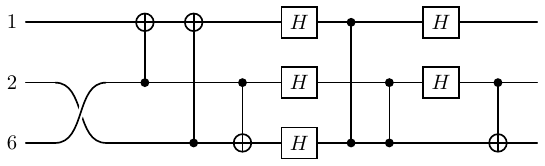}
    \caption{Physical circuit from LCS algorithm~\cite{Rengaswamy-tqe20} for $\overline{H_{1}}$.}
    \label{fig:lcs_642_singleH}
\end{figure}

\begin{figure}[t]
\centering

\includegraphics[scale=1,keepaspectratio]{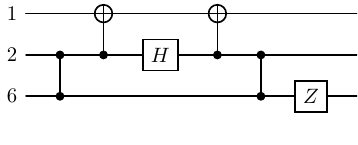}

\caption{\label{fig:solve-stitch_642_singleH}Physical circuit from solve-and-stitch for $\overline{H_{1}}$.}
\end{figure}

In this example, we consider realizing a single logical Hadamard gate, $\overline{H_{1}}$, on the $\llbr 6,4,2 \rrbr$ code instead of an entire Trotter block.
This highlights the generalizability of the solve-and-stitch approach beyond C-QSK circuits.
The non-trivial physical Pauli mappings are listed below:
\begin{IEEEeqnarray}{rClCrCl}
\label{eq:single_logical_Hadamard_constraints_physical}
X_1 X_2 & \mapsto & Z_{2}Z_{6} & \quad , \quad & Z_2 Z_6 & \mapsto & X_{1}X_{2}
\end{IEEEeqnarray}
Several circuits can satisfy the given Pauli constraints. 
In~\cite{Chao-npjqi18}, they proposed the circuit in Fig.~\ref{fig:Chao_642_singleH}, which exhibits certain symmetric properties and has depth just $8$. 
In contrast, the circuit generated by the LCS algorithm in Fig.~\ref{fig:lcs_642_singleH} has depth $11$ (assuming that the swap gate is physically implemented with a depth of $3$). 
Notably, it lacks the symmetric structure and appears more complex than the circuit in Fig.~\ref{fig:Chao_642_singleH}. 
However, our root qubit idea proves effective and yields a simpler circuit with depth just $6$, as shown in Fig.~\ref{fig:solve-stitch_642_singleH}.

\begin{figure}[h]
\begin{subfigure}[t]{0.5\textwidth}
\centering

\includegraphics[scale=1,keepaspectratio]{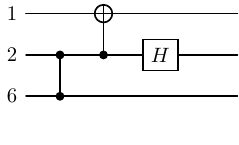}

\caption{\label{fig:XX_IZIIIZ}$X_{1}{X_{2}}\mapsto Z_{2}Z_{6}$}
\end{subfigure}%
\begin{subfigure}[t]{0.5\textwidth}
\centering

\includegraphics[scale=1,keepaspectratio]{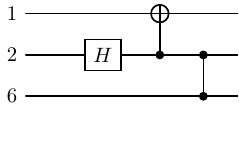}

\caption{\label{fig:IZIIIZ_XX}$Z_{2}{Z_{6}}\mapsto X_{1}X_{2}$}
\end{subfigure}
\caption{\label{fig:logical_H_rooted_circuits} Rooted circuits for realizing $\overline{H_{1}}$ on $\llbr 6,4,2 \rrbr$ code}
\end{figure}

For the mapping $X_{1}{X_{2}} \mapsto Z_{2}Z_{6}$ in Fig.~\ref{fig:XX_IZIIIZ}, we choose the second qubit as the root. 
We first apply $\text{CZ}_{2 \rightarrow 6}$ to transform $X_{2}$ into $X_{2}Z_{6}$, then implement $\text{CNOT}_{2 \rightarrow 1}$ to transform $X_{1}X_{2}$ into $X_{2}$, and finally apply a Hadamard on the second qubit to transform $X_{2}$ into $Z_{2}$. 
For the mapping $Z_{2}{Z_{6}} \mapsto X_{1}X_{2}$, we reverse this process, as shown in Fig.~\ref{fig:IZIIIZ_XX}.

In the construction of the physical circuit in Fig.~\ref{fig:solve-stitch_642_singleH}, we leverage the symmetric property of Pauli mappings. 
We drop a Hadamard gate and seamlessly stitch the two rooted circuits together. 
This process is slightly different from other examples provided previously since $\text{CNOT}_{2 \rightarrow 1}$ and $\text{CZ}_{2 \rightarrow 6}$ appear twice in the physical circuit. 
Moreover, we add $Z_1$ at the end to ensure that the stabilizers remain preserved exactly. 
The reduction in circuit depth emphasizes the efficacy of the root qubit idea in simplifying the physical circuit.

\subsection{Solve-and-Stitch on Hypergraph Product Code}

To further illustrate the generalizability of the solve-and-stitch construction, we provide a concrete example using a $\llbr 20,4,2 \rrbr$ hypergraph product code. The logical Clifford Trotter circuit we consider corresponds to the structure depicted in Fig.~\ref{fig:qsk_4qubit_evenH}, applied to four logical qubits.

The logical operators for this code are:
\begin{IEEEeqnarray}{rClCrCl}
\begin{aligned}
\bar{X}_1 &= X_1 X_3, \quad & \bar{Z}_1 &= Z_1 Z_9, \\
\bar{X}_2 &= X_2 X_4, \quad & \bar{Z}_2 &= Z_2 Z_{10}, \\
\bar{X}_3 &= X_5 X_7, \quad & \bar{Z}_3 &= Z_5 Z_{13}, \\
\bar{X}_4 &= X_6 X_8, \quad & \bar{Z}_4 &= Z_6 Z_{14}.
\end{aligned}
\label{eq:logical-operators-hypergraph}
\end{IEEEeqnarray}

The 16 stabilizers of the code are given by:
\begin{equation}
\begin{aligned}
&S_1 = Z_1 Z_3 Z_{17},\quad S_2 = X_1 X_9 X_{17},\quad S_3 = Z_2 Z_4 Z_{18}, \\
&S_4 = X_2 X_{10} X_{18}, \quad S_5 = Z_5 Z_7 Z_{19},\quad S_6 = X_4 X_{12} X_{18}, \\
&S_7 = Z_6 Z_8 Z_{20},\quad S_8 = X_4 X_{12} X_{18}, \quad S_9 = Z_9 Z_{11} Z_{17},\\
&S_{10} = X_5 X_{13} X_{19},\quad S_{11} = Z_{10} Z_{12} Z_{18},\quad S_{12} = X_6 X_{14} X_{20}, \\
&S_{13} = Z_{13} Z_{15} Z_{19},\quad S_{14} = X_7 X_{15} X_{19},\\
&S_{15} = Z_{14} Z_{16} Z_{20},\quad S_{16} = X_8 X_{16} X_{20}.
\end{aligned}
\label{eq:stabilizers-hypergraph}
\end{equation}

Given these stabilizers and logical operators, we derive the physical circuit by solving the Pauli propagation constraints induced by the logical Trotter circuit. These constraints are expressed as follows:
\begin{align}
\label{eq:logical_qsk_3qubit_constraints_physical}
X_1 X_3 & \mapsto Y_{1} X_{2} X_{2}X_{3}X_{4}X_{5}Z_{6}X_{7}Z_{9}Z_{14} \ , \nonumber \\ 
Z_1 Z_9 & \mapsto Z_1 Z_9 \ , \nonumber \\
X_2 X_4 & \mapsto X_2 X_4 \ , \nonumber \\
Z_2 Z_{10} & \mapsto Z_{1} Y_{2} X_{4} X_{5} Z_{6} X_{7}Z_{9}Z_{13}Z_{14} \ , \nonumber \\
X_5 X_7 & \mapsto X_{5} X_{7} \ , \nonumber \\ 
Z_5 Z_{13} & \mapsto Z_1 X_2 X_4 Y_5 Z_6 X_7 Z_9 Z_{13} Z_{14} \ , \nonumber \\
X_6 X_8 & \mapsto Z_1 X_2 X_4 X_5 Y_6 X_7 X_8 Z_9 Z_{14} , \nonumber \\ 
Z_6 Z_{14} & \mapsto Z_6 Z_{14}  \ .
\end{align}

\begin{figure}
    \centering
    \includegraphics[width=1\linewidth]{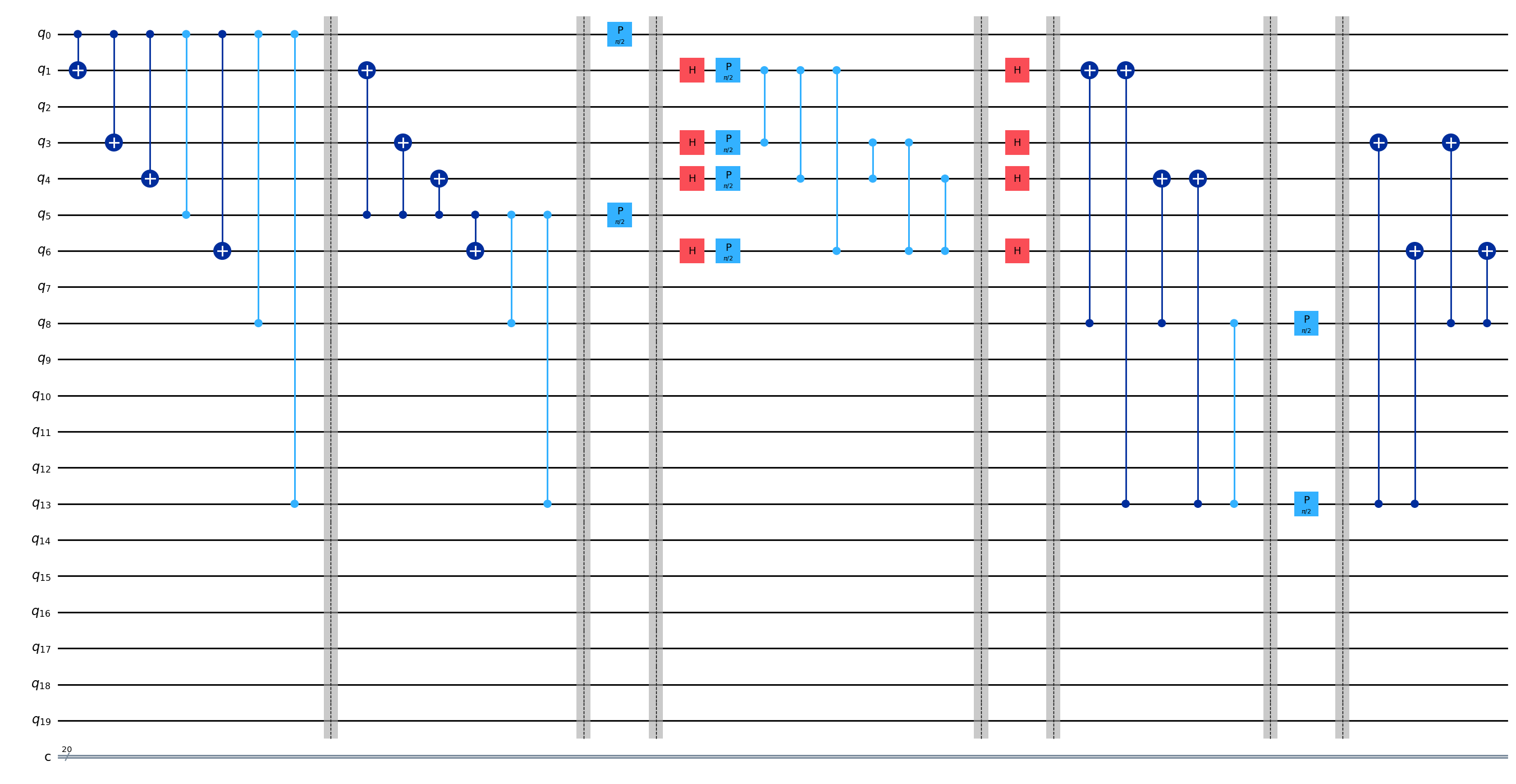}
    \caption{Physical implementation of the four-qubit logical Clifford Trotter circuit in Fig.~\ref{fig:qsk_4qubit_evenH} on the $\llbr 20,4,2 \rrbr$ hypergraph product code using the solve-and-stitch framework. The circuit is constructed by stitching together physical gadgets derived from Pauli propagation constraints, based on root qubit selection as described in the main text. The final circuit includes additional layers at the end to preserve all stabilizers. Gray dashed lines indicate manually enforced ordering of gates in IBM Qiskit to follow the intended solve-and-stitch structure. These annotations override automatic gate reordering by the visualization tool, ensuring that the modular block pattern is preserved visually.}
    \label{fig:hypergraph_physical_circuit}
\end{figure}

From the solution to these Pauli propagation constraints, we determine the root qubits for each logical gate gadget following the same procedure described in the main text. For the logical $X$ component, we examine the first Pauli constraint and observe that $X_1$ maps to $Y_1$; among all outputs, qubit 1 is the only one with a $Y$ term, which uniquely identifies it as a suitable root qubit. Similarly, for another non-trivial Pauli constraint, we select qubit 6 as the root qubit for the same reason. For the logical $Z$ component, we choose qubits 2 and 5 as root qubits based on the corresponding propagation patterns. Using these root qubits, we construct the physical gadgets for each logical Clifford gate using the solve-and-stitch method. These small circuits are then combined to form the full physical circuit. To preserve the code stabilizers across the entire circuit, we append additional gates at the end of the circuit. The resulting physical circuit has depth 33 and preserves the modular block structure introduced by solve-and-stitch. An illustration of the complete circuit is shown in Fig.~\ref{fig:hypergraph_physical_circuit}.
This example demonstrates that the solve-and-stitch framework extends beyond the $\llbr n,n-2,2 \rrbr$ family and applies to more general stabilizer codes such as hypergraph product codes. The process of propagating Pauli constraints and selecting root qubits remains consistent and modular, supporting the broader claim of generalizability.

\section{Fault-Tolerant QSK via Flag Qubits}
\label{sec:ft_qsk_flags}

\begin{figure}[b]
   \centering

\includegraphics[scale=1,keepaspectratio]{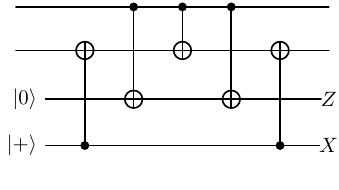}

\caption{\label{fig:flag_cx} Flag gadget on a CNOT to detect errors $XI$, $IZ$, $XZ$, $XX$, $IZ$, $XY$, $YI$, $YX$, $ZZ$ from a single fault.}
\end{figure}

\begin{figure}[b]
   \centering

\includegraphics[scale=1,keepaspectratio]{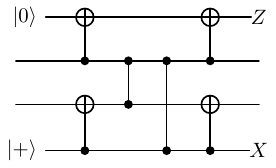}

\caption{\label{fig:flag_cz} Flag gadget on a CZ to detect errors $XX$, $YY$, $ZZ$ from a single fault.}
\end{figure}


The physical circuits constructed using the solve-and-stitch approach are not inherently fault-tolerant, as a single fault from a two-qubit gate can propagate to other qubits. 
However, by applying \emph{flag gadgets}~\cite{Chao-npjqi18} to the physical circuits on the $\llbr 6,4,2 \rrbr$ code, we observe that two additional flag qubits make the circuit fault-tolerant, i.e., they enable the detection of correlated errors arising from a single fault in the circuit.
Due to the distance-$2$ nature of the code, it is sufficient to show error detection for a single fault. 
We posit that a similar approach with flag qubits could be applied effectively to the $\llbr n,n-2,2 \rrbr$ code family in general.

Table~\ref{table:single_fault_errors} provides a comprehensive list of output errors resulting from a single fault in two-qubit gates in Fig.~\ref{fig:solve-stitch_642_qsk_evenH}. 
The ``Location'' in the first row denotes the position of the two-qubit gate in the quantum circuit, where ``$1$'' refers to the quantum wires at the circuit's beginning. 
Each single-fault-induced error in the first row occurs at the output of the gates in the first column, where the left qubit is the control and the right qubit is the target. 
Such an error at the gate output results in an error at the circuit's output, which is given as the corresponding cell entry. 
Output errors that commute with all stabilizers cannot be detected by syndrome measurements. 

The undetectable errors caused by single faults, such as $XI$, $IZ$, $XZ$ on $\text{CNOT}_{2\rightarrow3}$ or $XX$, $IZ$, $XY$ on $\text{CNOT}_{2\rightarrow4}$ or $YI$, $XZ$, $ZZ$ on $\text{CNOT}_{5\rightarrow3}$ or $YX$, $XY$, $ZZ$ on $\text{CNOT}_{5\rightarrow4}$ are observed in Table~\ref{table:single_fault_errors}. 
To address potential undetectable errors, we employ flag gadgets for each two-qubit gate~\cite{Chao-npjqi18}, necessitating the inclusion of two additional flag qubits. 
We introduce a gadget for each CNOT gate in Fig.~\ref{fig:solve-stitch_642_qsk_evenH}. 
The gadget illustrated in Fig.~\ref{fig:flag_cx} effectively captures all aforementioned faults. 
Additionally, errors such as $XX$, $YY$, $ZZ$ on $\text{CZ}_{25}$ and $\text{CZ}_{34}$ can be detected using the gadget presented in Fig.~\ref{fig:flag_cz}. 
Incorporating these gadgets for each two-qubit gate in Fig.~\ref{fig:solve-stitch_642_qsk_evenH} results in a fault-tolerant circuit, where all errors arising from single faults can now be detected.

\begin{figure}[t]
   \centering

\includegraphics[scale=1,keepaspectratio]{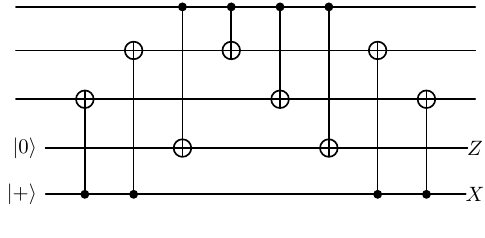}

\caption{\label{fig:flag_cx_many}Merged gadgets to catch errors arising from a single fault on multiple CNOT gates.}
\end{figure}

\begin{figure}[t]
   \centering

\includegraphics[scale=1,keepaspectratio]{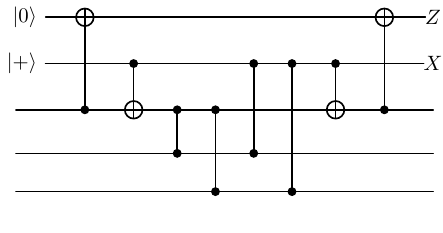}

\caption{\label{fig:flag_cz_many}Merged gadgets to catch errors arising from a single fault on multiple CZ gates.}
\end{figure}

In fact, the gadgets can be merged to minimize the use of two-qubit gates on flag qubits, as shown in Figs.~\ref{fig:flag_cx_many} and~\ref{fig:flag_cz_many}. 
By adjusting the positions of CNOT and CZ gates in Fig.~\ref{fig:solve-stitch_642_qsk_evenH}, it is possible to group CNOT gates together and CZ gates together. 
The resulting circuit in Appendix~\ref{sec:solve-stitch_642_qsk_fault-tolerant} with merged gadgets achieves a fault-tolerant implementation of the logical C-QSK circuit. 
Crucially, this construction generalizes naturally to any member of the $\llbr n,n{-}2,2 \rrbr$ family. Since the solve-and-stitch framework produces a modular and repetitive physical layout in Fig.~\ref{fig:qsk_distance2_physical_full}, the same flag gadget templates for CNOT and CZ groups in Fig.~\ref{fig:flag_cx_many} and Fig.~\ref{fig:flag_cz_many} can be reused across larger codes. Therefore, combining solve-and-stitch with reusable flag gadgets yields a systematic, low-depth, and fault-tolerant realization of logical C-QSK circuits for arbitrary $n$.
Hence, we conclude that the solve-and-stitch approach combined with flag gadgets enables a systematic, optimal-depth, fault-tolerant realization of logical C-QSK circuits on this code family.

\section{Conclusion and Future Work}
\label{sec:conclusion}

The general approach to fault-tolerance via quantum error correction is to design general-purpose codes and decoders and demonstrate fault-tolerant realizations of a universal set of logical gates on the codes.
Then, in principle, any quantum algorithm on the logical qubits can be executed fault-tolerantly by composing these realizations appropriately.
In this work, we were motivated by the fact that quantum computers are expected to show major advantages over classical computers only in specific algorithms and applications.
Hence, we began to explore a new path where compilers for universal fault-tolerant quantum computers tailor the codes to execute a given target algorithm efficiently and fault-tolerantly.
As an illustrative algorithm, we considered Trotter circuits for quantum simulation, which is a key application of quantum computers.
We developed a systematic solve-and-stitch approach to synthesize optimal-depth realizations of Clifford Trotter circuits on the well-known $\llbr n,n-2,2 \rrbr$ code family.
We analyzed the approach rigorously and described the steps in Fig.~\ref{fig:qsk_distance2_physical_full} and Algorithm~\ref{alg:solve-and-stitch}.
We ensured fault-tolerance by embedding flag gadgets in the synthesized physical circuits.
Furthermore, we discussed the potential for the approach to provide simplified circuits in a few different settings.

Obviously, this code family can only detect errors but not correct them.
In future work, we will explore the generalization of the approach to better code families.
We will leverage the insights developed in Theorem~\ref{thm:logical_X_Z_stitching} to understand the code properties necessary to realize Clifford Trotter circuits efficiently and fault-tolerantly.
This will bring us closer to truly tailoring code design for this application.

Naturally, we will investigate how these structured constructions can be used to design decoders that perform much better than structure-agnostic ones.
We will also explore the extension to Trotter circuits with finer angle $Z$-rotations.
Ultimately, we will determine the gains from the best tailored approaches and compare them with the surface code-based benchmarks for fault-tolerant quantum simulation~\cite{Su-prxq21}.

It will also be very interesting to investigate the tailored realization of other quantum algorithms that are shown to provide a quantum advantage.
In the near-term, variational quantum algorithms form one of the widely applicable use-cases of quantum computers~\cite{AuYeung-arxiv23,Dalzell-arxiv23}.
Hence, a fundamental understanding of the necessary code structure to efficiently implement error-corrected versions of these algorithms will likely result in valuable practical gains.
Instead of relying on long-term general purpose codes, this pursuit could identify small codes that still provide advantages in the near term.
We anticipate that these explorations will lead to intriguing design problems for both classical and quantum coding theorists.

\section*{Acknowledgment}
The work of N.~R. was partially supported by the National Science Foundation under Grant no. 2106189.
We thank Swayangprabha Shaw for help with IBM Qiskit and Fig.~\ref{fig:hypergraph_physical_circuit}.


\bibliographystyle{IEEEtran}
\bibliography{references}

\begin{thebibliography}{10}
\providecommand{\url}[1]{#1}
\csname url@samestyle\endcsname
\providecommand{\newblock}{\relax}
\providecommand{\bibinfo}[2]{#2}
\providecommand{\BIBentrySTDinterwordspacing}{\spaceskip=0pt\relax}
\providecommand{\BIBentryALTinterwordstretchfactor}{4}
\providecommand{\BIBentryALTinterwordspacing}{\spaceskip=\fontdimen2\font plus
\BIBentryALTinterwordstretchfactor\fontdimen3\font minus \fontdimen4\font\relax}
\providecommand{\BIBforeignlanguage}[2]{{%
\expandafter\ifx\csname l@#1\endcsname\relax
\typeout{** WARNING: IEEEtran.bst: No hyphenation pattern has been}%
\typeout{** loaded for the language `#1'. Using the pattern for}%
\typeout{** the default language instead.}%
\else
\language=\csname l@#1\endcsname
\fi
#2}}
\providecommand{\BIBdecl}{\relax}
\BIBdecl

\bibitem{Chen-qce24}
Z.~Chen and N.~Rengaswamy, ``Tailoring fault-tolerance to {T}rotter circuits,'' in \emph{IEEE International Conference on Quantum Computing and Engineering (QCE)}, 2024.

\bibitem{Postler-arxiv23}
\BIBentryALTinterwordspacing
L.~Postler, F.~Butt, I.~Pogorelov, C.~D. Marciniak, S.~Heu{\ss}en, R.~Blatt, P.~Schindler, M.~Rispler, M.~M{\"u}ller, and T.~Monz, ``Demonstration of fault-tolerant {S}teane quantum error correction,'' \emph{arXiv preprint arXiv:2312.09745}, 2023. [Online]. Available: \url{https://arxiv.org/abs/2312.09745}
\BIBentrySTDinterwordspacing

\bibitem{Bluvstein-nature24}
\BIBentryALTinterwordspacing
D.~Bluvstein, S.~J. Evered, A.~A. Geim, S.~H. Li, H.~Zhou, T.~Manovitz, S.~Ebadi, M.~Cain, M.~Kalinowski, D.~Hangleiter \emph{et~al.}, ``Logical quantum processor based on reconfigurable atom arrays,'' \emph{Nature}, vol. 626, no. 7997, pp. 58--65, 2024. [Online]. Available: \url{https://arxiv.org/abs/2312.03982}
\BIBentrySTDinterwordspacing

\bibitem{DaSilva-arxiv24}
\BIBentryALTinterwordspacing
M.~da~Silva, C.~Ryan-Anderson, J.~Bello-Rivas, A.~Chernoguzov, J.~Dreiling, C.~Foltz, J.~Gaebler, T.~Gatterman, D.~Hayes, N.~Hewitt \emph{et~al.}, ``Demonstration of logical qubits and repeated error correction with better-than-physical error rates,'' \emph{arXiv preprint arXiv:2404.02280}, 2024. [Online]. Available: \url{https://arxiv.org/abs/2404.02280}
\BIBentrySTDinterwordspacing

\bibitem{Ali-arxiv24}
\BIBentryALTinterwordspacing
H.~Ali, J.~Marques, O.~Crawford, J.~Majaniemi, M.~Serra-Peralta, D.~Byfield, B.~Varbanov, B.~M. Terhal, L.~DiCarlo, and E.~T. Campbell, ``Reducing the error rate of a superconducting logical qubit using analog readout information,'' \emph{arXiv preprint arXiv:2403.00706}, 2024. [Online]. Available: \url{https://arxiv.org/abs/2403.00706}
\BIBentrySTDinterwordspacing

\bibitem{Cohen-sciadv22}
\BIBentryALTinterwordspacing
L.~Z. Cohen, I.~H. Kim, S.~D. Bartlett, and B.~J. Brown, ``Low-overhead fault-tolerant quantum computing using long-range connectivity,'' \emph{Science Advances}, vol.~8, no.~20, p. eabn1717, 2022. [Online]. Available: \url{https://arxiv.org/abs/2110.10794}
\BIBentrySTDinterwordspacing

\bibitem{Wang-arxiv23}
\BIBentryALTinterwordspacing
Y.-F. Wang, Y.~Wang, Y.-A. Chen, W.~Zhang, T.~Zhang, J.~Hu, W.~Chen, Y.~Gu, and Z.-W. Liu, ``Efficient fault-tolerant implementations of non-{C}lifford gates with reconfigurable atom arrays,'' \emph{arXiv preprint arXiv:2312.09111}, 2023. [Online]. Available: \url{https://arxiv.org/abs/2312.09111}
\BIBentrySTDinterwordspacing

\bibitem{Zhu-arxiv23}
\BIBentryALTinterwordspacing
G.~Zhu, S.~Sikander, E.~Portnoy, A.~W. Cross, and B.~J. Brown, ``Non-{C}lifford and parallelizable fault-tolerant logical gates on constant and almost-constant rate homological quantum {LDPC} codes via higher symmetries,'' \emph{arXiv preprint arXiv:2310.16982}, 2023. [Online]. Available: \url{https://arxiv.org/abs/2310.16982}
\BIBentrySTDinterwordspacing

\bibitem{Montanaro-npjqi16}
\BIBentryALTinterwordspacing
A.~Montanaro, ``Quantum algorithms: an overview,'' \emph{npj Quantum Information}, vol.~2, no.~1, pp. 1--8, 2016. [Online]. Available: \url{https://arxiv.org/abs/1511.04206}
\BIBentrySTDinterwordspacing

\bibitem{AuYeung-arxiv23}
\BIBentryALTinterwordspacing
R.~Au-Yeung, B.~Camino, O.~Rathore, and V.~Kendon, ``Quantum algorithms for scientific applications,'' \emph{arXiv preprint arXiv:2312.14904}, 2023. [Online]. Available: \url{https://arxiv.org/abs/2312.14904}
\BIBentrySTDinterwordspacing

\bibitem{Dalzell-arxiv23}
\BIBentryALTinterwordspacing
A.~M. Dalzell, S.~McArdle, M.~Berta, P.~Bienias, C.-F. Chen, A.~Gily{\'e}n, C.~T. Hann, M.~J. Kastoryano, E.~T. Khabiboulline, A.~Kubica \emph{et~al.}, ``Quantum algorithms: {A} survey of applications and end-to-end complexities,'' \emph{arXiv preprint arXiv:2310.03011}, 2023. [Online]. Available: \url{https://arxiv.org/abs/2310.03011}
\BIBentrySTDinterwordspacing

\bibitem{Whitfield-molphy11}
\BIBentryALTinterwordspacing
J.~D. Whitfield, J.~Biamonte, and A.~Aspuru-Guzik, ``Simulation of electronic structure {H}amiltonians using quantum computers,'' \emph{Molecular Physics}, vol. 109, no.~5, pp. 735--750, 2011. [Online]. Available: \url{https://arxiv.org/abs/1001.3855}
\BIBentrySTDinterwordspacing

\bibitem{Daley-nature22}
A.~J. Daley, I.~Bloch, C.~Kokail, S.~Flannigan, N.~Pearson, M.~Troyer, and P.~Zoller, ``Practical quantum advantage in quantum simulation,'' \emph{Nature}, vol. 607, no. 7920, pp. 667--676, 2022.

\bibitem{Li-asplos22}
\BIBentryALTinterwordspacing
G.~Li, A.~Wu, Y.~Shi, A.~Javadi-Abhari, Y.~Ding, and Y.~Xie, ``Paulihedral: a generalized block-wise compiler optimization framework for quantum simulation kernels,'' in \emph{Proceedings of the 27th ACM International Conference on Architectural Support for Programming Languages and Operating Systems}, 2022, pp. 554--569. [Online]. Available: \url{https://arxiv.org/abs/2109.03371}
\BIBentrySTDinterwordspacing

\bibitem{Dehaene-pra03}
\BIBentryALTinterwordspacing
J.~Dehaene and B.~De~Moor, ``Clifford group, stabilizer states, and linear and quadratic operations over {GF}(2),'' \emph{Physical Review A}, vol.~68, no.~4, p. 042318, 2003. [Online]. Available: \url{https://arxiv.org/abs/quant-ph/0304125}
\BIBentrySTDinterwordspacing

\bibitem{Gottesman-phd97}
\BIBentryALTinterwordspacing
D.~Gottesman, ``Stabilizer codes and quantum error correction,'' Ph.D. dissertation, 1997. [Online]. Available: \url{https://arxiv.org/abs/quant-ph/9705052}
\BIBentrySTDinterwordspacing

\bibitem{Chao-npjqi18}
R.~Chao and B.~W. Reichardt, ``Fault-tolerant quantum computation with few qubits,'' \emph{NPJ Quantum Information}, vol.~4, no.~1, p.~42, 2018.

\bibitem{Rengaswamy-tqe20}
N.~Rengaswamy, R.~Calderbank, S.~Kadhe, and H.~D. Pfister, ``Logical {C}lifford synthesis for stabilizer codes,'' \emph{IEEE Transactions on Quantum Engineering}, vol.~1, pp. 1--17, 2020.

\bibitem{Calderbank-it98}
\BIBentryALTinterwordspacing
A.~R. Calderbank, E.~M. Rains, P.~M. Shor, and N.~J. Sloane, ``Quantum error correction via codes over {GF}(4),'' \emph{IEEE Transactions on Information Theory}, vol.~44, no.~4, pp. 1369--1387, 1998. [Online]. Available: \url{https://arxiv.org/abs/quant-ph/9608006}
\BIBentrySTDinterwordspacing

\bibitem{Aaronson-pra04}
\BIBentryALTinterwordspacing
S.~Aaronson and D.~Gottesman, ``Improved simulation of stabilizer circuits,'' \emph{Physical Review A}, vol.~70, no.~5, p. 052328, 2004. [Online]. Available: \url{https://arxiv.org/abs/quant-ph/0406196}
\BIBentrySTDinterwordspacing

\bibitem{Tillich-it13}
\BIBentryALTinterwordspacing
J.-P. Tillich and G.~Z{\'e}mor, ``Quantum {LDPC} codes with positive rate and minimum distance proportional to the square root of the blocklength,'' \emph{IEEE Transactions on Information Theory}, vol.~60, no.~2, pp. 1193--1202, 2013. [Online]. Available: \url{https://arxiv.org/abs/0903.0566}
\BIBentrySTDinterwordspacing

\bibitem{Panteleev-stoc22}
\BIBentryALTinterwordspacing
P.~Panteleev and G.~Kalachev, ``Asymptotically good quantum and locally testable classical {LDPC} codes,'' in \emph{Proceedings of the 54th Annual ACM SIGACT Symposium on Theory of Computing}, 2022, pp. 375--388. [Online]. Available: \url{https://arxiv.org/abs/2111.03654}
\BIBentrySTDinterwordspacing

\bibitem{Leverrier-focs22}
\BIBentryALTinterwordspacing
A.~Leverrier and G.~Z{\'e}mor, ``Quantum {T}anner codes,'' in \emph{2022 IEEE 63rd Annual Symposium on Foundations of Computer Science (FOCS)}.\hskip 1em plus 0.5em minus 0.4em\relax IEEE, 2022, pp. 872--883. [Online]. Available: \url{https://arxiv.org/abs/2202.13641}
\BIBentrySTDinterwordspacing

\bibitem{Kuehnke-arxiv25}
\BIBentryALTinterwordspacing
E.~J. Kuehnke, K.~Levi, J.~Roffe, J.~Eisert, and D.~Miller, ``Hardware-tailored logical {C}lifford circuits for stabilizer codes,'' \emph{arXiv preprint arXiv:2505.20261}, 5 2025. [Online]. Available: \url{https://arxiv.org/abs/2505.20261v1}
\BIBentrySTDinterwordspacing

\bibitem{Hatano-qaoom05}
\BIBentryALTinterwordspacing
N.~Hatano and M.~Suzuki, ``Finding exponential product formulas of higher orders,'' in \emph{Quantum annealing and other optimization methods}.\hskip 1em plus 0.5em minus 0.4em\relax Springer, 2005, pp. 37--68. [Online]. Available: \url{https://arxiv.org/abs/math-ph/0506007}
\BIBentrySTDinterwordspacing

\bibitem{Eastin-prl09}
\BIBentryALTinterwordspacing
B.~Eastin and E.~Knill, ``Restrictions on transversal encoded quantum gate sets,'' \emph{Physical review letters}, vol. 102, no.~11, p. 110502, 2009. [Online]. Available: \url{https://arxiv.org/abs/0811.4262}
\BIBentrySTDinterwordspacing

\bibitem{Su-prxq21}
\BIBentryALTinterwordspacing
Y.~Su, D.~W. Berry, N.~Wiebe, N.~Rubin, and R.~Babbush, ``Fault-tolerant quantum simulations of chemistry in first quantization,'' \emph{PRX Quantum}, vol.~2, no.~4, p. 040332, 2021. [Online]. Available: \url{https://arxiv.org/abs/2105.12767}
\BIBentrySTDinterwordspacing

\end{thebibliography}

\newpage

\appendices

\section{Preliminaries (Continued)}
\label{sec:prelim_contd}

\subsection{Pauli Group and Stabilizer Codes}
\label{sec:pauli}

For a single qubit, the Hermitian Pauli operators are denoted by $I,X,Y,Z$.
For $n \geq 1$ qubit(s), the \emph{Pauli group} is given by
\begin{align}
\mathcal{P}_n \coloneqq \langle \imath^{\kappa} I, X, Y, Z \, ; \, \kappa \in \{ 0,1,2,3 \} \rangle^{\otimes n}.
\end{align}
We denote an $n$-qubit Hermitian Pauli operator as $E(a,b)$, where $a,b \in \{ 0,1 \}^n$.
When $a_i = 1$ (resp. $b_i = 1$), the $X$ (resp. $Z$) gate is applied to qubit $i$, and when $a_i = b_i = 1$, the $Y$ gate is applied to qubit $i$, where $i \in \{ 1,2,\ldots,n \}$.
For example, for $n=3$, $E([101,011]) = X \otimes Z \otimes Y$.
The \emph{weight} of a Pauli operator is the number of non-identity elements in its Kronecker product, e.g., $X \otimes Z \otimes Y$ has weight $3$.

Any pair of single-qubit non-identity Pauli operators anti-commute, e.g., $XZ=-ZX$.
For multiple qubits, since the Kronecker product satisfies the property $(A \otimes B)(C \otimes D) = (AC) \otimes (BD)$, two non-identity Pauli operators can commute, e.g., $X \otimes X$ and $Z \otimes Z$.
We will represent such operators using qubit subscripts for convenience, e.g., $E_1 E_2' \equiv E \otimes E'$ for $E,E' \in \{ I,X,Y,Z \}$.
On $n$ qubits, this means $E_i \equiv I_1 I_2 \cdots I_{i-1} E_i I_{i+1} \cdots I_n = I \otimes \cdots \otimes I \otimes E \otimes I \otimes \cdots \otimes I$. 

Let $N \coloneqq 2^n$ for an $n$-qubit system.
A \emph{stabilizer group}, $\mathcal{S}$, is a group generated by commuting Hermitian Pauli operators such that $-I_N \notin \mathcal{S}$.
The associated \emph{stabilizer code}, $\mathcal{Q}_{\mathcal{S}}$, is the subspace of unit vectors in $\mathbb{C}^N$ that are the common $+1$-eigenvectors of operators in $\mathcal{S}$.
In other words, each $\ket{\psi} \in \mathcal{Q}_{\mathcal{S}}$ satisfies $S \ket{\psi} = \ket{\psi}$ for all $S \in \mathcal{S}$.
The \emph{logical Pauli operators} of a code are Hermitian Pauli operators that commute with the stabilizers but are not stabilizers themselves.
If $\mathcal{S}$ has $r = n-k$ independent generators and the minimum weight of any logical Pauli operator is $d$, then the code is an $\llbr n,k,d \rrbr$ code.
Such a code encodes $k$ logical qubits into $n$ physical qubits. 
It can detect up to $(d-1)$ Pauli errors but can correct only up to $\lfloor \frac{d-1}{2} \rfloor$ Pauli errors using a maximum-likelihood decoder.
A \emph{CSS} code is a stabilizer code whose stabilizer group can be generated by purely $X$-type and purely $Z$-type operators.

\subsection{The $\llbr n,n-2,2 \rrbr$ Code Family}
\label{sec:distance_2_family}

This is a family of CSS codes with exactly two stabilizer generators, $X_1 X_2 \cdots X_n$ and $Z_1 Z_2 \cdots Z_n$, where $n$ is set to be even~\cite{Gottesman-phd97,Chao-npjqi18}.
It is easily checked that these generators commute.
The generators of logical-$X$ and $Z$ operators are of the form $\overline{X_i} = X_1 X_{i+1}$ and $\overline{Z_i} = Z_{i+1} Z_n$, where $i = 1,2,\ldots,n-2$.
Hence, the distance of the codes in this family is always $2$.
The family is only error-detecting because it can detect up to $1$ error but cannot correct any errors.
As an example, for $n=6$, the $\llbr 6,4,2 \rrbr$ is described by the stabilizer group $\mathcal{S} = \langle X_1 \cdots X_6, Z_1 \cdots Z_6 \rangle$ and the following generators of the logical Pauli group:
\begin{align}
\overline{X_1} = X_1 X_2 \ &, \ \overline{Z_1} = Z_2 Z_6 \ ; \nonumber \\
\overline{X_2} = X_1 X_3 \ &, \ \overline{Z_2} = Z_3 Z_6 \ ; \nonumber \\
\overline{X_3} = X_1 X_4 \ &, \ \overline{Z_3} = Z_4 Z_6 \ ; \nonumber \\
\overline{X_4} = X_1 X_5 \ &, \ \overline{Z_4} = Z_5 Z_6 \ .
\end{align}
We will use this code as the running example but the results apply to the entire code family.

\section{Fault Tolerant Logical C-QSK on the $\llbr 6,4,2 \rrbr$ Code}
\label{sec:solve-stitch_642_qsk_fault-tolerant}

\begin{figure}
\centering

\includegraphics[scale=0.87,keepaspectratio]{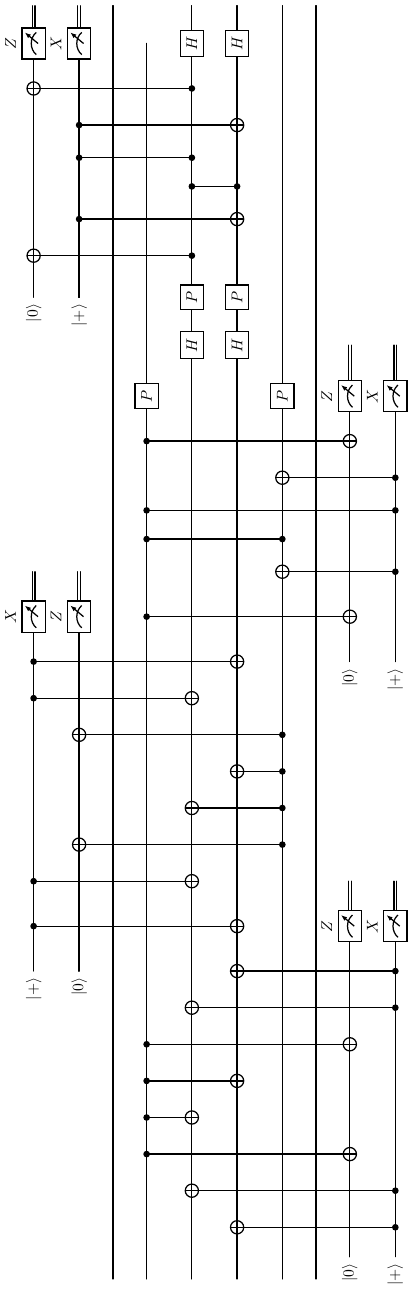}

\caption{\label{fig:solve-stitch_642_qsk_fault-tolerant} Fault-tolerant execution of the physical circuit in Fig.~\ref{fig:solve-stitch_642_qsk_evenH} by embedding flag gadgets.}
\end{figure}

Table~\ref{table:single_fault_errors} shows the errors at the output of Fig.~\ref{fig:solve-stitch_642_qsk_evenH} resulting from a single fault at the output of a two-qubit gate.
Fig.~\ref{fig:solve-stitch_642_qsk_fault-tolerant} shows the fault-tolerant execution of the physical circuit in Fig.~\ref{fig:solve-stitch_642_qsk_evenH} by embedding flag gadgets.

\setlength\rotFPtop{0pt plus 1fil}
\begin{sidewaystable}
    \centering
    \caption{\label{table:single_fault_errors} Errors at the output of Fig.~\ref{fig:solve-stitch_642_qsk_evenH} resulting from a single fault at the output of a two-qubit gate.}
    \scalebox{0.6}{
   \begin{tabular}{c|c|c|c|c|c|c|c|c|c|c|c|c|c|c|c|c|c|c}
         \toprule\toprule
         &  & Location  & $IX$ & $IY$ & $IZ$ & $XI$ & $XX$ & $XY$ & $XZ$ & $YI$ & $YX$ & $YY$ & $YZ$ & $ZI$ & $ZX$ & $ZY$ & $ZZ$ &\\
         \toprule\toprule
         & $\text{CNOT}_{2\rightarrow3}$  & 2 & $X_{3}$  & $Z_{3}X_{4}Z_{5}$ & \textcolor{red}{$-Y_{3}X_{4}Z_{5}$} & \textcolor{red}{$Y_{2}X_{4}Z_{5}$} & $Y_{2}X_{3}X_{4}Z_{5}$ & $Y_{2}Z_{3}$ & \textcolor{red}{$-Y_{2}Y_{3}$} & $-X_{2}X_{4}Z_{5}$ & $-X_{2}X_{3}X_{4}Z_{5}$ & $-X_{2}Z_{3}$ & $X_{2}Y_{3}$ & $Z_{2}$ & $Z_{2}X_{3}$ & $Z_{2}Z_{3}X_{4}Z_{5}$ &$-Z_{2}Y_{3}X_{4}Z_{5}$ &\\
         \midrule
         & $\text{CNOT}_{2\rightarrow4}$ & 3 & $X_{4}$ & $X_{3}Z_{4}Z_{5}$ & \textcolor{red}{$-X_{3}Y_{4}Z_{5}$} & $Y_{2}Z_{5}$ & \textcolor{red}{$Y_{2}X_{4}Z_{5}$} & \textcolor{red}{$Y_{2}X_{3}Z_{4}$} & $-Y_{2}X_{3}Y_{4}$ & $-X_{2}Z_{5}$ & $-X_{2}X_{4}Z_{5}$ & $-X_{2}X_{3}Z_{4}$ & $X_{2}X_{3}Y_{4}$ & $Z_{2}$ & $Z_{2}X_{4}$ & $Z_{2}X_{3}Z_{4}Z_{5}$ & $-Z_{2}X_{3}Y_{4}Z_{5}$ & \\
         \midrule
         & $\text{CZ}_{2\rightarrow5}$ & 4 & $X_{3}X_{4}Y_{5}$ & $-X_{3}X_{4}X_{5}$ & $Z_{5}$ & $Y_{2}$ & \textcolor{red}{$Y_{2}X_{3}X_{4}Y_{5}$} & $-Y_{2}X_{3}X_{4}X_{5}$ & $Y_{2}Z_{5}$ & $-X_{2}$ & $-X_{2}X_{3}X_{4}Y_{5}$ & \textcolor{red}{$X_{2}X_{3}X_{4}X_{5}$} & $-X_{2}Z_{5}$ & $Z_{2}$ & $Z_{2}X_{3}X_{4}Y_{5}$ & $-Z_{2}X_{3}X_{4}X_{5}$ & \textcolor{red}{$Z_{2}Z_{5}$}  &\\
         \midrule
         & $\text{CNOT}_{5\rightarrow3}$ & 5 & $X_{3}$ & $Z_{3}X_{4}$ & $-Y_{3}X_{4}$ & $X_{4}Y_{5}$ & $X_{3}X_{4}Y_{5}$ & $Z_{3}Y_{5}$ & \textcolor{red}{$-Y_{3}Y_{5}$} & \textcolor{red}{$-X_{4}X_{5}$} & $-X_{3}X_{4}X_{5}$ & $-Z_{3}X_{5}$ & $Y_{3}X_{5}$ & $Z_{5}$ & $X_{3}Z_{5}$ & $Z_{3}X_{4}Z_{5}$ & \textcolor{red}{$-Y_{3}X_{4}Z_{5}$} &\\
         \midrule
         & $\text{CNOT}_{5\rightarrow4}$ & 6 & $X_{4}$ & $X_{3}Z_{4}$ & $-X_{3}Y_{4}$ & $Y_{5}$ & $X_{4}Y_{5}$ & \textcolor{red}{$X_{3}Z_{4}Y_{5}$} & $-X_{3}Y_{4}Y_{5}$ & $-X_{5}$ & \textcolor{red}{$-X_{4}X_{5}$}  & $-X_{3}Z_{4}X_{5}$ & $X_{3}Y_{4}X_{5}$ & $Z_{5}$ & $X_{4}Z_{5}$ & $X_{3}Z_{4}Z_{5}$ & \textcolor{red}{$-X_{3}Y_{4}Z_{5}$} &\\
         \midrule
         & $\text{CZ}_{3\rightarrow4}$ & 10 & $Z_{4}$ & $-Y_{4}$ & $X_{4}$ & $Z_{3}$ & \textcolor{red}{$Z_{3}Z_{4}$} & $-Z_{3}Y_{4}$ & $Z_{3}X_{4}$ & $-Y_{3}$ & $-Y_{3}Z_{4}$ & \textcolor{red}{$Y_{3}Y_{4}$} & $-Y_{3}X_{4}$ & $X_{3}$ & $X_{3}Z_{4}$ & $-X_{3}Y_{4}$ & \textcolor{red}{$X_{3}X_{4}$} &\\
         \bottomrule
    \end{tabular}
    }
\end{sidewaystable}

\section{Even number of Hadamards for $\llbr 6,4,2 \rrbr$ code}
\label{sec:Example of [[6,4,2]] even}

\begin{figure}
\centering
\vspace*{-10pt}

\includegraphics[scale=1,keepaspectratio]{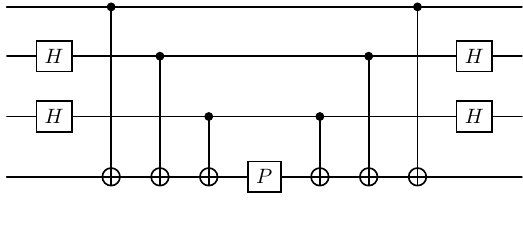}

\vspace*{-10pt}

\caption{\label{fig:qsk_4qubit_evenH} C-QSK circuit with even number of Hadamard gates.}

\end{figure}

For our first scenario, we assume that the logical C-QSK circuit consists of an even number of Hadamard gates.
The $\llbr 6,4,2 \rrbr$ code serves as an excellent testbed for our exploration, in which case the logical circuit comprises 4 qubits. 
As an example, consider the circuit in Fig.~\ref{fig:qsk_4qubit_evenH}, where the Hadamard gates are applied to the second and third qubits.
The circuit implements the exponentiated Pauli operator $\exp\left(-\imath \frac{\pi}{4} Z_1 X_2 X_3 Z_4 \right)$.
A similar approach applies for Pauli operators with $Y$ entries, where $H$ gates are replaced with $H_y$ gates (see Section~\ref{sec:qsk}).

This C-QSK circuit dictates the following input-output mappings of logical Pauli operators:
\begin{IEEEeqnarray}{rClCrCl}
\label{eq:logical_qsk_4qubit_evenH_constraints}
\overline{X_1} & \mapsto & \overline{Y_1} \, \overline{X_2} \, \overline{X_3} \, \overline{Z_4} & \quad , \quad & \overline{Z_1} & \mapsto & \overline{Z_1} \ , \nonumber \\
\overline{X_2} & \mapsto & \overline{X_2} & \quad , \quad & \overline{Z_2} & \mapsto & - \overline{Z_1} \, \overline{Y_2} \, \overline{X_3} \, \overline{Z_4} \ , \nonumber \\
\overline{X_3} & \mapsto & \overline{X_3} & \quad , \quad & \overline{Z_3} & \mapsto & - \overline{Z_1} \, \overline{X_2} \, \overline{Y_3} \, \overline{Z_4} \ , \nonumber \\
\overline{X_4} & \mapsto & \overline{Z_1} \, \overline{X_2} \, \overline{X_3} \, \overline{Y_4} & \quad , \quad & \overline{Z_4} & \mapsto & \overline{Z_4} \ .
\end{IEEEeqnarray}
Substituting for the logical operators of the $\llbr 6,4,2 \rrbr$ code, we obtain the following mappings of physical Pauli operators:
\begin{IEEEeqnarray}{rClCrCl}
\label{eq:logical_qsk_4qubit_evenH_constraints_physical}
X_1 X_2 & \mapsto & X_1 Y_2 X_3 X_4  Z_5 & \quad , \quad & Z_2 Z_6 & \mapsto & Z_2 Z_6 \ , \nonumber \\
X_1 X_3 & \mapsto & X_1 X_3 & \quad , \quad & Z_3 Z_6 & \mapsto & - Z_2 Y_3 X_4 Z_5 Z_6 \ , \nonumber \\
X_1 X_4 & \mapsto & X_1 X_4 & \quad , \quad & Z_4 Z_6 & \mapsto & - Z_2 X_3 Y_4 Z_5 Z_6 \ , \nonumber \\
X_1 X_5 & \mapsto & X_1 Z_2 X_3 X_4 Y_5 & \quad , \quad & Z_5 Z_6 & \mapsto & Z_5 Z_6 \ .
\end{IEEEeqnarray}
Additionally, we require that the two stabilizer generators $X_1 \cdots X_6$ and $Z_1 \cdots Z_6$ are mapped to themselves, even though it is sufficient to map them to a pair of equivalent stabilizers, i.e., normalize the stabilizer group.
The minimum depth solution produced by the LCS algorithm for this case is shown below in Fig.~\ref{fig:lcs_642_qsk}. 
This physical circuit satisfies all stipulated Pauli constraints above while preserving stabilizers. 
Observe that the (na{\"i}ve) depth of this circuit is $10$, which is comparable to the depth of the logical circuit in Fig.~\ref{fig:qsk_4qubit_evenH}.

\begin{figure}
\centering
\includegraphics[scale=0.82,keepaspectratio]{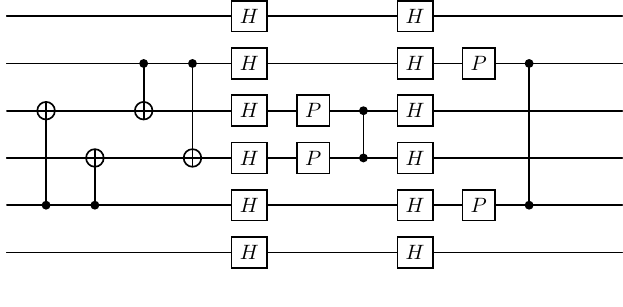}


\caption{\label{fig:lcs_642_qsk}The minimum depth solution of LCS algorithm for realizing the logical circuit in Fig.~\ref{fig:qsk_4qubit_evenH} on the $\llbr 6,4,2 \rrbr$ code. Note that the algorithm does not perform circuit simplification.}

\end{figure}

For the $\llbr n,n-2,2 \rrbr$ code family, we have $r = n-k = 2$, so the number of symplectic solutions (for a fixed set of Pauli constraints) is only $2^{r(r+1)/2} = 8$.
Hence, it is easy to enumerate the solutions and identify the minimum depth circuit.
However, this becomes impractical even for $r = 6$.
Besides, the LCS algorithm does not enable us to predict the minimum depth until we obtain all solutions and decompose each of them into elementary Clifford gates as in Fig.~\ref{fig:lcs_642_qsk}.

To address these limitations, we propose an alternative method for constructing physical circuits by identifying a \emph{root} qubit for each (physical) Pauli constraint. 
For a non-trivial constraint, we identify qubits that appear in both the input and output, and then eliminate the qubits with an unchanged Pauli gate.
From the remaining root qubits, we choose one as the root.
For instance, consider the mapping of $X_{1}X_{2}$ to $X_{1}Y_{2}X_{3}X_{4}Z_{5}$. 
In this case, we choose the second qubit as the root rather than the first qubit because $X_{1}$ remains unchanged, requiring no additional gates in the circuit. 
For (non-root) qubits that are absent in the input Pauli but present in the output of the constraint, we introduce two-qubit gates controlled by the root qubit because single-qubit gates like $H$ and $P$ cannot affect the Pauli on other qubits.

\begin{figure}
\begin{subfigure}[t]{0.5\textwidth}
\centering
\includegraphics[scale=1,keepaspectratio]{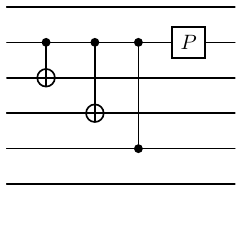}


\caption{\label{fig:XX_XYXXZ}$X_{1}{X_{2}}\mapsto X_{1}Y_{2}X_{3}X_{4}Z_{5}$}
\end{subfigure}%
\begin{subfigure}[t]{0.5\textwidth}
\centering
\includegraphics[scale=1,keepaspectratio]{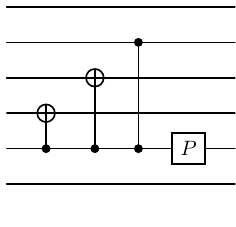}


\caption{\label{fig:XIIIX_XZXXY}$X_{1}{X_{5}}\mapsto X_{1}Z_{2}X_{3}X_{4}Y_{5}$}
\end{subfigure}
\caption{Rooted circuit construction for logical-$X$ constraints.}
\label{fig:local_circuit_X_642}
\end{figure}

Specifically, for the above example, since CNOT gate maps $XI$ to $XX$, we apply $\text{CNOT}_{2\rightarrow 3}$ and $\text{CNOT}_{2\rightarrow 4}$ for the third and fourth qubits with the second qubit as the root (control). 
Since the CZ gate maps $XI$ to $XZ$, we apply $\text{CZ}_{25}$ for the fifth qubit to satisfy its mapping. 
The output term of the second qubit is a $Y$ gate, so we directly introduce a Phase gate to achieve the desired mapping at the end. 
This ``local'' circuit construction is illustrated in Fig.~\ref{fig:XX_XYXXZ}.

For mappings involving $\overline{X_{2}}$ and $\overline{X_{3}}$, where the input gates are the same as the output gates, we have verified that these mappings remain unchanged as the gates propagate through the above rooted circuit we constructed. 
Consequently, there is no need to construct new rooted circuits for these trivial mappings. 
Additionally, for the Pauli constraint of $\overline{X_{4}}$, we choose the fifth qubit as the root and apply CNOT or CZ gates to other qubits based on the input-output Pauli relations. 
A Phase gate is then applied to the root qubit to complete the rooted circuit in  Fig.~\ref{fig:XIIIX_XZXXY} for mapping $X_1 X_5$ to $X_1 Z_2 X_3 X_4 Y_5$.

\begin{figure}
\centering
\begin{subfigure}[t]{0.35\textwidth}
\centering

\includegraphics[scale=1,keepaspectratio]{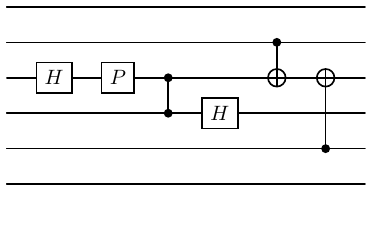}

\vspace*{-12pt}
\caption{\label{fig:IIZIIZ_IZYXZZ}$Z_{3}{Z_{6}} \mapsto Z_{2}Y_{3}X_{4}Z_{5}Z_{6}$ (ignoring sign)}
\end{subfigure}

\vspace*{10pt}

\begin{subfigure}[t]{0.35\textwidth}
\centering

\includegraphics[scale=1,keepaspectratio]{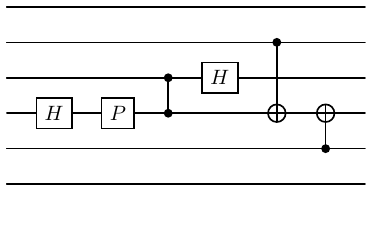}

\vspace*{-12pt}
\caption{\label{fig:IIIZIZ_IZXYZZ}$Z_{4}{Z_{6}}\mapsto Z_{2}X_{3}Y_{4}Z_{5}Z_{6}$ (ignoring sign)}
\end{subfigure}

\caption{Rooted circuit construction for logical-$Z$ constraints.}
\label{fig:local_circuit_Z_642}

\vspace*{-15pt}
\end{figure}

For the mappings corresponding to the logical-$Z$ constraints, the construction of rooted circuits has some variations. 
For example, consider the mapping $Z_{3} Z_{6} \mapsto Z_{2}Y_{3}X_{4}Z_{5}Z_{6}$. 
Here, just as before, we designate the third qubit as the root rather than the sixth qubit because $Z_{6}$ remains unchanged. 
However, unlike the logical-$X$ constraints, where the root's mapping is processed at the end, here we process this qubit at the beginning.
To transform $Z_{3}$ to $Y_{3}$ in the output, we apply $H$ followed by $P$ on the third qubit. 
Next, we observe that the output of the fourth qubit is $X_{4}$. 
While it might seem intuitive to apply $\text{CNOT}_{3\rightarrow 4}$, which maps $Y_{3}I_{4}$ to $Y_{3}Z_{4}$, we realize that the constraint $Z_{4} Z_{6} \mapsto Z_{2}X_{3}Y_{4}Z_{5}Z_{6}$ would later require a reversed CNOT, i.e., $\text{CNOT}_{4\rightarrow 3}$.
Since the CNOT is not symmetric with respect to swapping the control and target qubits, we apply a CZ gate on the third and fourth qubits instead, followed by $H$ on these qubits. 
Subsequently, we apply CNOTs for qubits $2$ and $5$, with the root qubit as the target and themselves as controls. 
Finally, as for the logical-$X$ constraints, we verify that these rooted circuits satisfy the trivial mappings of $\overline{Z_{1}}$ and $\overline{Z_{4}}$ as well.
The final rooted circuits for logical-$Z$ constraints are illustrated in Figs.~\ref{fig:IIZIIZ_IZYXZZ} and~\ref{fig:IIIZIZ_IZXYZZ}.

To consolidate the rooted circuits satisfying individual Pauli constraints into a comprehensive physical circuit that simultaneously meets all constraints, we follow a systematic ``stitching'' approach, initially addressing logical-$X$s and subsequently logical-$Z$s. 
We begin with the rooted circuit for $\overline{X_{1}}$, which we have verified adheres to the trivial mappings of $\overline{X_{2}}$ and $\overline{X_{3}}$. 
Subsequently, we seamlessly append the rooted circuit for $\overline{X_{4}}$ at the end. 
During this process, we ensure smooth integration of shared gates; for example, since $\text{CZ}_{25}$, is already present, we incorporate it without duplication. 
Then we optimize the depth by consolidating all $P$ gates into a single stage. 

\begin{figure}[]
\centering
\includegraphics[scale=0.9,keepaspectratio]{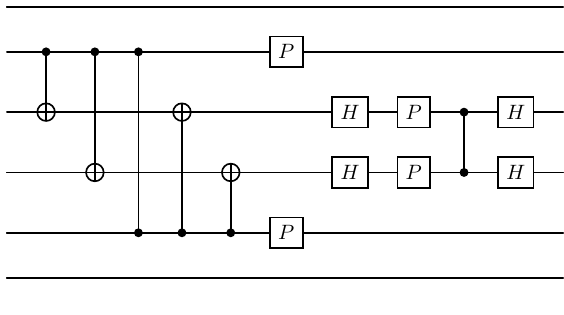}


\caption{\label{fig:solve-stitch_642_qsk_evenH}Physical realization of Fig.~\ref{fig:qsk_4qubit_evenH} on the $\llbr 6,4,2 \rrbr$ code produced by the solve-and-stitch approach using root qubits. Signs for $\overline{Z_2}$ and $\overline{Z_3}$ can be fixed with Pauli gates at the end.}

\end{figure}

Upon checking all mappings of the logical-$X$ part, we confirm the circuit's validity.
Moving to the rooted circuits for the logical-$Z$ constraints in Fig.~\ref{fig:local_circuit_Z_642}, we observe that the two CNOT gates in Fig.~\ref{fig:IIZIIZ_IZYXZZ} have already been incorporated in the logical-$X$ part. 
Consequently, we exclude them and proceed to add $H_{3}, P_{3}, \text{CZ}_{34}$, and $H_{4}$ into the final circuit. 
At this stage, we verify that all prior Pauli constraints are still satisfied.
Next, in Fig.\ref{fig:IIIZIZ_IZXYZZ}, we observe a similar $H$-$P$-CZ-$H$ structure, sharing the same CZ gate. 
This structure is consolidated into the physical circuit, as depicted in Fig.~\ref{fig:solve-stitch_642_qsk_evenH}. 
Once again, the CNOT gates in Fig.~\ref{fig:IIIZIZ_IZXYZZ} are already present in previous rooted circuits, so we do not duplicate them. 
Through this process, we validate that all Pauli constraints are satisfied simultaneously, the two stabilizer generators are preserved, and the depth aligns with the circuit generated by the LCS algorithm in Fig.~\ref{fig:lcs_642_qsk}. 
The notable differences lie in the repositioning of certain gates and the cancellation of some Hadamard gates.
Hence, we have a bottom-up approach to derive the best output of the LCS algorithm in this case, while being able to track the structure of the circuit at each stage of the process.
This directly addresses the question we posed at the beginning of Section~\ref{sec:qsk_distance_2_family}.

While it may appear that we must check for past constraints when stitching each rooted circuit to the existing physical circuit, this was done for pedagogical reasons.
Later, we prove rigorously that the stitching procedure always satisfies all constraints for logical C-QSK circuits on this code family.
Therefore, the validity of the circuits is established analytically, and for all members of the code family.

\section{Even number of Hadamards for $\llbr 8,6,2 \rrbr$ code}
\label{sec:Example of [[8,6,2]] even}

\begin{figure}
\centering

\includegraphics[scale=0.72,keepaspectratio]{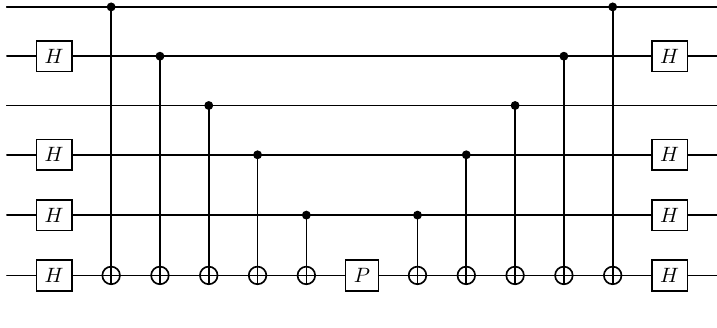}

\caption{\label{fig:qsk_6qubit_evenH_many} C-QSK circuit with even Hadamards on many qubits.}

\end{figure}

We have successfully implemented the solve-and-stitch approach on the $\llbr 6,4,2 \rrbr$ code by identifying root qubits for different non-trivial constraints. 
Here we show that the approach extends seamlessly to a larger code size, such as the $\llbr 8,6,2 \rrbr$ code. 
In this extended scenario, we incorporate Hadamard gates on multiple qubits in the logical circuit, as shown in Fig.~\ref{fig:qsk_6qubit_evenH_many}. 
The root qubit idea continues to be effective, demonstrating its utility in tailoring circuit synthesis for this code family. 
The circuits are constructed similar to the $\llbr 6,4,2 \rrbr$ code.

\begin{figure}
\begin{subfigure}[h]{0.45\textwidth}
\centering

\includegraphics[scale=0.8,keepaspectratio]{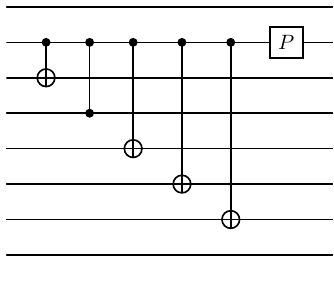}

\caption{\label{fig:XX_XYXZXXX}$\overline{X_1} \mapsto X_{1}Y_{2}X_{3}Z_{4}X_{5}X_{6}X_{7}$}
\end{subfigure}%
\begin{subfigure}[h]{0.65\textwidth}
\centering

\includegraphics[scale=0.8,keepaspectratio]{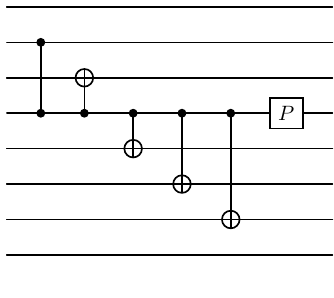}

\caption{\label{fig:XIIX_XZXYXXX}$\overline{X_3} \mapsto X_{1}Z_{2}X_{3}Y_{4}X_{5}X_{6}X_{7}$}
\end{subfigure}

\caption{\label{fig:local_circuit_X_862}Rooted circuit construction for logical-$X$ constraints.}
\end{figure}


As previously discussed in the $\llbr 6,4,2 \rrbr$ code scenario, when encountering an $X$ gate in the output of a logical-$Z$ constraint, a CZ gate is applied followed by $H$, instead of implementing a single CNOT gate. 
It is important to note that this results in a distinctive $H$-$P$-CZ-$H$ ``special'' structure, which may include multiple CZ gates (depending upon the number of $X$ terms in the output of the constraint).
Towards the end of these rooted circuits, CNOT gates are introduced, which would have already appeared in previous logical-$X$ rooted circuits.

\begin{figure}
\centering

\begin{subfigure}[h]{0.45\textwidth}
\centering

\includegraphics[scale=0.8,keepaspectratio]{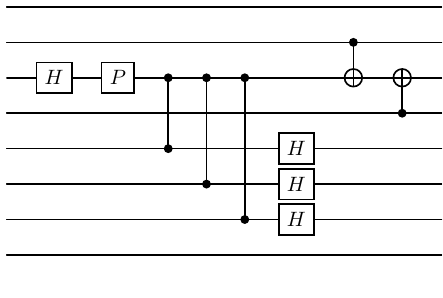}

\caption{\label{fig:IIZIIIIZ_IZYZXXXZ}$Z_{3}{Z_{8}}\mapsto Z_{2}Y_{3}Z_{4}X_{5}X_{6}X_{7}Z_{8}$ (ignoring sign)}
\end{subfigure}
\hfill

\begin{subfigure}[h]{0.45\textwidth}
\centering

\includegraphics[scale=0.8,keepaspectratio]{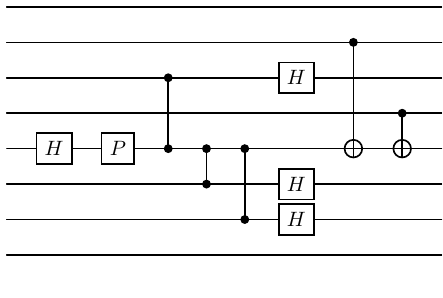}

\caption{\label{fig:IIIIZIIZ_IZXZYXXZ}$Z_{5}{Z_{8}}\mapsto Z_{2}X_{3}Z_{4}Y_{5}X_{6}X_{7}Z_{8}$ (ignoring sign)}
\end{subfigure}


\begin{subfigure}[h]{0.45\textwidth}
\centering

\includegraphics[scale=0.8,keepaspectratio]{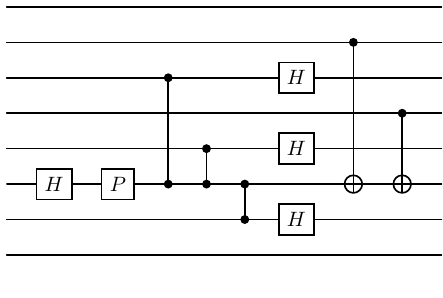}

\caption{\label{fig:IIIIIZIZ_IZXZXYXZ}$Z_{6}{Z_{8}}\mapsto Z_{2}X_{3}Z_{4}X_{5}Y_{6}X_{7}Z_{8}$ (ignoring sign)}
\end{subfigure}
\hfill

\begin{subfigure}[h]{0.45\textwidth}
\centering

\includegraphics[scale=0.8,keepaspectratio]{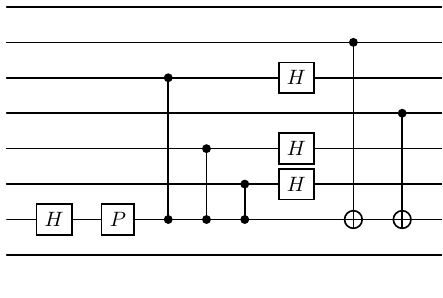}

\caption{\label{fig:IIIIIIZZ_IZXZXXXZ}$Z_{7}{Z_{8}}\mapsto Z_{2}X_{3}Z_{4}X_{5}X_{6}Y_{7}Z_{8}$ (ignoring sign)}
\end{subfigure}

\caption{\label{fig:local_circuit_Z_862} Rooted circuit construction for logical-$Z$ constraints.}

\end{figure}

\begin{figure}
\centering
\includegraphics[scale=0.73,keepaspectratio]{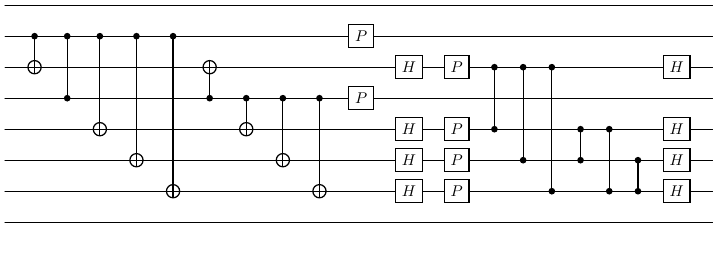}


\caption{\label{fig:solve-stitch_862_qsk_evenH}Physical realization of Fig.~\ref{fig:qsk_6qubit_evenH_many} on the $\llbr 8,6,2 \rrbr$ code produced by the solve-and-stitch approach using root qubits. Signs for $\overline{Z_i}$ can be fixed with Pauli gates at the end.}


\end{figure}

Fig.~\ref{fig:solve-stitch_862_qsk_evenH} illustrates the construction of the complete physical circuit on $\llbr 8,6,2 \rrbr$ code using the solve-and-stitch approach. 
As before, when incorporating new rooted circuits, it is crucial to ensure that these circuits do not duplicate existing gates in the physical circuit. 
It can be readily verified that the complete circuit satisfies all logical Pauli constraints and preserves the stabilizer group.
Of course, further simplifications are possible.

\section{Proofs for all results}

\subsection{Proof of Theoerem~\ref{thm:transversal_H}}
\label{sec:transversal_H_proof}

\begin{IEEEproof}
\normalfont
For convenience, we focus on the 3-qubit logical C-QSK circuit in Fig.~\ref{fig:qsk_3qubit} (with a fixed angle $\theta = \frac{\pi}{2}$), but the proof strategy generalizes to C-QSK circuits comprising an arbitrary number of qubits and combinations of $H$ and $H_y$.
According to Section~\ref{sec:qsk}, the mapping rules for Fig.~\ref{fig:qsk_3qubit} are:
\begin{IEEEeqnarray}{rClCrCl}
\label{eq:logical_qsk_3qubit_constraints}
\overline{X_1} & \mapsto & \overline{Y_1} \, \overline{X_2} \, \overline{Z_3} & \quad , \quad & \overline{Z_1} & \mapsto & \overline{Z_1} \ , \nonumber \\
\overline{X_2} & \mapsto & \overline{X_2} & \quad , \quad & \overline{Z_2} & \mapsto & - \overline{Z_1} \, \overline{Y_2} \, \overline{Z_3} \ , \nonumber \\
\overline{X_3} & \mapsto & \overline{Z_1} \, \overline{X_2} \, \overline{Y_3} & \quad , \quad & \overline{Z_3} & \mapsto & \overline{Z_3} \ .
\end{IEEEeqnarray}
To make the analysis easier, we can represent the logical Pauli operators as binary vectors according to Section~\ref{sec:pauli}.
Let $x_i,z_i,a_i,b_i \in \{ 0,1 \}^{n}$ for $i=1,2,3$.
Then the logical Pauli operators can be expressed as follows:
\begin{IEEEeqnarray}{rClCrCl}
\overline{X_{1}} & \equiv & E(x_{1},z_{1}) & \quad , \quad & \overline{Z_{1}} & \equiv & E(a_{1},b_{1}) \ , \nonumber \\
\overline{X_{2}} & \equiv & E(x_{2},z_{2}) & \quad , \quad & \overline{Z_{2}} & \equiv & E(a_{2},b_{2}) \ , \nonumber \\
\overline{X_{3}} & \equiv & E(x_{3},z_{3}) & \quad , \quad & \overline{Z_{3}} & \equiv & E(a_{3},b_{3}) \ .
\end{IEEEeqnarray}
The mapping rules of a single Hadamard gate are $HXH^\dagger = Z, HZH^\dagger = X$. 
Thus, the general mapping rule of transversal Hadamard on $n$ qubits can be expressed as follows:
\begin{align}
H^{\otimes n} E(x,z)  (H^{\otimes n})^\dagger &= E(z,x). \label{eq:1}
\end{align}
Assume that transversal Hadamard realizes the 3-qubit logical C-QSK circuit on a stabilizer code with the above logical Pauli operators. 
Then it must satisfy the mappings in Eq.~\eqref{eq:logical_qsk_3qubit_constraints}:
\begin{align}
& H^{\otimes n} \overline{X_{1}} (H^{\otimes n})^\dagger \nonumber \\
  &= E(z_1,x_1) \\
  &= \overline{Y_{1}} \, \overline{X_{2}} \, \overline{Z_{3}} \\
  & \propto E(x_1 \oplus a_1 \oplus x_2 \oplus a_3,z_1 \oplus b_1 \oplus z_2 \oplus b_3).
\end{align}
Hence, we obtain the following relations from $\overline{X_1}$:
\begin{align}
\overline{X_1} \colon z_1 &= x_1 \oplus a_1 \oplus x_2 \oplus a_3, \label{eq:3}\\
                      x_1 &= z_1 \oplus b_1 \oplus z_2 \oplus b_3. \label{eq:4}
\end{align}
Similarly, from the other constraints in Eq.~\eqref{eq:logical_qsk_3qubit_constraints} we obtain:
\begin{align}
\overline{Z_1} \colon a_1 &= b_1, \\ 
\overline{X_2} \colon x_2 &= z_2, \\
\overline{Z_2} \colon b_2 &= a_1 \oplus x_2 \oplus a_2 \oplus a_3, \\
                      a_2 &= b_1 \oplus z_2 \oplus b_2 \oplus b_3, \\
\overline{X_3} \colon z_3 &= a_1 \oplus x_2 \oplus x_3 \oplus a_3, \\
                      x_3 &= b_1 \oplus z_2 \oplus z_3 \oplus b_3, \\
\overline{Z_3} \colon a_3 &= b_3.
\end{align}
Combining the relations on $z_1$ and $b_2$ we get
\begin{align}
a_2 \oplus b_2 &= x_1 \oplus z_1. \label{eq:12}
\end{align}
Similarly, combining the relations on $a_2$ and $x_3$ we have
\begin{align}
x_3 \oplus z_3 &= a_2 \oplus b_2. \label{eq:13}
\end{align}
Thus, we have the chain of relations
\begin{align}
x_1 \oplus z_1 = a_2 \oplus b_2 = x_3 \oplus z_3. \label{eq:14}
\end{align}
It can be shown that two Hermitian Pauli operators satisfy
\begin{align}
E(a,b) E(c,d) &= (-1)^{\langle[a,b],[c,d]\rangle_{s}} E(c,d) E(a,b), \label{eq:15}
\end{align}
where $\langle[a,b],[c,d]\rangle_{s} \coloneqq ad^{T} \oplus bc^{T}$ is called the \emph{symplectic inner product}.
Therefore, two Pauli matrices $E(a,b)$, $E(c,d)$ commute if and only if $\langle[a,b],[c,d]\rangle_{s} = 0$. 
We know that logical-$X$ and $Z$ operators must anti-commute on the same logical qubit and must commute on different logical qubits, i.e., $\overline{X_i} \, \overline{Z_j} = (-1)^{\delta_{ij}} \overline{Z_j} \, \overline{X_i}$, where $\delta_{ij}$ is the Kronecker delta function. 
Thus, we have the following constraints:
\begin{align}
x_1  b_1^{T} \oplus z_1  a_1^{T} &= 1, \label{eq:17} \\
x_1  b_3^{T} \oplus z_1  a_3^{T} &= 0, \label{eq:18}\\
x_1  z_2^{T} \oplus z_1  x_2^{T} &= 0, \label{eq:19}\\
a_1  b_2^{T} \oplus b_1  a_2^{T} &= 0. \label{eq:20}
\end{align}
From previously established relations we know that $a_1=b_1, x_2=z_2, a_3=b_3$, and $x_1 \oplus z_1 = a_2 \oplus b_2 = x_3 \oplus z_3$. 
Thus, the last constraint above can be rewritten as follows:
\begin{align}
a_1 b_2^{T} \oplus b_1  a_2^{T} &= b_1  b_2^{T} \oplus b_1 a_2^{T} \\
  &= b_1 (a_2^{T} \oplus b_2^{T}) \\
  &= b_1 (x_1^{T} \oplus z_1^{T}) \\
  &= b_1 x_1^{T} \oplus b_1 z_1^{T} \\
  &= x_1 b_1^{T} \oplus z_1 a_1^{T} \\
  &= 1, 
\end{align}
which is a contradiction. 
Therefore, it is impossible to implement the $3$-qubit logical C-QSK circuit in Fig.~\ref{fig:qsk_3qubit} using a transversal Hadamard on any stabilizer code.
\end{IEEEproof}

\subsection{Proof of Theorem~\ref{thm:transversal_P}}
\label{sec:transversal_P_proof}
\begin{IEEEproof}
The mapping rules of a single Phase gate are ${PXP^\dagger} = Y, PZP^\dagger = Z$. 
Thus, the general mapping rule of transversal Phase on $n$ qubits can be expressed as follows:
\begin{align}
P^{\otimes n} E(x,z)  (P^{\otimes n})^\dagger &= E(x,x \oplus z). 
\end{align}
Assume that transversal Phase realizes the 3-qubit logical C-QSK circuit in Fig.~\ref{fig:qsk_3qubit} on a stabilizer code with logical $X$ (resp. $Z$) operators $E(x_i,z_i)$ (resp. $E(a_i,b_i)$).
Then it must satisfy:
\begin{IEEEeqnarray}{rClCrCl}
\label{eq:logical_qsk_3qubit_constraints_2}
\overline{X_1} & \mapsto & \overline{Y_1} \, \overline{X_2} \, \overline{Z_3} & \quad , \quad & \overline{Z_1} & \mapsto & \overline{Z_1} \ , \nonumber \\
\overline{X_2} & \mapsto & \overline{X_2} & \quad , \quad & \overline{Z_2} & \mapsto & - \overline{Z_1} \, \overline{Y_2} \, \overline{Z_3} \ , \nonumber \\
\overline{X_3} & \mapsto & \overline{Z_1} \, \overline{X_2} \, \overline{Y_3} & \quad , \quad & \overline{Z_3} & \mapsto & \overline{Z_3} \ .
\end{IEEEeqnarray}
For trivial mappings, we have relations such as
\begin{align}
P^{\otimes n} \overline{Z_{i}} (P^{\otimes n})^\dagger &= E(a_i,a_i \oplus b_i) = \overline{Z_{i}} \propto E(a_i,b_i).
\end{align}
Hence, we obtain the following constraints:
\begin{align}
\overline{Z_1} \colon a_1 =\textbf{0}, \ 
\overline{X_2} \colon x_2 =\textbf{0}, \ 
\overline{Z_3} \colon a_3 =\textbf{0}.
\end{align}
For non-trivial constraints such as $\overline{X_{1}}$, we have 
\begin{align}
& P^{\otimes n} \overline{X_{1}} (P^{\otimes n})^\dagger \nonumber \\
  &= E(x_1,x_1 \oplus z_1) \\
  &= \overline{Y_{1}} \, \overline{X_{2}} \, \overline{Z_{3}} \\
  & \propto E(x_1 \oplus a_1 \oplus x_2 \oplus a_3,z_1 \oplus b_1 \oplus z_2 \oplus b_3).\\
  & = E(x_1,z_1 \oplus b_1 \oplus z_2 \oplus b_3).
\end{align}
Hence, we obtain the following constraints from $\overline{X_1}, \overline{Z_2}, \overline{X_3}$:
\begin{align}
\overline{X_1} \colon x_1 &= b_1 \oplus z_2 \oplus b_3, \\
\overline{Z_2} \colon a_2 &= b_1 \oplus z_2 \oplus b_3, \\
\overline{X_3} \colon x_3 &= b_1 \oplus z_2 \oplus b_3.
\end{align}
Combining the relations on $a_1$, $x_2$ and $a_3$, we get
\begin{align}
a_1 = x_2 =a_3 =\textbf{0}.
\end{align}
Similarly, combining the relations on  $x_1$, $a_2$ and $x_3$ we have
\begin{align}
x_1 = a_2 =x_3 = b_1 \oplus z_2 \oplus b_3. \label{eq:83}
\end{align}
Since $\overline{X_{1}}$, $\overline{Z_{1}}$ anti-commute and $\overline{Z_{1}}$, $\overline{Z_{2}}$ commute, we have
\begin{align}
x_1  b_1^{T} \oplus z_1  a_1^{T} = x_1  b_1^{T} = 1, \label{eq:84}\\
a_1  b_2^{T} \oplus b_1  a_2^{T} = b_1  a_2^{T} = 0. \label{eq:85}
\end{align}
From Eqns.~\eqref{eq:83} and~\eqref{eq:85}, we can rewrite Eqn.~\eqref{eq:85} as
\begin{align}
b_1 a_2^{T} = b_1  x_1^{T} = x_1 b_1^{T} = 1,
\end{align}
which is a contradiction to Eqn.~\eqref{eq:85}. 
Therefore, it is impossible to implement the $3$-qubit logical C-QSK circuit in Fig.~\ref{fig:qsk_3qubit} using a transversal Phase on any stabilizer code.
The proof can be generalized to C-QSK circuits of larger sizes.
\end{IEEEproof}

\subsection{Proof of Lemma~\ref{lem:qsk_logical_Pauli_mappings}}
\label{Proof of qsk_logical_Pauli_mapping}
\begin{IEEEproof}
\normalfont
Consider the general C-QSK circuit where the first and last layers have $H_i$ gates for $i \in I_h$, the middle layers start with $\text{CNOT}_{j \rightarrow k}$ gates for $j \in [k-1]$, then perform $P = R_z(\frac{\pi}{2})$ on qubit $k$, and finally execute the prior CNOTs again but in the reverse order (see Fig.~\ref{fig:qsk_4qubit_oddH} for $k=4$ where $I_h = \{ 1,2,3 \}$).

Consider $i \notin I_{h}$.
Then the mappings of $\overline{Z_{i}}$ are trivial because $\overline{Z_{i}}$ is on the control qubit of CNOT gates, except for qubit $k$. 
For the last qubit, the symmetric positioning of CNOT gates ensures that this trivial mapping still holds.
However, $\overline{X_{i}}$ propagates into $\overline{X_i} \, \overline{X_k}$ since CNOT maps $XI$ to $XX$. 
When $\overline{X_{k}}$ encounters the Phase gate, it transforms into $\overline{Y_{k}}$, so we have the propagated operator $\overline{X_i} \, \overline{Y_k}$. 
Since CNOT gates map $XY$ to $YZ$, the operator becomes $\overline{Y_{i}} \, \overline{Z_{k}}$ after the second $\text{CNOT}_{i \rightarrow k}$. 
For the remaining qubits $j$, CNOT gates map $IY$ to $ZY$, so $\overline{Y_k}$ propagates into $\overline{Z_{j}} \, \overline{Y_k}$.
At the final layer, if a qubit $j \neq i$ contains a Hadamard gate, i.e., $j \in I_h$, then the $\overline{Z_{j}}$ becomes $\overline{X_{j}}$. 
Otherwise, $\overline{Z_{j}}$ remains unchanged.

Next, consider $i \in I_{h}$.
In this case, the mappings of $\overline{X_{i}}$ are trivial as $\overline{X_{i}} \mapsto \overline{Z_{i}} \mapsto \overline{X_{i}}$ through the two Hadamard gates at the first and last layers.
In between these gates, the $\overline{Z_{i}}$ does not propagate through $\text{CNOT}_{i \rightarrow k}$.
However, for $\overline{Z_{i}}$, it first changes into $\overline{X_{i}}$ through $H$ and then propagates into the last qubit after encountering $\text{CNOT}_{i \rightarrow k}$. 
The subsequent mappings are analogous to the scenario where $i \notin I_{h}$, with the distinction that as qubit $i \in I_h$, the output gate of qubit $i$ becomes $-\overline{Y_{i}}$ because the final $H$ gate maps $Y$ to $-Y$.
\end{IEEEproof}

\subsection{Proof of Lemma~\ref{lem:logical_X_stitching}}
\label{Proof of logical_X_stitching}
\begin{IEEEproof}
\normalfont
The $X$-mappings listed in Corollary~\ref{cor:qsk_physical_Pauli_mappings_evenH} and Corollary~\ref{cor:qsk_physical_Pauli_mappings_oddH} identify the CNOT and CZ gates present in the rooted circuits.
Recall that $\text{CNOT}_{i+1 \rightarrow j+1}$ maps $X_{i+1} \mapsto X_{i+1} X_{j+1}, X_{j+1} \mapsto X_{j+1}$, and $\text{CZ}_{i+1, j+1}$ maps $X_{i+1} \mapsto X_{i+1} Z_{j+1}, X_{j+1} \mapsto Z_{i+1} X_{j+1}$.
For the $i$-th logical-$X$ constraint (rooted circuit), the root is qubit $i+1$.
We only need to consider the case $i \notin I_h$ because the case $i \in I_h$ does not need a rooted circuit (the logical-$X$ mapping is trivial).

The mapping for $X_1 X_{i+1}$ implies that the CNOT gates are $\text{CNOT}_{i+1 \rightarrow j+1}$ for $j \in I_h$.
Since the non-root qubits $j+1$ are only targets, we always have $X_1 X_{j+1} \mapsto X_1 X_{j+1}$, so this rooted circuit satisfies the (trivial) $X$-mappings for $j \in I_h$.

Next, the mapping for $X_1 X_{i+1}$ also implies that the CZ gates are $\text{CZ}_{i+1, j+1}$ for $j \notin (I_h \cup \{ i \})$.
This time, these indices $j$ also have a similar non-trivial mapping as the root $i+1$.
Since $j \notin I_h$, the qubit $j+1$ is never involved in any CNOT with $i+1$, so we only need to check that the CZ does not violate the mapping for $X_1 X_{j+1}$.
This is indeed the case because the CZ also appears in the rooted circuit with root $j+1$, since $i \notin ( I_h \cup \{ j \})$ there.
For the mapping in Eqn.~\eqref{eq:qsk_physical_Pauli_mappings_oddH_1X}, a $\text{CZ}_{i+1, n}$ is added, but there is no constraint involving qubit $X_n$ at the input of the mapping.
As there is no other gate involving qubits $i+1$ and $j+1$ for $j \notin I_h$, we conclude that the rooted circuit with root $i+1$ does not affect other Pauli mappings.

Finally, the $\text{CNOT}_{i+1 \rightarrow 1}$ gates (for Eqn.~\eqref{eq:qsk_physical_Pauli_mappings_oddH_1X}) and the $P_{i+1}$ gates never affect other mappings because the root qubits $i+1$ are unique.
Hence, we can simply concatenate the individual rooted circuits for logical-$X$ constraints and drop duplicate CZ gates to satisfy all such constraints simultaneously.
\end{IEEEproof}

\subsection{Proof of Lemma~\ref{lem:logical_Z_stitching}}
\label{Proof of logical_Z_stitching}
\begin{IEEEproof}
\normalfont
The proof is analogous to the above case of logical-$X$ constraints, but we will highlight some subtle differences.
We only need to consider $i \in I_h$ because $i \notin I_h$ has trivial $Z$-mappings.
For these rooted circuits, we first transform $Z_{i+1}$ to $Y_{i+1}$ through a $H$ and $P$ on the root $i+1$.
Subsequently, we perform $\text{CZ}_{i+1, j+1}$ for $j \in (I_h \setminus \{ i \})$ rather than $\text{CNOT}_{i+1 \rightarrow j+1}$ because the rooted circuit for this $j$ will require $\text{CNOT}_{j+1 \rightarrow i+1}$, thereby doubling the number of two-qubit gates.
Instead, we follow the $\text{CZ}_{i+1, j+1}$ with $H_{j+1}$ to produce $X_{j+1}$ (and similarly $H_{i+1}$ in the rooted circuit for $j$).
Since the other rooted circuit also transforms $Z_{j+1} \mapsto Y_{j+1}$ before this CZ, its mapping is not violated. 
The $H_{j+1}$ at the end only changes the sign of $Y_{j+1}$ to $-1$, which helps satisfy the sign component of the corresponding mapping.

The $\text{CNOT}_{n \rightarrow i+1}$ gates to cancel $Z_n$ and the $\text{CZ}_{i+1,1}$ followed by $H_1$ to produce $X_1$ in Eqn.~\eqref{eq:qsk_physical_Pauli_mappings_oddH_2Z} are harmless for other $Z$-mappings.
Finally, the $\text{CNOT}_{j+1 \rightarrow i+1}$ gates to produce $Z_{j+1}$ for $j \notin I_h$ (from $Y_{i+1}$), which preserve the trivial mappings for $Z_{j+1} Z_n$ as qubit $j+1$ is the control.
\end{IEEEproof}

\subsection{Proof of theorem~\ref{thm:logical_X_Z_stitching}}
\label{Proof of logical_X_Z_stitching}
\begin{IEEEproof}
\normalfont
First, let us show that appending the logical-$X$ stitched circuit before the logical-$Z$ stitched circuit doesn't affect the logical-$Z$ Pauli mappings.
In effect, we need to track $Z_{i+1} Z_n$ through the combined circuit.
From Remark~\ref{rem:logical_Z_stitching_CNOTs_first} we know that the CNOTs at the end of the logical-$Z$ stitched circuit can be moved to its front, i.e., at the middle of the logical-$X$ and logical-$Z$ stitched circuits.
But Remark~\ref{rem:logical_Z_stitching_CNOTs_first} also shows that they already appear in the logical-$X$ stitched circuit (at the beginning).
It is easily checked that these CNOTs commute with other gates in the logical-$X$ stitched circuit because their target qubits are never roots and hence are never the control qubits for any CNOT.
Since the CNOTs in the middle of the two stitched circuits serve their purpose, these duplicated CNOTs can be dropped.
This leaves the logical-$X$ stitched circuit with only CZ and $P$ gates, which propagate $Z_{i+1} Z_n$ as is into the unchanged logical-$Z$ stitched circuit.
The only exception is the $\text{CNOT}_{i+1 \rightarrow 1}$ gates when the number of Hadamards is odd, but these do not affect $Z_{i+1} Z_n$.
Thus, the logical-$Z$ mappings are preserved by the combined circuit.

Second, let us show that the logical-$Z$ stitched circuit leaves the logical-$X$ mappings unaffected.
For this argument, we will group the CNOTs in the middle of the stitched circuits with the logical-$X$ stitched circuit. 
This means that the remainder of the combined circuit only consists of the $H$-$P$-CZ-$H$ symmetric structure.
If the number of Hadamards is odd, then by Lemma~\ref{lem:logical_Z_stitching} this structure is preceded by the set of $\text{CNOT}_{n \rightarrow i+1}$ gates, where $i \in I_h$.
But Corollary~\ref{cor:qsk_physical_Pauli_mappings_oddH} implies that the Pauli term on qubit $i+1$ at the output of logical-$X$ mappings is always $X_{i+1}$ (see Eqns.~\eqref{eq:qsk_physical_Pauli_mappings_oddH_1X} and~\eqref{eq:qsk_physical_Pauli_mappings_oddH_2X}).
This term does not propagate through $\text{CNOT}_{n \rightarrow i+1}$ because it belongs to the target of the CNOT, so we only need to consider the ensuing symmetric structure.
However, Eqn.~\eqref{eq:qsk_physical_Pauli_mappings_evenH_2Z} and Lemma~\ref{lem:logical_Z_stitching} imply that this structure also involves qubits $i+1$ for $i \in I_h$. 
For odd number of Hadamards, Eqn.~\eqref{eq:qsk_physical_Pauli_mappings_oddH_2Z} implies that qubit $1$ is involved as well.
Note that Lemma~\ref{lem:logical_Z_stitching} only introduces $\text{CZ}_{i+1,1}$ and then a $H_1$.
But we must precede the CZ with a $H_1$ and $P_1$ in the combined circuit to ensure that Eqn.~\eqref{eq:qsk_physical_Pauli_mappings_oddH_2X} is satisfied.
In either case, by the arguments earlier, the $X$-mappings at the end of the logical-$X$ stitched circuit always have an $X$ term in these qubits.
This $X_1$ or $X_{i+1}$ comes out unchanged when propagated through the $H$-$P$-CZ-$H$ structure since $H$ maps $X \mapsto Z$, which commutes through $P$ and CZ, and maps back to $X$ after $H$.
Hence, the combined circuit also preserves the logical-$X$ mappings.

Finally, we must ensure that the combined circuit preserves the stabilizer group of the codes.
In fact, we will show the stricter result that the stabilizer generators $X_1 X_2 \cdots X_n$ and $Z_1 Z_2 \cdots Z_n$ commute individually through the combined circuit.
Assume that the number of Hadamards is even.
Then, from Eqns.~\eqref{eq:qsk_physical_Pauli_mappings_evenH_1X} and~\eqref{eq:qsk_physical_Pauli_mappings_evenH_2X}, we observe that multiplying all the mappings for $i \in [n-2]$ we obtain $X_2 X_2 \cdots X_{n-1}$ at the input.
By including $X_1$ and $X_n$, this produces the $X$-stabilizer generator.
Since qubits $1$ and $n$ are not involved in the combined circuit, $X_1$ and $X_n$ are left unchanged.
We are left to check that the result of the above multiplication also produces $X_2 X_2 \cdots X_{n-1}$ at the output of the mapping.
Since $|I_h|$ is even, the product of Eqn.~\eqref{eq:qsk_physical_Pauli_mappings_evenH_2X} for all $i \in I_h$ results in $\prod_{i \in I_h} X_{i+1}$.
So it suffices to show that the product of Eqn.~\eqref{eq:qsk_physical_Pauli_mappings_evenH_1X} for all $i \notin I_h$ results in $\prod_{i \notin I_h} X_{i+1}$.
We have
\begin{align}
& \prod_{i \notin I_h} X_{1} Y_{i+1} \left( \prod_{j \in I_{h}} X_{j+1} \right) \left( \prod_{j \in [n-2] \setminus (I_{h} \cup \{ i \})} Z_{j+1} \right) \\
& = \left[ X_1^{|I_h^c|} \prod_{j \in I_{h}} X_{j+1}^{|I_h^c|} \right] \prod_{i \notin I_h} \left[ Y_{i+1} \prod_{j \notin (I_{h} \cup \{ i \})} Z_{j+1} \right] \\
& = \prod_{i \notin I_h} \left[ Y_{i+1} \prod_{j \notin (I_{h} \cup \{ i \})} Z_{j+1} \right],
\end{align}
where $I_h^c = [n-2] \setminus I_h$ also has $|I_h^c|$ even.
There are $|I_h^c|$ terms in the product and each term has $|I_h^c|$ single-qubit Pauli components in the product.
Taking the product over all terms, we observe that the first Pauli component is of the form $YZZ \cdots Z = \imath X$, the second component is of the form $ZYZZ\cdots Z = -\imath X$, the third is of the form $ZZYZZ\cdots Z = \imath X$, and so on.
Hence, each $\imath$ pairs with a $-\imath$ to form $1$ and the final term is $\prod_{i \notin I_h} X_{i+1}$ as required.
This proves that $X_1 \cdots X_n$ is preserved. 
An analogous argument shows that $Z_1 \cdots Z_n$ is preserved too.

Next, assume that the number of Hadamards is odd.
Again, multiplying Eqns.~\eqref{eq:qsk_physical_Pauli_mappings_oddH_1X} and~\eqref{eq:qsk_physical_Pauli_mappings_oddH_2X} with $X_1$ and $X_n$ produces the stabilizer $X_1 \cdots X_n$ at the input.
However, the combined circuit does involve qubits $1$ and $n$, so we cannot ignore the propagation of $X_1$ and $X_n$.
For this argument, we will group the CNOTs in the middle with the logical-$X$ part of the combined circuit to form the logical-$X$ stitched circuit.
At the end of this stitched circuit, all logical-$X$ mappings are satisfied, so the product of Eqns.~\eqref{eq:qsk_physical_Pauli_mappings_oddH_1X} and~\eqref{eq:qsk_physical_Pauli_mappings_oddH_2X} for all $i \in [n-2]$ represents the propagation of $X_2 X_3 \cdots X_{n-1}$:
\begin{align}
& \left[ \prod_{i \in I_h} X_1 \right] \prod_{i \notin I_h} \left[ Y_{i+1} \prod_{j \in [n-2] \setminus (I_h \cup \{ i \})} Z_{j+1} \right] Z_n \\
& = X_1 \prod_{i \notin I_h} \left[ Y_{i+1} \prod_{j \notin (I_h \cup \{ i \})} Z_{j+1} \right] Z_n.
\end{align}
From a similar argument as in the case of even $|I_h|$, the middle product has terms $YZZ\cdots Z = Y, ZYZZ\cdots Z = -Y, ZZYZZ\cdots Z = Y$, and so on.
There will be an even number of negative signs, so the final term is $X_1 Z_n \prod_{i \notin I_h} Y_{i+1}$.
It is easy to check that $X_1$ commutes through the logical-$X$ stitched circuit, which together with the above calculation yields the mapping $X_1 X_2 \cdots X_{n-1} \mapsto Z_n \prod_{i \notin I_h} Y_{i+1}$.
But $X_n$ propagates into $X_n Z_{i+1}$ through each $\text{CZ}_{i+1,n}$ from Lemma~\ref{lem:logical_X_stitching}, resulting in $X_n \prod_{i \notin I_h} Z_{i+1}$.
Multiplying this with the above mapping produces 
\begin{align}
X_1 X_2 \cdots X_n & \mapsto (Z_n X_n) \prod_{i \notin I_h} ( Y_{i+1} Z_{i+1} ) \\
                   & = (\imath Y_n) \left( \pm\imath \prod_{i \notin I_h} X_{i+1} \right) \\
                   & = \pm Y_n \prod_{i \notin I_h} X_{i+1}.
\end{align}
We introduce a $P_n$ or $P_n^{\dagger}$ gate after the logical-$X$ stitching circuit to map $\pm Y_n \mapsto X_n$.
Then we add a $\text{CNOT}_{n \rightarrow 1}$ gate to map $X_n \prod_{i \notin I_h} X_{i+1} \mapsto X_1 X_n \prod_{i \notin I_h} X_{i+1}$.
It is easily verified that both these gates do not affect the logical-$X$ and logical-$Z$ mappings.
The remainder of the combined circuit only consists of $\text{CNOT}_{n \rightarrow i+1}$ for $i \in I_h$ (from Lemma~\ref{lem:logical_Z_stitching}) and the $H$-$P$-CZ-$H$ symmetric structure.
The $X_n$ propagates through the CNOTs to produce $X_n \prod_{i \in I_h} X_{i+1}$, which commutes through the structure.
The $\prod_{i \notin I_h} X_{i+1}$ term above is untouched by the CNOTs and the symmetric structure, so we have the desired mapping $X_1 \cdots X_n \mapsto X_1 \cdots X_n$.
Thus, the $X$-stabilizer generator is preserved.

For the $Z$-stabilizer generator, we take a similar approach.
The product of Eqns.~\eqref{eq:qsk_physical_Pauli_mappings_oddH_1Z} and~\eqref{eq:qsk_physical_Pauli_mappings_oddH_2Z} produces the mapping
\begin{align}
Z_2 Z_3 \cdots Z_{n-1} & \mapsto - X_1 Z_n \prod_{i \in I_h} \left[ Y_{i+1} \prod_{j \in I_h \setminus \{ i \} } X_{j+1} \right] \\
  & = - X_1 Z_n \prod_{i \in I_h} Y_{i+1}.
\end{align}
It is easy to see that $Z_n \mapsto Z_n$ in the combined circuit, so we can multiply this with the above to obtain $Z_2 \cdots Z_n \mapsto - X_1 \prod_{i \in I_h} Y_{i+1}$.
We must introduce $Z_1$ at the input to make the $Z$-stabilizer.
The $Z_1$ propagates into $Z_1 \prod_{i \notin I_h} Z_{i+1}$ via the $\text{CNOT}_{i+1 \rightarrow 1}$ gates from Lemma~\ref{lem:logical_X_stitching}, and the product $\prod_{i \notin I_h} Z_{i+1}$ is unaffected by the rest of the combined circuit.
Then $Z_1$ propagates into $Z_1 Z_n$ due to the $\text{CNOT}_{n \rightarrow 1}$ introduced above for the $X$-stabilizer.
Finally, through the symmetric structure, $Z_1 \mapsto Y_1$ after $H_1$ and $P_1$, then maps to $Y_1 \prod_{i \in I_h} Z_{i+1}$ after the $\text{CZ}_{i+1,1}$ gates, and finally becomes $- Y_1 \prod_{i \in I_h} X_{i+1}$ after the Hadamards.
Combining with the $Z_n \prod_{i \notin I_h} Z_{i+1}$ and multiplying with the mapping $Z_2 \cdots Z_n \mapsto - X_1 \prod_{i \in I_h} Y_{i+1}$ we obtain
\begin{align}
Z_1 Z_2 \cdots Z_n & \mapsto (Y_1 X_1) \prod_{i \in I_h} \left( X_{i+1} Y_{i+1} \right) \prod_{i \notin I_h} \left( Z_{i+1} \right) Z_n \\
  & = (-\imath Z) \prod_{i \in I_h} (\imath Z) \prod_{i \notin I_h} Z_{i+1} Z_n \\
  & = \pm Z_1 Z_2 \cdots Z_n.
\end{align}
A sign discrepancy can be fixed using a layer of Pauli gates, which can be found efficiently using the binary representation of Pauli operators, e.g., see~\cite{Rengaswamy-tqe20}.
Hence, the $Z$-stabilizer generator is also preserved, which completes the proof.
\end{IEEEproof}

\subsection{Proof of Theorem~\ref{thm:qsk_depth_distance2}}
\label{Proof of qsk_depth_distance2}
\begin{IEEEproof}
\normalfont
For the case of even $h$, we only need to consider the non-highlighted blocks and gates in Fig.~\ref{fig:qsk_distance2_physical_full}.
The $\text{CZ}_{i+1,j+1}$ gates for all $i,j \notin I_h$ have depth $\binom{k-h}{2}$.
Then the layer of $P_{i+1}$ gates for $i \notin I_h$ increases depth by $1$.
The $\text{CNOT}_{i+1 \rightarrow j+1}$ gates for $i \notin I_h, j \in I_h$ have depth $(k-h) h$.
The layers of $H_{i+1}$ and $P_{i+1}$ gates for $i \in I_h$ increase depth by $2$.
The $\text{CZ}_{i+1,j+1}$ gates for all $i,j \in I_h$ have depth $\binom{h}{2}$.
The final layer of $H_{i+1}$ gates for $i \in I_h$ adds $1$ to the depth.
Hence, 
\begin{align}
\text{Depth} &= \binom{k-h}{2} + 1 + (k-h) h + 2 + \binom{h}{2} + 1 \\
  &= \frac{(k-h) (k-h-1) + 2h (k-h) + h(h-1)}{2} + 4 \\
  &= \frac{(k-h) (k+h-1) + h(h-1)}{2} + 4 \\
  &= \frac{k^2 + kh - k -hk - h^2 + h + h^2 - h}{2} +4 \\
  &= \frac{k(k-1)}{2} + 4. 
\end{align}
For the case of odd $h$, we need to include all the highlighted blocks and gates in Fig.~\ref{fig:qsk_distance2_physical_full}.
The $\text{CZ}_{i+1,n}$ gates for all $i \notin I_h$ add depth of $k-h$.
The $P_n$ or $P_n^{\dagger}$ can be included in the layer of $P_{i+1}$ gates for $i \notin I_h$.
The $\text{CNOT}_{n \rightarrow 1}$ and $\text{CNOT}_{i+1 \rightarrow 1}$ gates for all $i \notin I_h$ add depth of $k-h+1$.
The $\text{CNOT}_{n \rightarrow i+1}$ gates for $i \in I_d$ add depth of $h$.
The $H_1$ and $P_1$ can be included in the layers of $H_{i+1}$ and $P_{i+1}$ gates for $i \in I_h$.
The $\text{CZ}_{i+1,1}$ gates for all $i \in I_h$ add depth of $h$.
The final $H_1$ can be included in the layer of $H_{i+1}$ gates for $i \in I_h$.
The new additions introduce a combined depth of 
\begin{align}
(k-h) + (k-h+1) + h + h = 2k + 1.
\end{align}
Hence, the total depth of the complete physical circuit is 
\begin{align}
\frac{k(k-1)}{2} + 4 + \frac{4k+2}{2} = \frac{(k+2)(k+1)}{2} + 4.
\end{align}
By the proof of Theorem~\ref{thm:logical_X_Z_stitching}, note that a final layer of Pauli gates might be necessary to fix signs, which can increase the depth by $1$.
This applies to both even and odd $h$.
\end{IEEEproof}

\subsection{Proof of Theorem~\ref{thm:qsk_depth_distance2_logical_id}}
\label{Proof of qsk_depth_distance2_logical_id}
\begin{IEEEproof}
\normalfont
Consider the case when $h$ is even.
The logical identity gadget has $\binom{n}{2} = \binom{k+2}{2} = \frac{(k+2)(k+1)}{2}$ CZ gates. 
The circuit in Fig.~\ref{fig:qsk_distance2_physical_full} begins with $\binom{k-h}{2} = \frac{(k-h)(k-h-1)}{2}$ CZ gates, which can now be canceled.
The $P_{i+1}$ gates are also canceled, but the remaining $P$ gates from the gadget still contribute $1$ to the depth. 
Hence, the number of remaining CZ gates becomes 
\begin{align}
& \frac{k^2 + 3k + 2}{2}-\frac{(k^2 + h^2 - 2kh) -(k-h)}{2} \nonumber \\
&= \frac{4k + 2 - (1-2k) h - h^2}{2}. 
\end{align}
Therefore, the total depth of the optimized physical circuit is
\begin{align}
& \frac{4k + 2 - (1 - 2k) h - h^2}{2} + (k-h) h + \frac{h^2 - h}{2} + 5 \nonumber \\
  &= \frac{2hk - 2h^2 + 4k + 2 - h + 2kh - h^2 + h^2 - h}{2} + 5 \\ 
  &= \frac{4hk - 2h^2 + 4k + 2 - 2h}{2} + 5 \\
  &= 2hk - h^2 + 2k + 1 - h + 5 \\
  &= (2 + 2h) k - h^2 - h + 6.
\end{align}
Next, consider the case of odd $h$.
Beyond the CZ cancellations for even $h$, we observe from Fig.~\ref{fig:qsk_distance2_physical_full} that the $\text{CZ}_{i+1,n}$ gates for all $i \notin I_h$ can also be cancelled.
This reduces the optimized depth above for even $h$ by $k-h$.
Together with the remaining highlighted blocks and gates in Fig.~\ref{fig:qsk_distance2_physical_full} for odd $h$, the optimized depth can be calculated as
\begin{align}
& (2 + 2h)k - h^2 - h + 6 - (k - h) + (k - h + 1) + h + h \nonumber \\
&= (2 + 2h)k - h^2 + h + 7.
\end{align}
Hence, the difference in optimized depths is $2h+1$.
\end{IEEEproof}

\subsection{Proof of Corollary~\ref{cor:qsk_depth_distance2_logical_id_H}}
\label{Proof of qsk_depth_distance2_logical_id_H}
\begin{IEEEproof}
\normalfont
Let us fix $k$. 
To understand the effect of increasing $k$, let us replace the set $I_h$ with its complement $I_h^c = [n-2] \setminus I_h$ and calculate the new depth.
This corresponds to sandwiching the logical circuit between two layers of Hadamard gates on all $k$ qubits, which can be realized on the code as discussed above.
For even $h$, the new depth is given by
\begin{align}
& (2 + 2(k-h)) k - (k-h)^2 - (k-h) + 6 \nonumber \\
& = k^2 + k - h^2 + h + 6.
\end{align}
This depth is smaller than the expression in Theorem~\ref{thm:qsk_depth_distance2_logical_id} when
\begin{align}
k^2 + k - h^2 + h + 6 & < (2 + 2h) k - h^2 - h + 6 \\
\Rightarrow k^2 - k + 2h - 2hk & < 0 \\
\Rightarrow k(k-1) - 2h(k-1) & < 0 \\
\Rightarrow (k-1) (k-2h) & < 0 \\
\Rightarrow h & > \frac{k}{2},
\end{align}
as claimed.
Similarly, for odd $h$, the new depth is 
\begin{align}
& (2 + 2(k-h)) k - (k-h)^2 + (k-h) + 7 \nonumber \\
& = k^2 + 3k - h^2 - h + 7.
\end{align}
Once again, this depth is smaller than Theorem~\ref{thm:qsk_depth_distance2_logical_id} when
\begin{align}
k^2 + 3k - h^2 - h + 7 & < (2 + 2h) k - h^2 + h + 7 \\
\Rightarrow k^2 + k - 2hk - 2h & < 0 \\
\Rightarrow k(k+1) - 2h(k+1) & < 0 \\
\Rightarrow (k+1) (k-2h) & < 0 \\
\Rightarrow h & > \frac{k}{2},
\end{align}
leading us to the same conclusion.
In both cases, the layer of physical transversal Hadamard and swap of qubits $1$ and $n$ adds $3$ to the depth, $2$ at the beginning of the complete physical circuit and $1$ at the end due to the existing layer of Hadamards in Fig.~\ref{fig:qsk_distance2_physical_full}.
This completes the proof.
\end{IEEEproof}

\section{Examples of physical implementation of logical C-QSK with mixed Hadamard and $H_y$ gates} 
\label{sec:Hy examples}

This section provides examples for the physical implementation of logical C-QSK circuits with mixed Hadamard and $H_y$ gates according to different parities of them. 
The complete procedure is provided in Algorithm~\ref{alg:solve-and-stitch-Hy}.

\subsection{Odd Hadamard, Odd $H_y$ on logical C-QSK}
\label{subsec:odd H, odd Hy}

\begin{figure}
\centering

\includegraphics[scale=1,keepaspectratio]{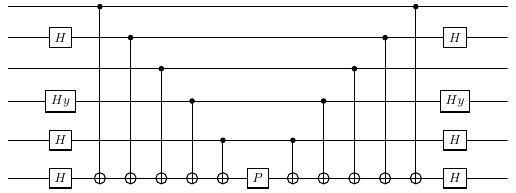}
\caption{\label{fig:odd_H_odd_Hy_logical}C-QSK circuit with odd number of Hadamard gates and odd number of $H_y$ gates.}
\end{figure}

This C-QSK circuit in Fig.~\ref{fig:odd_H_odd_Hy_logical} dictates the following input-output mappings of logical Pauli operators:
\begin{IEEEeqnarray}{rClCrCl}
\label{eq:odd_H_odd_Hy_logical}
\overline{X_1} & \mapsto & \overline{Y_1} \, \overline{X_2} \, \overline{Z_3} \, \overline{Y_4} \, \overline{X_5} \, \overline{X_6} & \quad , \quad & \overline{Z_1} & \mapsto & \overline{Z_1} \ , \nonumber \\
\overline{X_2} & \mapsto & \overline{X_2} & \quad , \quad & \overline{Z_2} & \mapsto &  \overline{Z_1} \, \overline{Y_2} \, \overline{Z_3} \, \overline{Y_4} \, \overline{X_5} \, \overline{X_6} \ , \nonumber \\
\overline{X_3} & \mapsto & \overline{Z_1} \, \overline{X_2} \, \overline{Y_3} \, \overline{Y_4} \, \overline{X_5} \, \overline{X_6} & \quad , \quad & \overline{Z_3} & \mapsto & \overline{Z_3} \ , \nonumber \\
\overline{X_4} & \mapsto & -\overline{Z_1} \, \overline{X_2} \, \overline{Z_3} \, \overline{Z_4} \, \overline{X_5} \, \overline{X_6} & \quad , \quad & \overline{Z_4} & \mapsto & \- \overline{Z_1} \, \overline{X_2} \, \overline{Z_3} \, \overline{Y_4} \, \overline{Y_5} \, \overline{X_6} \ , \nonumber \\ 
\overline{X_5} & \mapsto & \overline{X_5} & \quad , \quad & \overline{Z_5} & \mapsto & \overline{Z_1} \, \overline{X_2} \, \overline{Z_3} \, \overline{Y_4} \, \overline{Y_5} \, \overline{X_6} \ , \nonumber \\
\overline{X_6} & \mapsto & \overline{X_6} & \quad , \quad & \overline{Z_6} & \mapsto & \overline{Z_1} \, \overline{X_2} \, \overline{Z_3} \, \overline{Y_4} \, \overline{X_5} \, \overline{Y_6} \ .  
\end{IEEEeqnarray}

\begin{figure}
\centering

\includegraphics[scale=1,keepaspectratio]{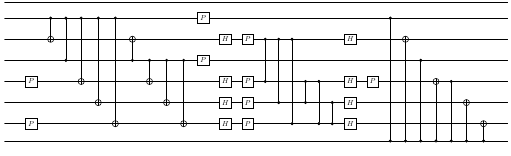}
\caption{\label{fig:odd_H_odd_Hy}Physical implementation of logical C-QSK circuit in Fig.~\ref{fig:odd_H_odd_Hy_logical}.}
\end{figure}

Substituting for the logical operators of the $\llbr 8,6,2 \rrbr$ code, we obtain the following mappings of physical Pauli operators:
{\small
\begin{IEEEeqnarray}{rClCrCl}
\label{eq:odd_H_odd_Hy}
X_1 X_2 & \mapsto & X_1 Y_2 X_3 Z_4 Y_5 X_6 X_7 Z_8 & \quad , \quad & Z_2 Z_8 & \mapsto & Z_2 Z_8 \ , \nonumber \\
X_1 X_3 & \mapsto & X_1 X_3 & \quad , \quad & Z_3 Z_8 & \mapsto & Z_2 Y_3 Z_4 Y_5 X_6 X_7 \ , \nonumber \\
X_1 X_4 & \mapsto & X_1 Z_2 X_3 Y_4 Y_5 X_6 X_7 Z_8 & \quad , \quad & Z_4 Z_8 & \mapsto & Z_4 Z_8 \ , \nonumber \\
X_1 X_5 & \mapsto & X_1 Z_2 X_3 Z_4 Z_5 X_6 X_7 Z_8 & \quad , \quad & Z_5 Z_8 & \mapsto & Z_2 X_3 Z_4 X_5 X_6 X_7 \ , \nonumber \\
X_1 X_6 & \mapsto & X_1 X_6 & \quad , \quad & Z_6 Z_8 & \mapsto & Z_2 X_3 Z_4 Y_5 Y_6 X_7 \ , \nonumber \\
X_1 X_7 & \mapsto & X_1 X_7 & \quad , \quad & Z_7 Z_8 & \mapsto & Z_2 X_3 Z_4 Y_5 X_6 Y_7\ . \nonumber
\end{IEEEeqnarray}
}
Similar to the even-even case in the main text, we replace all $H_y$ gates into Hadamard gates in logical circuit, run Algorithm~\ref{alg:solve-and-stitch}, add $P_5$ at the beginning and the end of the physical circuit, then apply $\text{CZ}_{8\rightarrow2}$, $\text{CNOT}_{8\rightarrow3}$, $\text{CNOT}_{8\rightarrow6}$, $\text{CNOT}_{8\rightarrow7}$. Finally, we apply $\text{CNOT}_{8\rightarrow5}$ and $\text{CZ}_{8\rightarrow5}$ simultaneously since $H_y$ gate appears in the forth qubit in Fig.~\ref{fig:odd_H_odd_Hy_logical}. The full physical circuit is displayed in Fig.~\ref{fig:odd_H_odd_Hy}.

\subsection{Odd Hadamard, Even $H_y$ on logical C-QSK}
\label{subsec:Odd H, Even Hy}

\begin{figure}
\centering

\includegraphics[scale=1,keepaspectratio]{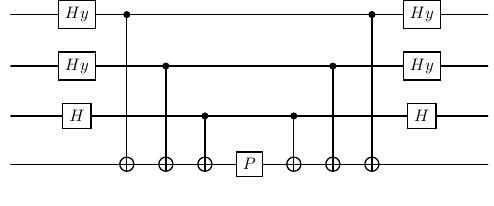}
\caption{\label{fig:odd_H_even_Hy_logical}C-QSK circuit with odd number of Hadamard gates and even number of $H_y$ gates.}
\end{figure}

This C-QSK circuit in Fig.~\ref{fig:odd_H_even_Hy_logical} dictates the following input-output mappings of logical Pauli operators:
\begin{IEEEeqnarray}{rClCrCl}
\label{eq:odd_H_even_Hy_logical}
\overline{X_1} & \mapsto & -\overline{Z_1} \, \overline{Y_2} \, \overline{X_3} \, \overline{Z_4}  & \quad , \quad & \overline{Z_1} & \mapsto & \overline{X_1} \, \overline{Y_2} \, \overline{X_3} \, \overline{Z_4} \ , \nonumber \\
\overline{X_2} & \mapsto & -\overline{Y_1} \, \overline{Z_2} \, \overline{X_3} \, \overline{Z_4} & \quad , \quad & \overline{Z_2} & \mapsto &  \overline{Y_1} \, \overline{X_2} \, \overline{X_3} \, \overline{Z_4}  \ , \nonumber \\
\overline{X_3} & \mapsto & \overline{X_3}  & \quad , \quad & \overline{Z_3} & \mapsto & -\overline{Y_1} \, \overline{Y_2} \, \overline{Y_3} \, \overline{Z_4} \ , \nonumber \\
\overline{X_4} & \mapsto & \overline{Y_1} \, \overline{Y_2} \, \overline{X_3} \, \overline{Y_4} & \quad , \quad & \overline{Z_4} & \mapsto & \overline{Z_4} \ .  
\end{IEEEeqnarray}

\begin{figure}
\centering

\includegraphics[scale=1,keepaspectratio]{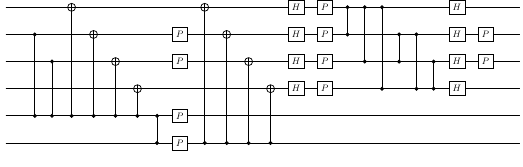}
\caption{\label{fig:odd_H_even_Hy}Physical implementation of logical C-QSK circuit in Fig.~\ref{fig:odd_H_even_Hy_logical}.}
\end{figure}

Substituting for the logical operators of the $\llbr 6,4,2 \rrbr$ code, we obtain the following mappings of physical Pauli operators:
\begin{IEEEeqnarray}{rClCrCl}
\label{eq:odd_H_even_Hy}
X_1 X_2 & \mapsto & Z_2 Y_3 X_4 Z_5 Z_6 & \quad , \quad & Z_2 Z_6 & \mapsto & -X_1 X_2 Y_3 X_4 Z_5 \ , \nonumber \\
X_1 X_3 & \mapsto & Y_2 Z_3 X_4 Z_5 Z_6 & \quad , \quad & Z_3 Z_6 & \mapsto & -X_1 Y_2 X_3 X_4 Z_5\ , \nonumber \\
X_1 X_4 & \mapsto & X_1 X_4 & \quad , \quad & Z_4 Z_6 & \mapsto & -X_1 Y_2 Y_3 Y_4 Z_5 \ , \nonumber \\
X_1 X_5 & \mapsto & Y_2 Y_3 X_4 Y_5 Z_6 & \quad , \quad & Z_5 Z_6 & \mapsto & Z_5 Z_6 \ . 
\end{IEEEeqnarray}

To get the physical circuit, again we replace all $H_y$ gates into Hadamard gates in logical circuit, run Algorithm~\ref{alg:solve-and-stitch}. Add $\text{CZ}_{52}$ and $\text{CZ}_{53}$ at the beginning,  then apply $P_{2}$ and $P_{3}$ at the position of original Phase gates in the physical circuit gained from Algorithm~\ref{alg:solve-and-stitch}, then duplicate these $P_{2}$ and $P_{3}$ at the end as well. The full physical circuit is displayed in Fig.~\ref{fig:odd_H_even_Hy}.

\subsection{Even Hadamard, Odd $H_y$ on logical C-QSK}
\label{subsec:Even H, Odd Hy}

\begin{figure}
\centering

\includegraphics[scale=1,keepaspectratio]{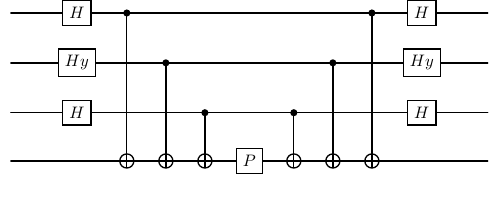}
\caption{\label{fig:even_H_odd_Hy_logical}C-QSK circuit with even number of Hadamard gates and odd number of $H_y$ gates.}
\end{figure}

This C-QSK circuit in Fig.~\ref{fig:even_H_odd_Hy_logical} dictates the following input-output mappings of logical Pauli operators:
\begin{IEEEeqnarray}{rClCrCl}
\label{eq:even_H_odd_Hy_logical}
\overline{X_1} & \mapsto & \overline{X_1}  & \quad , \quad & \overline{Z_1} & \mapsto & -\overline{Y_1} \, \overline{Y_2} \, \overline{X_3} \, \overline{Z_4} \ , \nonumber \\
\overline{X_2} & \mapsto & -\overline{X_1} \, \overline{Z_2} \, \overline{X_3} \, \overline{Z_4} & \quad , \quad & \overline{Z_2} & \mapsto &  \overline{X_1} \, \overline{X_2} \, \overline{X_3} \, \overline{Z_4}  \ , \nonumber \\
\overline{X_3} & \mapsto & \overline{X_3}  & \quad , \quad & \overline{Z_3} & \mapsto & -\overline{X_1} \, \overline{Y_2} \, \overline{Y_3} \, \overline{Z_4} \ , \nonumber \\
\overline{X_4} & \mapsto & \overline{X_1} \, \overline{Y_2} \, \overline{X_3} \, \overline{Y_4} & \quad , \quad & \overline{Z_4} & \mapsto & \overline{Z_4} \ . 
\end{IEEEeqnarray}

\begin{figure}
\centering

\includegraphics[scale=1,keepaspectratio]{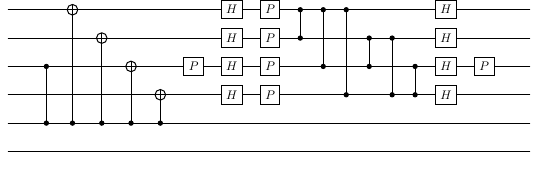}
\caption{\label{fig:even_H_odd_Hy}Physical implementation of logical C-QSK circuit in Fig.~\ref{fig:even_H_odd_Hy_logical}.}
\end{figure}

Substituting for the logical operators of the $\llbr 6,4,2 \rrbr$ code, we obtain the following mappings of physical Pauli operators:
\begin{IEEEeqnarray}{rClCrCl}
\label{eq:even_H_odd_Hy}
X_1 X_2 & \mapsto & X_1 X_2 & \quad , \quad & Z_2 Z_6 & \mapsto & X_1 Y_2 Y_3 X_4 Z_5 Z_6 \ , \nonumber \\
X_1 X_3 & \mapsto & X_2 Z_3 X_4 Z_5 & \quad , \quad & Z_3 Z_6 & \mapsto & X_1 X_2 X_3 X_4 Z_5 Z_6 \ , \nonumber \\
X_1 X_4 & \mapsto & X_1 X_4 & \quad , \quad & Z_4 Z_6 & \mapsto & X_1 X_2 Y_3 Y_4 Z_5 Z_6 \ , \nonumber \\
X_1 X_5 & \mapsto & -X_2 Y_3 X_4 Y_5 & \quad , \quad & Z_5 Z_6 & \mapsto & Z_5 Z_6 \ . 
\end{IEEEeqnarray}

For the physical circuit, after we replace all $H_y$ gates into Hadamard gates in logical circuit and run Algorithm~\ref{alg:solve-and-stitch}, we delete all CZ gates and Phase gates in logical-$X$ component of this physical circuit, then delete all the gates involving in the last qubit. Then apply $\text{CZ}_{53}$ at the beginning, then apply additional $P_3$ just before the logical-$Z$ component of the circuit, duplicate this $P_3$ at the end of the circuit as well. The final physical circuit in this scenario is shown in Fig.~\ref{fig:even_H_odd_Hy}.

\end{document}